\documentclass[%
 reprint, twocolumn,
 superscriptaddress,
 nofootinbib,
 amsmath,amssymb,
 aps, prd,
]{revtex4-2} 

\usepackage[table,svgnames,dvipsnames]{xcolor}
\usepackage[modulo,mathlines]{lineno}
\usepackage[normalem]{ulem}
\usepackage{aas_macros}
\usepackage{subfigure}
\usepackage{graphicx} 
\usepackage{graphics}
\usepackage{dcolumn} 
\usepackage{bm} 
\usepackage{cases} 
\usepackage{booktabs}
\usepackage{comment}
\usepackage{multirow}
\usepackage{makecell}
\usepackage{siunitx}
\usepackage{tabularx}
\usepackage{xspace}
\usepackage{soul} 
\graphicspath{{figures/}}
\usepackage{fontawesome}
\usepackage{hyperref} 
\hypersetup{colorlinks=true,citecolor=blue,filecolor=blue,urlcolor=blue,linkcolor=blue}
\usepackage{adjustbox}
\usepackage{orcidlink} 
\usepackage{hanging} 

\def\be{\begin{equation}}
\def\ee{\end{equation}}

\def\ba#1\ea{\begin{align*}#1\end{align*}}

\renewcommand{\emph}[1]{\textit{#1}}

\newcommand{\bgs}{{\tt BGS}}
\newcommand{\elgo}{{\tt ELG1}}
\newcommand{\elgt}{{\tt ELG2}}

\newcommand{\lrgo}{{\tt LRG1}}
\newcommand{\lrgt}{{\tt LRG2}}
\newcommand{\lrgth}{{\tt LRG3}}
\newcommand{\lrgs}{{\tt LRG}s}

\newcommand{\lrgelg}{{\tt LRG3$+$ELG1}}
\newcommand{\qso}{{\tt QSO}}

\newcommand{\ion}[2]{#1\thinspace{}#2\xspace}

\usepackage[nameinlink,noabbrev]{cleveref}
\crefname{equation}{Eq.}{Eqs.}
\crefname{section}{Section}{Sections}
\crefname{figure}{Figure}{Figures}
\crefname{table}{Table}{Tables}
\crefname{appendix}{Appendix}{Appendices}
\Crefname{figure}{Figure}{Figures}
\Crefname{equation}{Equation}{Equations}
\Crefname{section}{Section}{Sections}
\Crefname{table}{Table}{Tables}
\crefformat{equation}{\eqA eq.~\eqB #2#1#3)}
\newcommand{\eqA}{}
\newcommand{\eqB}{(}
\DeclareRobustCommand{\pcref}[1]{%
  \begingroup
  \renewcommand{\eqA}{(}\renewcommand{\eqB}{}%
  \cref{#1}%
  \endgroup
}

\newcommand{\mksym}[1]{\ifmmode {\rm #1}\else #1\fi}
\newcommand{\dataplus}{\allowbreak+}

\newcommand{\leftparbox}[2]{\parbox{#1}{\begin{flushleft} #2 \end{flushleft}}}
\newcommand{\oneonesig}[4][5cm]{
\begin{equation}
\left.
#2 \quad\mbox{\text{\leftparbox{#1}{(#3)#4}}}
  \right.
\end{equation}
}
\newcommand{\onetwosig}[4][5cm]{
\begin{equation}
\left.
  #2 \quad \text{(95\%,~#3)#4}
  \right.
\end{equation}
}

\newcommand{\twoonesig}[4][\pbwidth]{
\begin{equation}
\left.
 \begin{aligned}
#2 \\ #3
 \end{aligned}
\ \right\} \ \ \mbox{\text{\leftparbox{#1}{#4}}}
\end{equation}
}

\newcommand{\Om}{\Omega_\mathrm{m}}
\newcommand{\Ocdm}{\Omega_\mathrm{c}}
\newcommand{\Ob}{\Omega_\mathrm{b}}
\newcommand{\Ok}{\Omega_\mathrm{K}}

\newcommand{\Ode}{\Omega_\mathrm{DE}} 
\newcommand{\ob}{\omega_\mathrm{b}}
\newcommand{\obc}{\omega_\mathrm{bc}}
\newcommand{\ocdm}{\omega_\mathrm{c}}

\newcommand{\Neff}{N_{\mathrm{eff}}}
\newcommand{\lcdm}{$\Lambda$CDM} 
\newcommand{\wcdm}{$w$CDM} 
\newcommand{\wowacdm}{$w_0\wa$CDM} 
\newcommand{\lya}{Ly$\alpha$\xspace}

\newcommand{\aiso}{\alpha_\mathrm{iso}}
\newcommand{\aper}{\alpha_\perp}
\newcommand{\apar}{\alpha_{||}}
\newcommand{\DVrd}{D_\mathrm{V}/r_\mathrm{d}}
\newcommand{\DMrd}{D_\mathrm{M}/r_\mathrm{d}}
\newcommand{\DHrd}{D_\mathrm{H}/r_\mathrm{d}}
\newcommand{\DM}{D_\mathrm{M}}
\renewcommand{\DH}{D_\mathrm{H}}
\newcommand{\DV}{D_\mathrm{V}}
\newcommand{\DL}{D_\mathrm{L}}
\newcommand{\Hrd}{H_0r_\mathrm{d}}
\newcommand{\rd}{r_\mathrm{d}}

\newcommand{\sumnu}{\sum m_\nu}
\newcommand{\Planck}{\emph{Planck}}
\newcommand{\zpiv}{z_\mathrm{p}}
\newcommand{\wpiv}{w_\mathrm{p}}
\newcommand{\wa}{w_a}
\newcommand{\dchisq}{\Delta\chi^2_\mathrm{MAP}}

\newcommand{\hinvmpc}{\,h^{-1}{\rm Mpc}}

\newcommand{\kmsMpc}{\,{\rm km\,s^{-1}\,Mpc^{-1}}}
\newcommand{\eV}{{\,\rm eV}}

\begin{document}

\title{DESI DR2 Results II: Measurements of Baryon Acoustic Oscillations and Cosmological Constraints}


\author{M.~Abdul Karim\orcidlink{0009-0000-7133-142X}}
\affiliation{IRFU, CEA, Universit\'{e} Paris-Saclay, F-91191 Gif-sur-Yvette, France}

\author{J.~Aguilar}
\affiliation{Lawrence Berkeley National Laboratory, 1 Cyclotron Road, Berkeley, CA 94720, USA}

\author{S.~Ahlen\orcidlink{0000-0001-6098-7247}}
\affiliation{Physics Dept., Boston University, 590 Commonwealth Avenue, Boston, MA 02215, USA}

\author{S.~Alam\orcidlink{0000-0002-3757-6359}}
\affiliation{Tata Institute of Fundamental Research, Homi Bhabha Road, Mumbai 400005, India}

\author{L.~Allen}
\affiliation{NSF NOIRLab, 950 N. Cherry Ave., Tucson, AZ 85719, USA}

\author{C.~Allende~Prieto\orcidlink{0000-0002-0084-572X}}
\affiliation{Departamento de Astrof\'{\i}sica, Universidad de La Laguna (ULL), E-38206, La Laguna, Tenerife, Spain}
\affiliation{Instituto de Astrof\'{\i}sica de Canarias, C/ V\'{\i}a L\'{a}ctea, s/n, E-38205 La Laguna, Tenerife, Spain}

\author{O.~Alves}
\affiliation{Department of Physics, University of Michigan, 450 Church Street, Ann Arbor, MI 48109, USA}

\author{A.~Anand\orcidlink{0000-0003-2923-1585}}
\affiliation{Lawrence Berkeley National Laboratory, 1 Cyclotron Road, Berkeley, CA 94720, USA}

\author{U.~Andrade\orcidlink{0000-0002-4118-8236}}
\affiliation{Leinweber Center for Theoretical Physics, University of Michigan, 450 Church Street, Ann Arbor, Michigan 48109-1040, USA}
\affiliation{Department of Physics, University of Michigan, 450 Church Street, Ann Arbor, MI 48109, USA}

\author{E.~Armengaud\orcidlink{0000-0001-7600-5148}}
\affiliation{IRFU, CEA, Universit\'{e} Paris-Saclay, F-91191 Gif-sur-Yvette, France}

% ...

\collaboration{DESI Collaboration}

\email{spokespersons@desi.lbl.gov}

\begin{abstract}
    We present baryon acoustic oscillation (BAO) measurements from more than 14 million galaxies and quasars drawn from the Dark Energy Spectroscopic Instrument (DESI) Data Release 2 (DR2), based on three years of operation. For cosmology inference, these galaxy measurements are combined with DESI Lyman-$\alpha$ forest BAO results presented in a companion paper. The DR2 BAO results are consistent with DESI DR1 and SDSS, and their distance-redshift relationship matches those from recent compilations of supernovae (SNe) over the same redshift range. The results are well described by a flat $\Lambda$CDM model, but the parameters preferred by BAO are in mild, $2.3\sigma$ tension with those determined from the cosmic microwave background (CMB), although the DESI results are consistent with the acoustic angular scale $\theta_*$ that is well-measured by Planck. This tension is alleviated by dark energy with a time-evolving equation of state parametrized by $w_0$ and $w_a$, which provides a better fit to the data, with a favored solution in the quadrant with $w_0>-1$ and $w_a<0$. This solution is preferred over $\Lambda$CDM at $3.1\sigma$ for the combination of DESI BAO and CMB data. When also including SNe, the preference for a dynamical dark energy model over $\Lambda$CDM ranges from $2.8-4.2\sigma$ depending on which SNe sample is used. We present evidence from other data combinations which also favor the same behavior at high significance. From the combination of DESI and CMB we derive 95\% upper limits on the sum of neutrino masses, finding $\sum m_\nu<0.064$ eV assuming $\Lambda$CDM and $\sum m_\nu<0.16$ eV in the $w_0w_a$ model. Unless there is an unknown systematic error associated with one or more datasets, it is clear that $\Lambda$CDM is being challenged by the combination of DESI BAO with other measurements and that dynamical dark energy offers a possible solution.
\end{abstract}

\maketitle

\tableofcontents

\section{Introduction}
\label{sec:intro}

Cosmic acceleration remains the most pressing problem in contemporary cosmology, implying a pervasive new form of energy with exotic physical properties, or a breakdown of Einstein gravity on cosmological scales, or perhaps both.  To probe the physics of acceleration, cosmologists seek to measure the history of cosmic expansion and the history of gravitational clustering with the greatest achievable precision over a wide span of redshift.  Baryon acoustic oscillations (BAO) provide a powerful tool for measuring the expansion history, using a characteristic scale that is imprinted on matter clustering by pressure waves that propagate in the coupled baryon-photon fluid of the pre-recombination Universe \cite{Eisenstein1998,2003ApJ...594..665B,2003ApJ...598..720S}.  This paper examines the cosmological implications of the BAO measurements from the second data release (DR2) of the Dark Energy Spectroscopic Instrument (DESI, \cite{Snowmass2013.Levi,DESI2016b.Instr,DESI2022.KP1.Instr}), consisting of data from the first three years of operation.

Since the first clear detections of BAO in the Sloan Digital Sky Survey and the Two-Degree Field Galaxy Redshift Survey \cite{2005ApJ...633..560E,20052dFBAO}, BAO measurements have played a central role in observational cosmology.  The key technical requirement is a large-volume spectroscopic survey with sufficient sampling density, and previous surveys designed with BAO measurements as a defining goal include WiggleZ \cite{2011MNRAS.418.1707B}, the Baryon Oscillation Spectroscopic Survey (BOSS) \cite{2013AJ....145...10D} of SDSS-III \cite{2011AJ....142...72E}, and its extension eBOSS \cite{2016AJ....151...44D} in SDSS-IV \cite{2017AJ....154...28B}.  In addition to galaxy and quasar redshifts, BOSS and eBOSS measured BAO in the \lya\ forest absorption spectra of $z>2$ quasars, an approach first proposed by \cite{White2003,2007PhRvD..76f3009M}. Transverse BAO can also be measured in photometric surveys (e.g., \cite{DESBAO2024}), though the precision obtained for a given number of tracers is much higher with spectroscopic redshifts. DESI is designed specifically to enable a spectroscopic BAO survey of unprecedented power and efficiency \cite{DESI2016a.Science}, as shown in its survey validation \cite{DESI2023a.KP1.SV} based on the early data release \cite{DESI2023b.KP1.EDR}. In its first year of observations \cite{DESI2024.I.DR1}, DESI already achieved BAO measurements competitive with those of all previous surveys combined \cite{DESI2024.III.KP4,DESI2024.IV.KP6}. Now, with redshifts of more than 30 million galaxies and quasars, and \lya\ forest spectra of more than 820,000 quasars, the DESI DR2 sample is by far the largest spectroscopic galaxy sample to date.

The BAO technique (reviewed in \S 4 of \cite{Weinberg:2013agg}), which is key to DESI, exploits the enhancement of clustering at the scale of the pre-recombination sound horizon,
\begin{equation}
    \rd = \int_{z_{\rm d}}^\infty \frac{c_s(z)}{H(z)} dz~.
    \label{eqn:rd}
\end{equation}    
Here $c_s(z)$ is the speed of sound in the photon-baryon fluid, and $z_{\rm d} \approx 1060$ \cite{Planck-2018-cosmology} is the redshift at which acoustic waves stall because photons no longer `drag' the baryons. Assuming standard pre-recombination physics, the sound horizon can be computed given the densities of baryons, cold dark matter (CDM), photons, and other relativistic species \cite{Brieden:2022heh}, 
\begin{align}  
    \begin{split}
    \rd ~=~ & 147.05\,{\rm Mpc} \,\times \\
        & \left(\frac{\ob}{0.02236}\right)^{-0.13} 
           \left(\frac{\obc}{0.1432}\right)^{-0.23}         
        \left(\frac{N_{\rm eff}}{3.04}\right)^{-0.1}~. 
    \end{split}
    \label{eqn:rdformula}
\end{align}
\cref{eqn:rdformula} is scaled to the best-fit values from \Planck\ \cite{Planck-2018-cosmology} of $\ob \equiv \Ob h^2$ and $\obc \equiv (\Ob+\Ocdm) h^2$ and to the energy content of three neutrino species that are fully relativistic at $z>z_{\rm d}$. Here $h \equiv H_0/(100\kmsMpc)$ and $\Ob$ and $\Ocdm$ denote the present day fractional energy densities relative to critical in baryons and cold dark matter, respectively.  A measurement of the BAO scale in the transverse direction at redshift $z$ constrains the transverse comoving distance, which is given by
\begin{equation}
    \DM(z) = \frac{c}{H_0\sqrt{\Ok}}\, \sinh\left[\sqrt{\Ok}\int_0^z \frac{dz^\prime}{H(z^\prime)/H_0}\right]\,,
  \label{eqn:DM}
\end{equation}
where $\Ok$ is the curvature density parameter, which converges to the flat universe case
\begin{equation}
    \DM(z) =  \frac{c}{H_0} \int_0^z \frac{dz^\prime}{H(z^\prime)/H_0}  \qquad \hbox{(flat universe)}
\end{equation}
in the limit $|\Ok| \ll 1$.
A measurement in the line-of-sight direction constrains the expansion rate $H(z)$ or the corresponding distance,
\begin{equation}
    \DH(z) = \frac{c}{H(z)}.
    \label{eqn:DH}
\end{equation}
Because the inferred distances are relative to the sound horizon, the directly constrained quantities are the ratios $\DM/\rd$ and $\DH/\rd$. 

\begin{figure*}
    \begin{center}\includegraphics[width=0.99\textwidth]{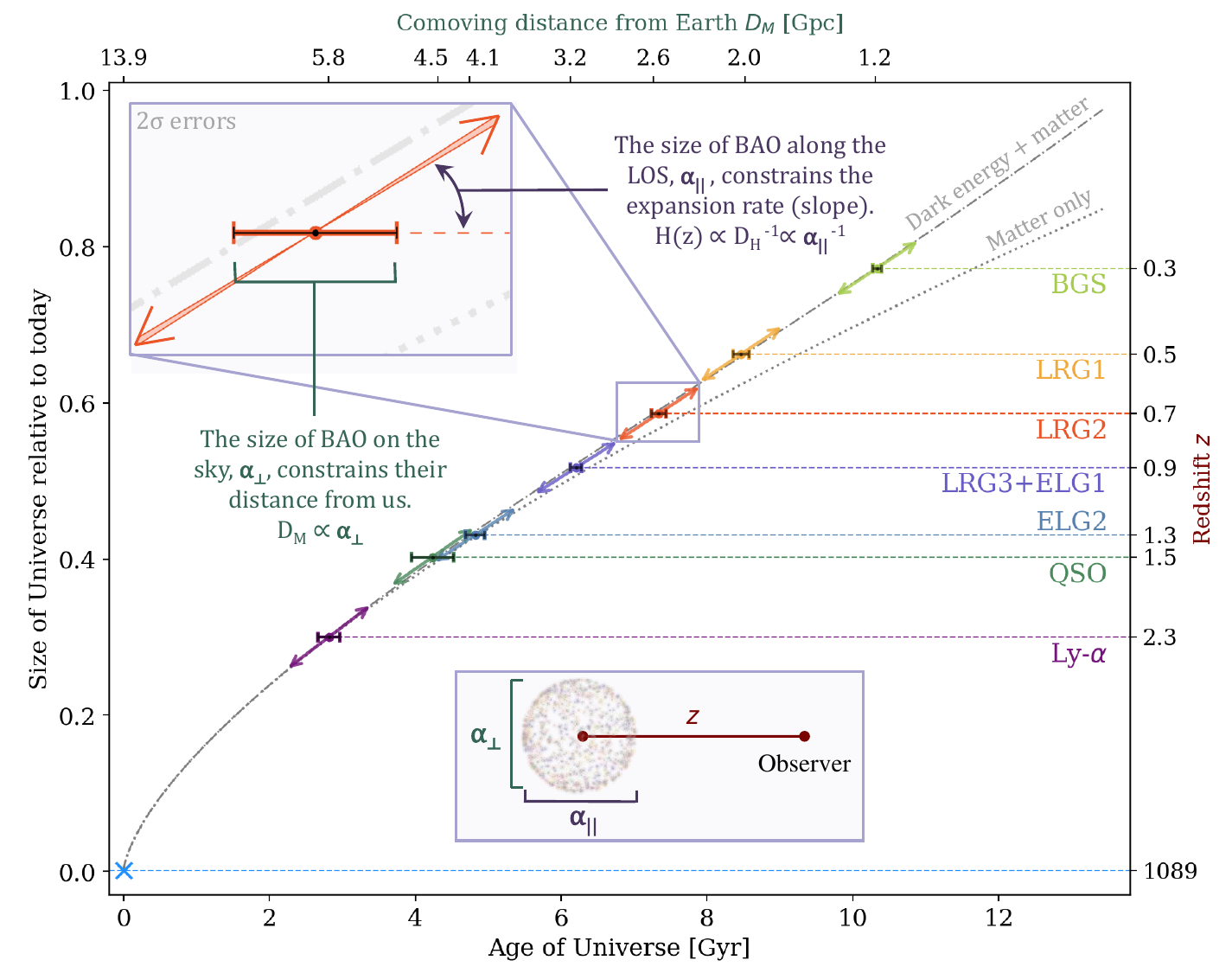}
    \end{center}
    \caption{\label{fig:BAOcosmology}An illustration of how BAO measurements from DESI constrain the expansion history of the Universe, shown here as scale factor versus time. Assuming a flat cosmology and $\rd=147.05$ Mpc \pcref{eqn:rdformula}, the angular size of BAO on the sky measures the comoving distance $\DM$ to the sample \pcref{eqn:DM}, which corresponds to the horizontal position of points on this plot.  The size of BAO along the line-of-sight (LOS)
    measures the Hubble distance $\DH$ \pcref{eqn:DH} and therefore the expansion rate, which corresponds to the slope on the curve. The scale factor is inferred from the effective redshift of the sample. Shown are the DESI DR2 BAO measurements, presented in this paper, for each of the seven tracer samples using 2$\sigma$ errors for the horizontal position and slope. The 2$\sigma$ uncertainty in the slope is only visible as the thicker slope range in the expanded insert.  The constraint from the CMB acoustic scale is plotted as a light blue cross. These measurements are compared to models of the universe with and without a cosmological constant, both assuming a flat cosmology and plotted with respect to $\DM$.  The late-time acceleration caused by dark energy is easily seen as favored by the BAO data. The bottom axis shows the age of the Universe as a function of $\DM$ in a flat \lcdm\ universe; it would be moderately different in the matter-only case.  DESI-fitted models with evolving dark energy are not visually distinguishable from the \lcdm\ model on this plot.}
\end{figure*}

The combination of BAO and cosmic microwave background (CMB) anisotropies is powerful for two reasons.  First, the CMB provides tight constraints on $\ob$ and $\obc$, leading to a 0.2\% determination of $\rd$ from \cref{eqn:rdformula} (for standard $\Neff=3.04$). As a result, the BAO+CMB combination allows absolute measurements of $\DM(z)$ and $\DH(z)$. Second, the same physics imprints both the BAO and the acoustic peaks in the CMB power spectrum \cite{HuDodelson2002}. The angular scale of these peaks, denoted $\theta_\ast$, is measured with exquisite precision (fractional error $\sim 10^{-4}$), giving a near-perfect measurement of $\DM(z_*)/r_\ast$, where $r_\ast$ is the comoving sound horizon at the end of recombination, at redshift $z_*\approx 1089$. \cref{fig:BAOcosmology} shows a pedagogical view of the cosmological role of these measurements of the expansion history.

Type Ia supernovae (SNe) are standardizable candles which serve as low-redshift probes in addition to BAO, measuring the luminosity distance $\DL(z) = (1+z)\DM(z)$. The SNe Hubble diagram provided the first direct evidence for cosmic acceleration \cite{SupernovaSearchTeam:1998fmf,SupernovaCosmologyProject:1998vns}, and subsequent cosmological surveys have obtained well measured light curves for many hundreds of supernovae out to $z=1$ and beyond (e.g., \cite{Scolnic:2021amr,Rubin:2023,DES:2024tys,DES:2024tys}, and references therein).
BAO and SNe measurements constrain dark energy and neutrino masses through their impact on the background evolution, thus determining $H(z)$. Assuming that general relativity (GR) correctly describes the dynamics of expansion, the evolution of $H(z)$ is governed by the Friedmann equation, which can be written
\begin{align}
    \begin{split}
    \frac{H(z)}{H_0} = \Big[&\Omega_\mathrm{bc}(1+z)^3 + \Omega_{\gamma} (1+z)^4 + 
                             \Ok(1+z)^2 + \\
                              &\Omega_\nu \frac{\rho_\nu(z)}{\rho_{\nu,0}} + 
                              \Ode \frac{\rho_{\mathrm{DE}}(z)}{\rho_{\mathrm{DE},0}} \Big]^{1/2}~.
    \end{split}
     \label{eqn:friedmann}
\end{align}
Here $\Omega_{\rm bc}=\Ob+\Ocdm$ and $\Omega_{\gamma}$, $\Ok$, $\Omega_\nu$ and $\Ode$ refer to the energy densities in radiation, curvature, neutrinos and dark energy, respectively. We refer to the fractional energy density in matter as $\Om$, which includes neutrinos when they are non-relativistic. The energy densities of baryons and cold dark matter scale as $(1+z)^3$. The scaling of the neutrino energy density transitions from $(1+z)^3$ to $(1+z)^4$ at high redshifts, when $(1+z) \sim (m_\nu/5\times 10^{-4}\eV)$ \cite{PDG:2022,Lesgourgues:2012uu}. The sum of neutrino masses determines the present day density \cite{Lesgourgues:2006}
\begin{equation}
    \Omega_\nu h^2 = \frac{\sumnu}{93.14\,\eV} ~.
    \label{eqn:Omeganu}
\end{equation}
For BAO cosmology, the important characteristic of `CDM' is that its energy density scales as $(1+z)^3$, both before and after recombination, and that it does not couple non-gravitationally to photons or baryons in a way that affects the scale of the acoustic oscillations.  Some variations such as self-interaction or `warm' thermal velocities would affect small scale clustering and galaxy formation but not BAO.  Decaying dark matter, on the other hand, would alter the $(1+z)^3$ energy scaling even if the dark matter is cold and non-interacting, thus affecting the BAO scale.

The \lcdm\ model assumes a cosmological constant dark energy ($\Lambda$) with energy density $c^2\rho_\mathrm{DE}$ that is constant in space and time.  If dark energy has an equation-of-state parameter $w(z) \equiv P(z)/(c^2 {\rho}_\mathrm{DE}(z))$ where $P(z)$ is its pressure, then its energy density evolves as 
\begin{equation}
    \frac{\rho_\mathrm{DE}(z)}{\rho_\mathrm{DE,0}} = \exp\left[3\int_0^z [1+w(z^\prime)] \frac{dz^\prime}{1+z^\prime}\right] ~.
  \label{eqn:DEevolution}
\end{equation}
For constant $w$, the r.h.s. of \cref{eqn:DEevolution} is simply $(1+z)^{3(1+w)}$, and a cosmological constant corresponds to $w=-1$.
A commonly used parametric model expresses $w(z)$ in terms of the expansion factor $a = (1+z)^{-1}$,
\begin{equation}
    w(a) = w_0 + \wa (1-a)~,
    \label{eqn:w0wa}
\end{equation}
so that $w$ evolves from a value $\sim(w_0+\wa)$ at high redshift to a present-day value of $w_0$.  This parametrization accurately represents the behavior of many physically motivated dark energy models \cite{Chevallier:2001,Linder2003,Linder2024}, though more complicated evolution is possible.  We refer to models that assume CDM and \cref{eqn:w0wa} as \wowacdm\ and models with constant $w$ (i.e., $\wa=0$) as \wcdm.  In the \wowacdm\ model the integral of \cref{eqn:DEevolution} can be evaluated analytically, yielding
\begin{equation}
    \frac{\rho_\mathrm{DE}(a)}{\rho_\mathrm{DE,0}} = a^{-3(1+w_0+\wa)} e^{-3\wa(1-a)} ~.
  \label{eqn:DEevolution2}
\end{equation}
Through most of this paper we will assume a flat universe and thus $\Ok=0$ in \cref{eqn:friedmann}, motivated by the tight constraints obtained on $\Ok$ when it is allowed to vary freely \cite{DESI2024.VI.KP7A}. 

The combination of BAO, CMB, and SNe data has allowed tight constraints on the energy density $\Ode$ and equation-of-state $w(z)$ of dark energy, on space curvature $\Ok$, on neutrino masses $\sumnu$, and on many possible departures from standard cosmology (see, e.g., \cite{2015PhRvD..92l3516A,2017MNRAS.470.2617A,Alam-eBOSS:2021,DES:BAO+SN}).  For reviews that explain the complementary constraining power of CMB, BAO, SNe, weak lensing, and other cosmological measurements, see \cite{Frieman:2008sn,Weinberg:2013agg,2022PTEP.2022h3C01W}. The analysis of the DESI DR1 measurements from \cite{DESI2024.III.KP4,DESI2024.IV.KP6} in \cite{DESI2024.VI.KP7A} provided tight constraints on parameters of the \lcdm\ model and intriguing hints of evolving dark energy, with significances ranging from $\sim2.5\sigma$ to $\sim3.9\sigma$ depending on the combination of datasets used for the analysis. The analysis of the full shape of the power spectrum measured with galaxies and quasars \cite{DESI2024.V.KP5,DESI2024.VII.KP7B} confirmed these findings and added new information on the amplitude of perturbations and modified gravity parameters \cite{KP7s1-MG}.

Rapidly evolving dark energy, with $|\wa|\sim 1$, would be an astounding discovery, and these results have inspired both enthusiastic theorizing and healthy skepticism.  In our analysis here of the DESI DR2 BAO results, we pay particular attention to the nature and statistical significance of the evidence for evolving dark energy and to how that evidence depends on the choice of datasets.  We also examine the constraints on $\sumnu$ from the DESI DR2 data in combination with CMB and SNe, for both \lcdm\ and \wowacdm.  When we refer to `DESI' alone in tables and figure legends, we treat the BAO as an uncalibrated standard ruler.  In some of our \lcdm\ analyses, we examine constraints that adopt a big bang nucleosynthesis (BBN) prior on $\ob$, with the value of $\obc$ in \cref{eqn:rdformula} coming from the model fit itself.  We achieve tighter constraints and sharper tests by combining DESI with CMB data that directly constrain $\ob$ and $\obc$ and add the precise measurement of $\theta_*$ at $z=z_*$.

\begin{table}[t]
    \begin{ruledtabular}
    \centering
    \begin{tabular}{l p{5.5cm} r}
        \textbf{Ref.}  &  \textbf{Topic} & \textbf{Section} \\
        \hline
        \cite{Y3.clust-s1.Andrade.2025} & Validation of the DESI DR2 Measurements of Baryon Acoustic Oscillations from Galaxies and Quasars & \cref{sec:bao_measurements} \\
        \cite{Y3.cpe-s1.Lodha.2025} & Extended Dark Energy analysis using DESI DR2 BAO measurements & \cref{sec:de_constraints} \\
        \cite{Y3.cpe-s2.Elbers.2025} & Constraints on Neutrino Physics from DESI DR2 BAO and DR1 Full Shape & \cref{sec:constraints_neutrinos} \\
    \end{tabular}
    \end{ruledtabular}
    \caption{Supporting papers relevant to this work and the corresponding sections where their results are discussed.}
    \label{tab:supportingpapers}
\end{table}

This work is accompanied by a set of supporting papers, highlighted in \cref{tab:supportingpapers}. The structure of this paper is as follows. In \cref{sec:data}, we describe the DESI DR2 data and large-scale structure catalogs. \Cref{sec:bao_measurements} presents the DESI DR2 distance measurements and internal consistency checks, and presents a comparison with SDSS. In \cref{sec:external_data}, we describe the external datasets that will be combined with DESI BAO. 
\cref{sec:inference_method} introduces our cosmological inference method.
\cref{sec:cosmologcal_constraints} presents cosmological parameter constraints in a $\Lambda$CDM framework. \Cref{sec:de_constraints} presents constraints in a more generalized dark energy framework and analyzes tensions with respect to the $\Lambda$CDM model. \cref{sec:constraints_neutrinos} presents constraints on the neutrino sector and explores the the role of neutrino masses in our results. Finally, \cref{sec:conclusions} presents a summary and our main conclusions.

\section{DESI Data}
\label{sec:data}

The DESI Collaboration has measured redshifts for over 30 million galaxies and quasars in just three years of operation, $\sim$14 million of which are used\footnote{The majority of the redshifts not used are from the low redshift bright time program, as described in the following subsection.} in this analysis, as described below. This extraordinarily high rate of data collection is possible because we built an extremely efficient instrument that can measure thousands of spectra in a single observation \cite{DESI2022.KP1.Instr}, combined with the light-gathering power of the 4-m Nicholas U. Mayall Telescope at the Kitt Peak National Observatory. DESI collects the light for 5000 spectra per observation with a robotic focal plane assembly \cite{FocalPlane.Silber.2023} that can quickly align the positions of fiber optics cables \cite{FiberSystem.Poppett.2024} across the seven square degree field of view of the prime focus corrector \cite{Corrector.Miller.2023}. 
For each observation, there is a custom focal plane configuration or `tile' that defines the DESI `target' \cite{TS.Pipeline.Myers.2023} associated with each robotic positioner. When repeated observations of a tile are required to obtain the minimum effective observing time \cite{SurveyOps.Schlafly.2023}, these maintain the same configuration. The light of each of these targets, along with calibration stars and sky spectra, are recorded from 360--980\,nm with ten bench-mounted spectrographs that are located in a climate-controlled enclosure. This configuration helps to enable the superb wavelength and flux calibration of the DESI data.

The DESI main survey started observations on 14 May 2021 after a period of survey validation \cite{DESI2023a.KP1.SV}. This paper presents the analysis of the main survey data that will be released with the second data release or DR2, which includes observations through 9 April 2024. The DESI spectroscopic reduction \cite{Spectro.Pipeline.Guy.2023} and redshift estimation (\texttt{Redrock} \cite{Redrock.Bailey.2024,Anand24redrock}) pipelines were applied to the DR2 dataset in a homogeneous processing run denoted as `Kibo'. An error in the processing involved in the co-addition of spectra from separate exposures was subsequently identified and fixed, and the pipeline was rerun and denoted `Loa'. Approximately 0.1\% of the measured redshifts change significantly between Kibo and Loa. The LSS catalogs for DR2 BAO measurements were produced for both Kibo and Loa. Decisions that affect the masking of the data (described further below) were made using the Kibo data and were not reconsidered with Loa. Both datasets will be released with DR2.  

The DESI survey has two main observing programs, `bright' and `dark,' that are defined based on the nighttime sky conditions, and each of these programs has distinct target classes \cite{SurveyOps.Schlafly.2023,BGS.TS.Hahn.2023,LRG.TS.Zhou.2023,ELG.TS.Raichoor.2023,QSO.TS.Chaussidon.2023,TS.Pipeline.Myers.2023}. The extragalactic sample for the bright program is the `bright galaxy sample' (BGS).\footnote{There are also secondary targets that are observed at lower priority \cite{TS.Pipeline.Myers.2023}.} Luminous red galaxies (LRGs), emission line galaxies (ELGs), and quasars (QSOs) are all observed during dark conditions. Dark- and bright-time targets are processed separately. At $z<2.1$, we measure BAO with the autocorrelation of the confirmed members of each target class, processed through `large-scale structure' (LSS) catalogs as described in~\cref{subsec:lss}. At higher redshifts, we measure the auto-correlation of the \lya forest absorption in the spectra of quasars and the cross-correlation of the forest absorption with quasar positions. The data samples used for these \lya measurements are described further in \cref{subsec:lya}. The analysis of the \lya data and BAO results are presented in the companion key paper~\cite{DESI.DR2.BAO.lya}. DR2 contains 6671 dark and 5171 bright tiles. These are respectively 2.4 times and 2.3 times the number released in DR1.

\subsection{Galaxy and quasar large-scale Structure Catalogs}
\label{subsec:lss}

For clustering science analyses, the catalogs of measured redshifts and details of the target selection are converted into large-scale structure (LSS) catalogs, as described in \cite{KP3s15-Ross}. For the specific choices involved in the analysis, we mostly match those decided in \cite{DESI2024.II.KP3}. New developments include the following:
\begin{itemize}
    \item The bad fiber list has been updated, applying the same methods as for DR1 \cite{KP3s3-Krolewski}, but using the Kibo redshift results.
    \item For correcting QSO imaging systematics, we switch from a random forest method to the linear regression method used for the LRG and BGS samples, motivated by the conclusions of \cite{ChaussidonY1fnl}.
    \item The BGS sample is more dense, as we choose to apply a less restrictive luminosity threshold. We discuss this further below.
\end{itemize}

\begin{figure*}
    \centering 
    \includegraphics[width=0.99\columnwidth]{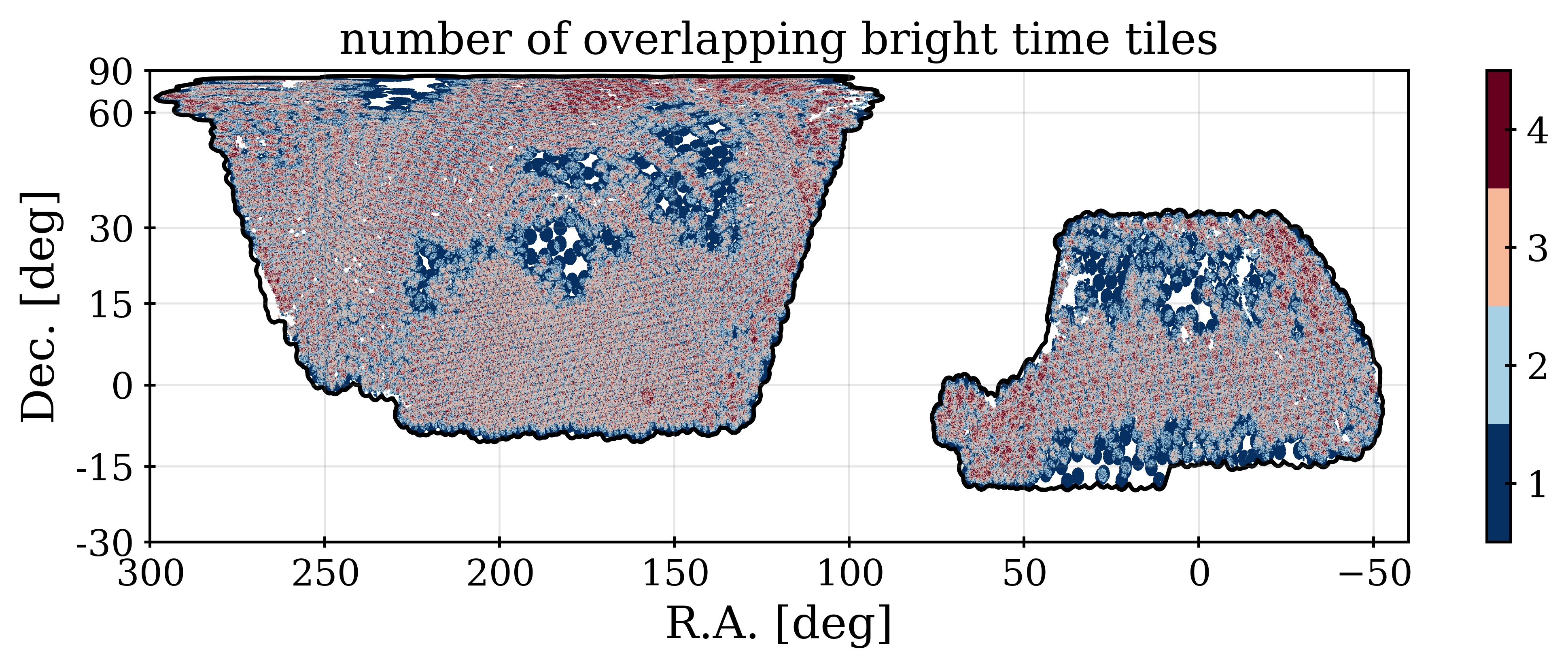}  
    \includegraphics[width=0.99\columnwidth]{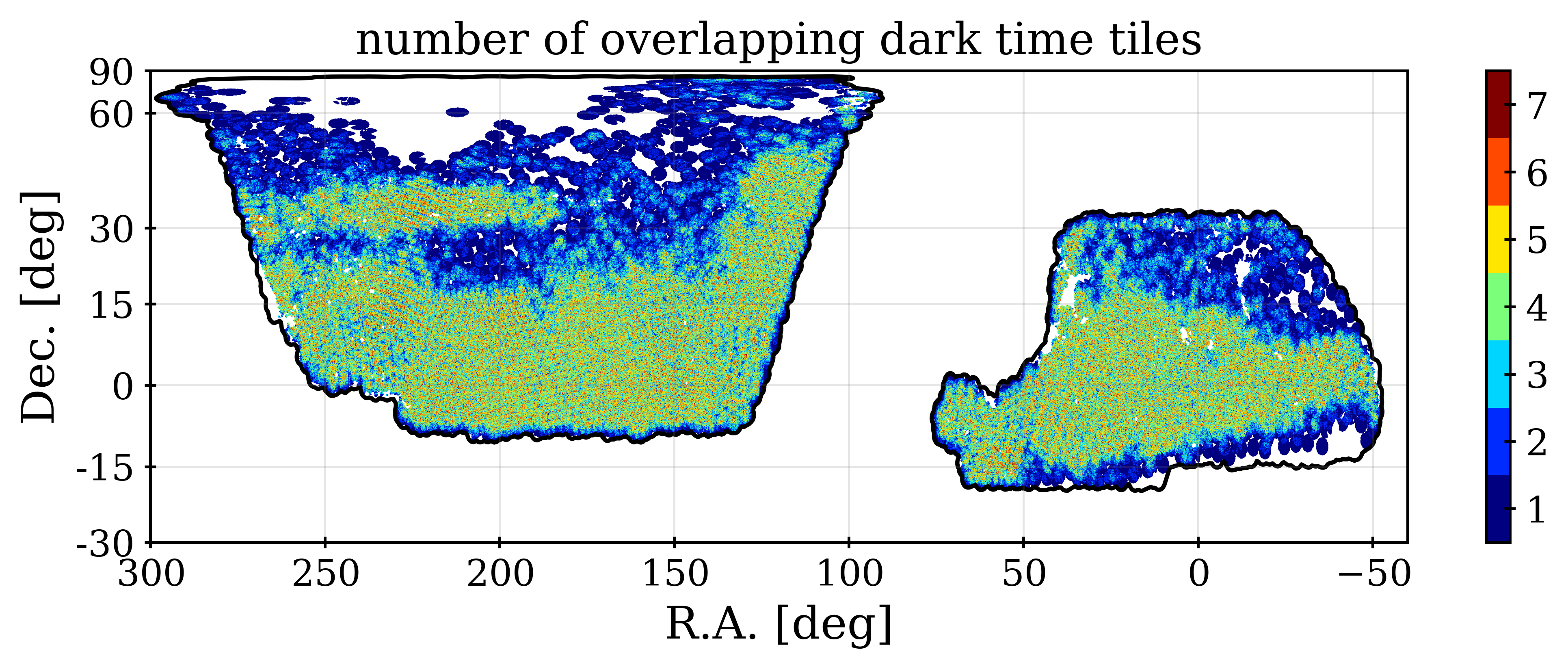} 
    \includegraphics[width=0.99\columnwidth]{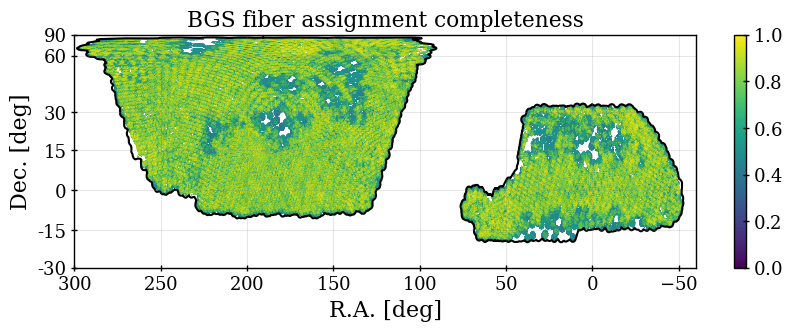}
    \includegraphics[width=0.99\columnwidth]{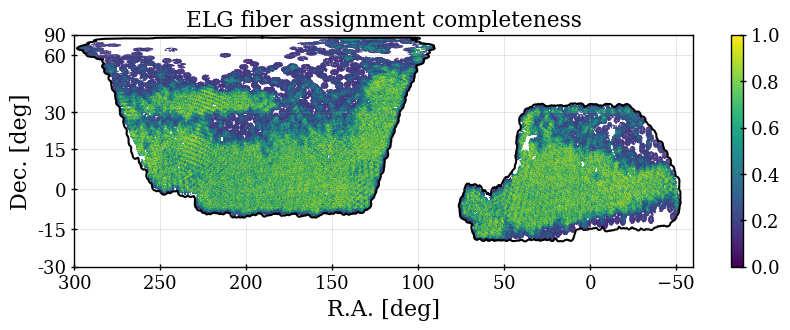}
    \includegraphics[width=0.99\columnwidth]{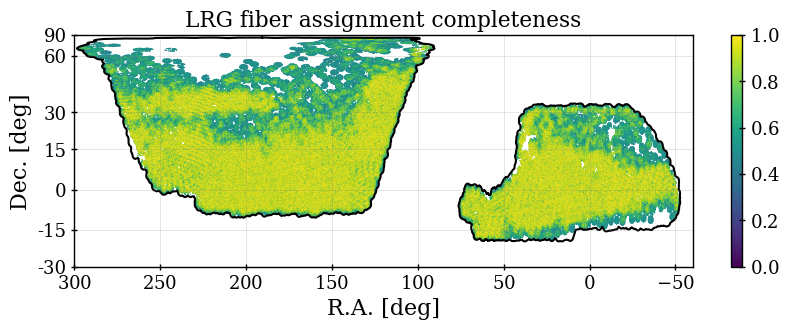}
    \includegraphics[width=0.99\columnwidth]{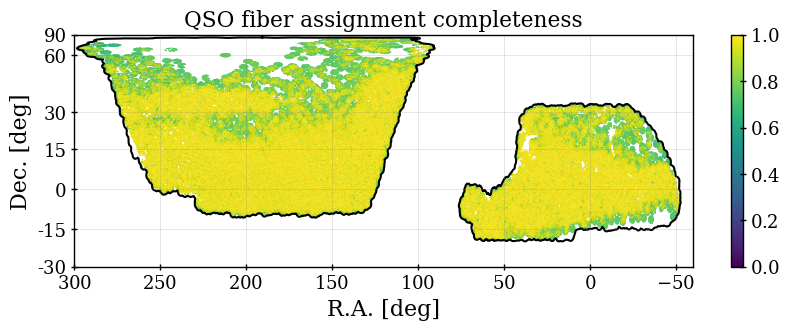}
    \caption{The top two panels show the number of overlapping tiles in DR2 for bright and dark time. The lower four panels show the assignment completeness of the four DESI samples, within unique tile groupings. One can observe that the patterns in the completeness correspond to the number of overlapping tiles. The black outline shows the edges of the area within which DESI dark or bright tiles have been defined, as of April 9th, 2024. The completeness maps can be compared to the values in \cref{tab:Y3lss}.}
    \label{fig:allcomp}
\end{figure*}

\begin{table}
    \begin{ruledtabular}
    \centering
    \resizebox{\columnwidth}{!}{
    \begin{tabular}{lccccr}
    Tracer & \# of good $z$ & $z$ range& Area [deg$^2$] & $C_{\rm assign}$ & $z$ succ. \\\hline
    BGS  & 1,188,526  & $0.1-0.4$ & 12,355 & 75.5\% & 98.8\%\\
    LRG  & 4,468,483 & $0.4-1.1$ & 10,031 & 82.6\% & 99.0\%\\
    ELG  & 6,534,844  & $0.8-1.6$ & 10,352 & 53.7\% & 73.9\% \\
    QSO & 2,062,839 & $0.8-3.5$ & 11,181 & 93.6\% & 68.0\% \\
    \end{tabular}
    }
    \end{ruledtabular}
    \caption{Statistics for each of DESI tracer types used for the BAO measurements described in this paper. We list the number of good redshifts included, the redshift range used, the sky area occupied, and the observational completeness within that area. The criteria for selecting good redshifts and determinations of the footprint area and completeness are the same as applied in \cite{DESI2024.II.KP3}. The spectroscopic success rates (`$z$ succ.'), area, and assignment completeness ($C_{\rm assign}$) are determined for the sample without any cuts on redshift. The area is different for different tracer classes due to priority vetoes (e.g., a QSO target can remove sky area from lower priority samples) and small differences in the imaging vetoes applied. The assignment completeness is the percentage of targets within the DR2 footprint that were observed.
    \label{tab:Y3lss}}
\end{table}

 \cref{tab:Y3lss} contains basic information about the content of the LSS catalogs used for this analysis. In addition to the increase in the raw number of redshifts, one can observe that the sky coverage and the completeness within that sky coverage have increased substantially compared to the DR1 LSS catalogs. Compared to DR1, the size of the ELG sample has grown by a factor of 2.7, while the LRG and QSO samples have grown by factors of 2.1 and 1.7, respectively. The difference in these factors is due to the fact that the coverage, in terms of the typical number of overlapping tiles, has increased substantially between DR1 and DR2. This can be seen by comparing the top panels of \cref{fig:allcomp} to the top panels of the equivalent DR1 figure (figure 2 in \cite{DESI2024.II.KP3}). The increase in coverage grows the size of the ELG sample the most, as ELG targets are assigned at the lowest priority (see \cite{TS.Pipeline.Myers.2023,KP3s15-Ross}). Their assignment completeness has increased by 53\% (from 35.2\% to 53.7\%), while the increases are more modest for LRG and QSO (19\% and 7\%).
  
 The increase in area from DR1 is between 54\% (QSO) and 75\% (LRG and ELG). The areas differ due to the tiles covered (bright vs. dark programs), the priority masking (primarily applied to LRG and ELG due to the influence of QSO) and imaging veto masks (these all differ slightly per tracer). The masks applied match the criteria defined in DR1 \cite{DESI2024.II.KP3}. The amount of area in the priority mask is considerably lower than in DR1, despite the greater overall footprint. As was the case for the assignment completeness, this is due to the increased typical number of overlapping tiles:  the priority mask is primarily caused by QSO targets receiving higher priority, which happens in the first observation of a tile but not in subsequent overlapping tiles except for quasars at $z>2.1$. The result is that the area in the LRG and ELG priority mask has reduced from 1667 to 1153 deg$^2$, thus narrowing the gap between the size of the QSO and LRG/ELG footprints. The remaining differences between the footprints of the dark time tracers are due to the application of the imaging veto masks, which remain the same as applied to DR1. The area of the BGS sample has grown by 65\% compared to DR1, and this is simply due to the footprint of the newly observed tiles.

The BGS sample we use is nearly 4$\times$ the size of the DR1 BGS sample used for BAO measurements. The increased area and completeness account for half of this increase. The rest of the increase is due to a relaxation of the absolute magnitude cut. First, we changed the definition of the absolute magnitudes used for the cut. In this analysis, we do not apply any $k$ or $e$ corrections and we instead determine the absolute $r$-band magnitudes, $M_r$ purely based on the apparent magnitude of the galaxy and the distance modulus to its redshift.\footnote{We compute distances in units of $\hinvmpc$ and use $h=1$ to define absolute magnitudes. Therefore the quantities we refer to as absolute magnitudes $M_r$ are really $M_r + 5\log h$, but this constant offset does not affect the sample definitions.} Based on this new definition, we also changed the absolute magnitude cut to the value at which no galaxies would be removed from the sample at its maximum redshift of 0.4. This was determined to be $M_r<-21.35$, whereas the cut applied in DR1 was $M_r<-21.5$. 

\begin{figure}
    \centering 
    \includegraphics[width=\columnwidth]{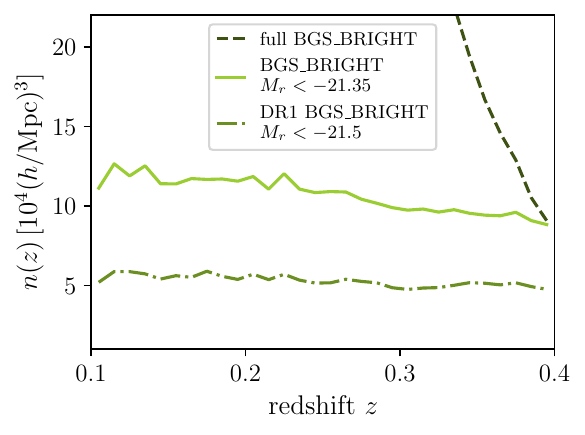} 
     \caption{A comparison of the comoving number density for different selections applied to the DESI BGS sample. This analysis uses the BGS\_BRIGHT $M_r<-21.35$ sample. Note that the DR1 cut shown here used a slightly different definition of the absolute magnitudes $M_r$.}
    \label{fig:nzbgs}
\end{figure}

\cref{fig:nzbgs} shows the comoving number density of the BGS sample (applying completeness corrections) that we use in solid light green. The sample used in our DR1 analysis is shown in dot-dashed dark green. The total \texttt{BGS\_BRIGHT} parent sample is shown with a dashed black curve. For the total sample, the number density quickly becomes greater than the range shown on the plot, because the target sample was chosen with a simple $r$-band flux cut \cite{BGS.TS.Hahn.2023}. The density of the sample chosen for this analysis matches that of the parent sample at the upper redshift limit $z=0.4$. The BGS sample we use has approximately double the number density of the one used in the DR1 analysis. Our lack of use of $k$ or $e$ corrections makes the number density evolve with redshift slightly more than for the DR1 BGS sample. However, the change of number density with redshift is $\sim$25\%, which is still lower than that of the other tracers used in the analysis. The DESI DR2 data contains more than 9.1 million BGS galaxies (considering both the \texttt{BGS\_BRIGHT} and \texttt{BGS\_ANY} samples \cite{BGS.TS.Hahn.2023}) with good redshifts $0.1<z<0.4$, of which we use only a small fraction (\cref{tab:Y3lss}). Using all of the BGS galaxies instead would increase the effective volume by approximately 20\%. The exact amount depends on the assumed clustering amplitude, which will change significantly with redshift, as the effective luminosity threshold of the sample decreases at lower redshift. This large change in the physical properties of the sample introduces significant modeling complexities as many elements of our analysis pipeline assume a  roughly constant clustering amplitude (e.g., reconstruction, covariance estimation, BAO fitting). We aim to utilize more of the BGS in future DESI cosmological analyses.

\begin{figure}
    \centering
    \includegraphics[width=\columnwidth]{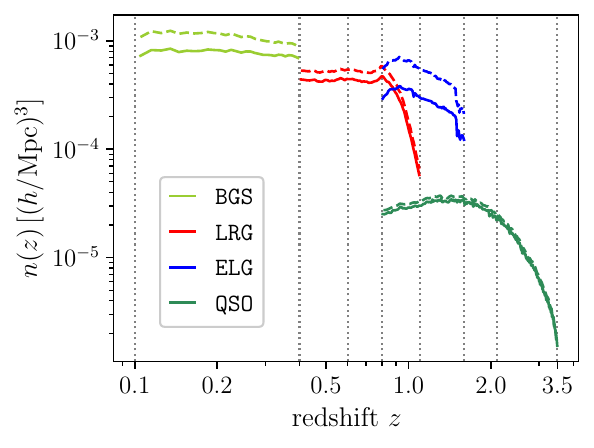}
    \caption{Comoving number density as a function of redshift for the different DESI DR2 tracers. Solid and dashed lines show results before and after applying sample completeness corrections. The vertical dotted lines delimit the redshift bins used for clustering measurements.}
    \label{fig:nz_all}
\end{figure}

\begin{table}
    \begin{ruledtabular}
    \centering
    \begin{tabular}{cccc}
    Tracer & Redshift range & $N_{\rm tracer}$  & $V_{\rm eff}$ (Gpc$^3$)\\
    \hline
    \bgs  & $0.1-0.4$ & 1,188,526  & 3.8  \\
    \lrgo  & $0.4-0.6$ & 1,052,151  & 4.9 \\
    \lrgt  & $0.6-0.8$ & 1,613,562  & 7.6  \\
    \lrgelg & $0.8-1.1$ & 4,540,343  & 14.8\\
    \elgt  & $1.1-1.6$ & 3,797,271  &  8.3\\
    \qso  & $0.8-2.1$ & 1,461,588  & 2.7 \\
    \lya  & $1.8-4.2$ & 1,289,874  &---\\
    \hline
    \lrgth  & $0.8-1.1$ & 1,802,770  & 9.8  \\
    \elgo & $0.8-1.1$ & 2,737,573  & 5.8 \\
    \end{tabular}
    \end{ruledtabular}
    \caption{\label{tab:Y3data} Statistics for each of the DESI tracer types used for the DESI DR2 BAO measurements presented in this paper. Redshift bins are non-overlapping, except for the shot-noise dominated QSO sample and the $0.8<z<1.1$ LRG and ELG. In this redshift range our baseline analysis uses the combined sample \lrgelg, but information for the individual \lrgth\ and \elgo\ samples is also provided in the final two rows.
    The effective volume calculation, $V_{\rm eff}$, provides a rough estimate for the relative amount of cosmological information in each redshift bin.
    }
\end{table}

Figure \cref{fig:nz_all} shows the raw and the completeness-corrected comoving number density for each of the samples used in this analysis. Dotted vertical lines denote the redshift binning we apply to obtain BAO measurements. The labels, number of redshifts, and effective volumes\footnote{These are calculated in the same way as for DR1, described in \cite{DESI2024.III.KP4}.} for the data in each of these redshift bins are provided in \cref{tab:Y3data}. The redshift binning is the same as applied to DR1 and one can thus compare directly to the values in table 2 of \cite{DESI2024.III.KP4}. The effective volume has increased compared to DR1 by a factor that varies from 1.8 (QSO) to 3.1 (\texttt{ELG2}).

The version of the LSS catalogs used in this analysis, DR2 \texttt{v1.1/BAO}, will be released with DESI DR2. 
Prior to producing the results presented in this paper, the DESI BAO analysis team further processed the LSS catalogs to blind the results and apply BAO reconstruction. The blinding was removed only after the analysis choices were validated and fixed. We provide more detail on each process below. 

\subsubsection{Blinding}
To minimize the risk of confirmation bias in our analysis, we used a blinding scheme to deliberately conceal the true position of the BAO peak observed from the data until all choices about our inference pipeline were finalized. This blinding was applied at the level of the LSS catalogs, using the same prescription as was applied to DR1 and detailed in \cite{KP3s9-Andrade}, but with a new random seed to produce unknown (but controlled) shifts to galaxy redshifts that conceal the true BAO peak position. A number of validation tests of both the data and the analysis pipeline were performed on the blinded data catalogs, as described in \cite{Y3.clust-s1.Andrade.2025}. Once these tests were passed and the pipeline frozen, the blinding was removed and true catalogs were first processed during the DESI winter collaboration meeting in December 2024.

\subsubsection{Reconstruction}
Non-linear gravitational evolution induces large-scale bulk flows that can degrade the precision and accuracy of BAO measurements. 
Density field reconstruction \citep{Eisenstein2007b} attempts to correct for this effect by estimating the displacement field from the observed galaxy distribution, using it to undo the gravitational flow and partially restore the acoustic peak to its linear regime shape. We `reconstructed' the LSS catalogs using the \texttt{IterativeFFT} reconstruction algorithm \citep{Burden:2015pfa} implemented in \texttt{pyrecon}\footnote{\url{https://github.com/cosmodesi/pyrecon}} and using the fiducial DR1 settings from \cite{KP4s4-Paillas}. This decision was informed by an extensive comparison of different reconstruction algorithms in the context of DESI \citep{KP4s3-Chen}. 
The reconstructed catalogs are used for the clustering and associated BAO measurements from galaxies and quasars presented throughout the rest of this paper.

\subsection{Lyman-$\alpha$ forest catalog}
\label{subsec:lya}

The DESI DR2 quasar catalog for the \lya forest analysis was constructed in a similar manner to the DR1 catalog described in \cite{DESI2024.IV.KP6}. The catalog contains high-redshift quasars and is built from the outputs of three automatic classifiers that analyze all quasar targets: the \texttt{Redrock} redshift estimator \citep{Redrock.Bailey.2024} that classifies based on templates, including quasar templates optimized for DESI quasars \citep{RedrockQSO.Brodzeller.2023,KP6s9-Martini}, a \ion{Mg}{II} afterburner that searches for broad \ion{Mg}{II} emission in quasar candidates classified as galaxies, and the \texttt{QuasarNet} neural network classifier developed by \citep{Busca18} that has been updated \cite{Green2025.QN} for DESI. The catalog also contains any other targets that are classified by \texttt{Redrock} as high-redshift quasars. 

We also catalog two types of absorption systems that are masked by our analysis pipeline: Damped \lya Absorption Systems (DLAs) and Broad Absorption Line (BAL) quasars. DLAs are systems with high ($N_\mathrm{H\,I} > 2 \times 10^{20}$\,cm$^{-2}$) column densities of neutral hydrogen in the intergalactic medium (IGM). These systems have somewhat higher clustering than typical \lya forest absorption and have very broad damping wings that can compromise a significant fraction of a forest spectrum. We use three analysis tools to identify DLAs in the DR2 dataset; the performance of these methods are described in \cite{Y3.lya-s2.Brodzeller.2025}. Approximately 20\% of the QSOs in the \lya forest sample are BAL quasars. These absorption systems contaminate the forest by adding absorption that is uncorrelated with the large-scale structure of the IGM. We identify BALs with the method described in \cite{KP6s9-Martini}. Further details about the quasar catalog for the \lya forest are described in \cite{DESI.DR2.BAO.lya}. 

The DR2 quasar catalog has over 1.2 million quasars at $z > 1.77$, including over 820,000  quasars at $z > 2.09$. We use Ly$\alpha$ forests from the spectra of quasars at $z > 2.09$, while quasars at $z > 1.77$ are used as discrete tracers in the computation of the cross-correlation with Ly$\alpha$. Both samples are close to a factor of two larger than the DR1 sample, which had 420,000 quasars at $z>2.09$ and over 700,000 quasars at $z > 1.77$. In addition to the larger sample size, the average signal-to-noise per pixel is higher, because 67\% of DR2 quasars have had at least two observations (out of the four planned for the \lya\ quasar sample), while only 43\% of DR1 quasars had been observed more than once.

\section{BAO Measurements}\label{sec:bao_measurements}

\subsection{Clustering from Galaxies and Quasars}

Our baseline BAO measurements from galaxies and quasars are derived from the two-point correlation function (2PCF) in redshift space. We measure it using the Landy-Szalay estimator \citep{Landy-Szalay:1993} (adapted to post-reconstruction measurements, as in \cite{Padmanabhan2012}) using \texttt{pycorr},\footnote{\url{https://github.com/cosmodesi/pycorr/}} which implements a wrapper around a modified version of the \texttt{Corrfunc} pair-counting code \citep{Sinha:2019reo}. We bin the correlation function in $s$ and $\mu = \cos\theta$, where $s$ is the redshift-space scalar separation and $\theta$ is the angle between the galaxy pair and the line of sight, using $s$-bins of $4 \hinvmpc$ width and 200 $\mu$-bins in $[-1, 1]$. We then decompose it into Legendre multipoles, focusing on the monopole and quadrupole moments for the analysis.

We estimate the errors of the correlation function measurements using the \texttt{RascalC} semi-analytical covariance matrix code\footnote{\url{https://github.com/oliverphilcox/RascalC}}~\cite{rascal,rascal-jackknife,RascalC,RascalC-legendre-3,2023MNRAS.524.3894R}, in the same way as for the DESI DR1 BAO analysis \cite{DESI2024.III.KP4}. The computation gives the contributions of the measured (non-linear) two-point correlation function (including the disconnected four-point function) with survey geometry and selection effects, and shot-noise rescaling calibrated with jackknife resampling to account for missing non-Gaussian contributions. This procedure has been validated using DESI DR1 mocks in \cite{KP4s7-Rashkovetskyi}, which also provides a more complete description of the algorithm. In general, covariance estimation for DR2 is less challenging than DR1 (e.g., the footprint is less complex and the samples have higher completeness). The DESI DR2 covariance pipeline is publicly available on \texttt{GitHub}.\footnote{\url{https://github.com/cosmodesi/RascalC-scripts/tree/DESI-DR2-BAO/DESI/Y3/post}}

The BAO fitting procedure---described in detail in \cite{DESI2024.III.KP4}---involves the use of a template of the correlation function multipoles in a fiducial cosmology (which are converted from Fourier space templates of the power spectrum, \cite{KP4s2-Chen}). During the parameter posterior sampling, the BAO features in these templates are shifted with respect to those seen in the data, and the amount of shifting is regulated by scaling parameters that are varied freely during the fit. The scaling parameters that shift the BAO features along and across the line of sight are related to the cosmology by
\begin{equation}
    \alpha_{||}(z) = \frac{D_{\rm H}(z)r^{\mathrm{fid}}_{\rm d}}{D_{\rm H}^{\mathrm{fid}}(z)r_{\rm d}}, \qquad \alpha_{\perp}(z) = \frac{D_{\rm M}(z)r^{\mathrm{fid}}_{\rm d}}{D^{\mathrm{fid}}_{\rm M}(z)r_{\rm d}},
    \label{eqn:alpha_defs}
\end{equation}
where the $^{\rm fid}$ superscript denotes quantities in the fiducial cosmology that is used to convert redshifts to distances and define the power spectrum template. These factors can be re-parametrized into a different basis,
\begin{equation}
    \alpha_{\rm iso} =  (\alpha_\parallel\alpha_\perp^2)^{1/3},
 \qquad \alpha_{\rm AP} = \alpha_\parallel/\alpha_\perp,
    \label{eqn:aisoap_defs}
\end{equation}
which modulate an isotropic ($\alpha_{\rm iso}$) or anisotropic ($\alpha_{\rm AP}$) shifting of the BAO feature. The measurements are less correlated in the latter basis, and we use it as our baseline when sampling the BAO model posterior. We also define the isotropic BAO distance $\DV(z)\equiv\left(z\DM(z)^2\DH(z)\right)^{1/3}$, for which we report results in later sections.

The templates are also allowed to vary in amplitude and are combined with a set of nuisance parameters that model the broadband shape of the correlation function. The reported BAO constraints are marginalized over these nuisance parameters so that the information coming from the position of the acoustic feature can be isolated. The broadband parametrization involves a piecewise-spline fitting basis that was introduced in the DR1 analysis, described in detail in \cite{KP4s2-Chen}.

The BAO model and inference pipeline are largely the same as used for DR1 and described in \cite{DESI2024.III.KP4}, with a few small modifications:
\begin{itemize}
    \item \textbf{Scale cut ($s_{\text{min}}$):} for the DR1 analysis, the minimum scale used in the BAO fits was $s_{\text{min}} = 50~\hinvmpc$. Reference \cite{KP4s2-Chen} showed through tests on mock catalogs that the recovered values of $\aiso$ and $\alpha_{\rm AP}$ and their errors are very stable against changes to the minimum scale in the range $50\;\hinvmpc \lesssim s_{\rm min}\lesssim 80\;\hinvmpc$. However, when fitting to the blinded DR2 data we found somewhat large $\chi^2$ values when using $s_{\rm min}=50\;\hinvmpc$ for some tracers and redshift bins, which improved significantly when changing the scale cuts to $s_{\rm min}=60\;\hinvmpc$ (with only very minor effects on the recovered $\alpha$ values). To guard against the possibility that this effect in the blinded data was due to some unknown systematic affecting the 2PCF in the range $50<s<60\;\hinvmpc$, we chose to set $s_{\text{min}} = 60~\hinvmpc$ as our baseline for our BAO fits before unblinding.

    \item \textbf{2D fits for \elgo\ and \qso:} 
    for tracers with a sufficiently large signal-to-noise ratio in the 2PCF, $\aiso$ and $\alpha_{\rm AP}$ can be simultaneously determined, using anisotropic clustering information by fitting both the monopole and quadrupole moments of the correlation function (a `2D fit'). However, when the signal-to-noise of the quadrupole measurement is lower, robust determination of $\alpha_{\rm AP}$ is harder. In these cases, we only fit the monopole, which depends on $\aiso$, to avoid the complications of including a weak non-Gaussian constraint on $\alpha_{\rm AP}$.     
    For the DR1 analysis, our baseline analysis for \bgs, \elgo, and \qso\ used only the monopole for BAO fits. In DR2, the signal-to-noise ratio of the \elgo\ and \qso\ samples has increased sufficiently to promote them to 2D fits for both $\aiso$ and $\alpha_{\rm AP}$, while for \bgs\ we continue to use the monopole only to measure $\aiso$.\footnote{At the low effective redshift $z_{\rm eff}=0.295$ of the \bgs\ sample, measurement of $\alpha_{\rm AP}$ also carries very little cosmological information.} This decision was made by verifying the stability of the constraints when fitting the blinded data and mock galaxy catalogs \cite{Y3.clust-s1.Andrade.2025}.
    
    \item \textbf{Systematic errors:} the systematic error treatment in the DESI DR2 analysis incorporates refinements to improve the robustness and accuracy of the results. Fiducial cosmology-related systematic errors were adjusted, increasing the systematic uncertainty on $\alpha_{\rm AP}$ from 0.1\% in DR1 to 0.18\%, based on including an evolving dark energy model motivated by the DR1 results in the test \cite{KP4s9-Perez-Fernandez}. Systematics related to unknown details of the small-scale galaxy-halo connection, which dominated the DR1 error budget \cite{KP4s10-Mena-Fernandez, KP4s11-Garcia-Quintero}, were refined to be tracer-specific. The contributions for LRG, ELG, QSO, and BGS tracers were updated to better capture effects dependent on redshift and galaxy type. The details of these choices are presented in Section V B of \cite{Y3.clust-s1.Andrade.2025}. The systematic error in $\aiso$ increases the total error budget in this parameter by a fractional amount between $1\%$ (for \bgs) and $9\%$ (\lrgelg), while for $\alpha_{\rm AP}$ the increase ranges between $0.1\%$ (\qso) and $2\%$ (\lrgelg).  Thus, in all cases, our error budget remains dominated by statistical errors associated with finite survey volume and sampling density. Additionally, we assessed the impact of including theoretically motivated systematic correlations across redshift bins and found no significant effect on our main results. 
    
\end{itemize}
The updated methods, systematic error estimates and the range of validation tests performed on blinded data before the pipeline was frozen are described in detail in the supporting publication \cite{Y3.clust-s1.Andrade.2025}.

\begin{figure*}
    \centering
    \begin{tabular}{ccc}
        \includegraphics[width=0.32\textwidth]{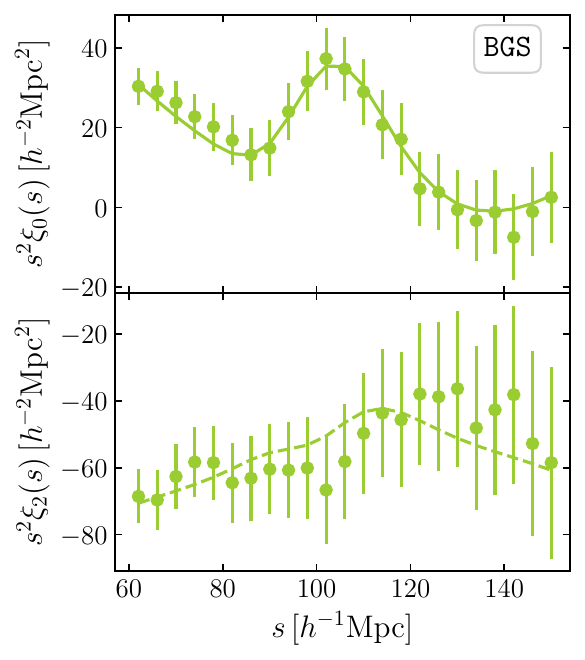} & 
        \includegraphics[width=0.32\textwidth]{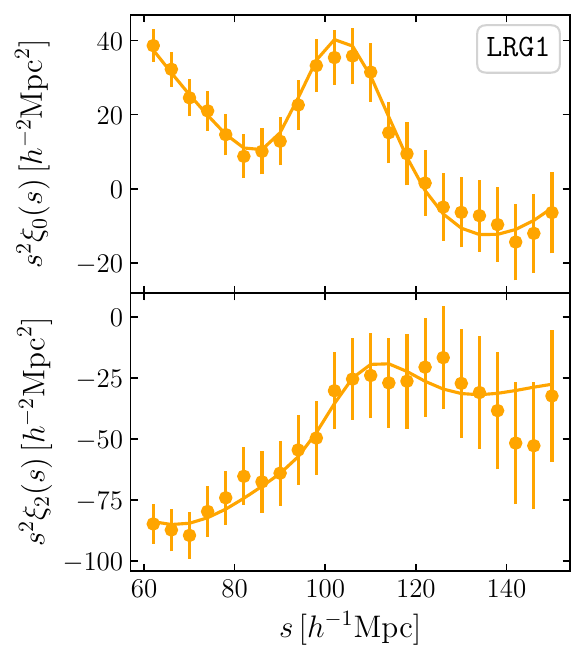} & 
        \includegraphics[width=0.32\textwidth]{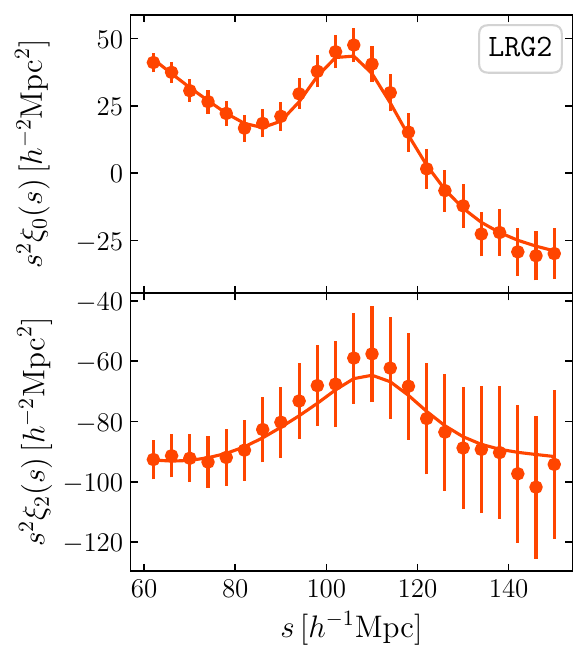} \\
        \includegraphics[width=0.32\textwidth]{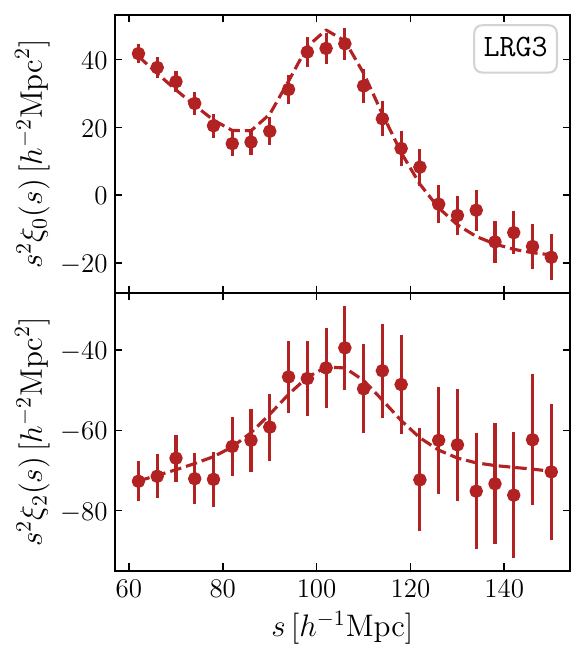} &
        \includegraphics[width=0.32\textwidth]{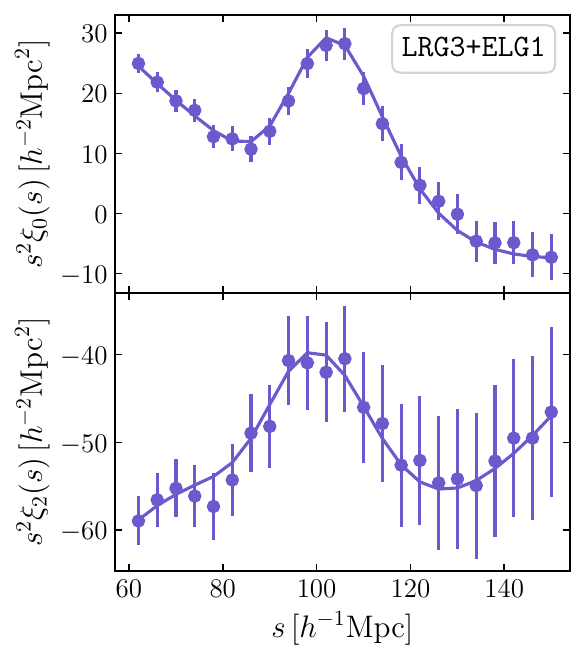} & 
        \includegraphics[width=0.32\textwidth]{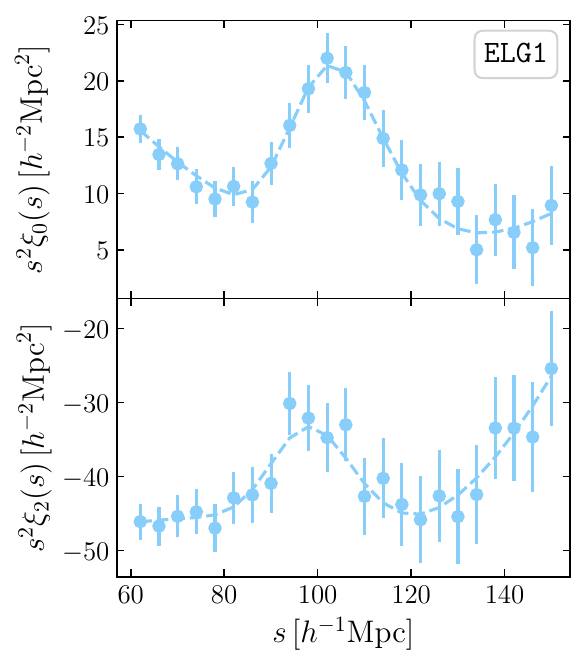} \\
        \includegraphics[width=0.32\textwidth]{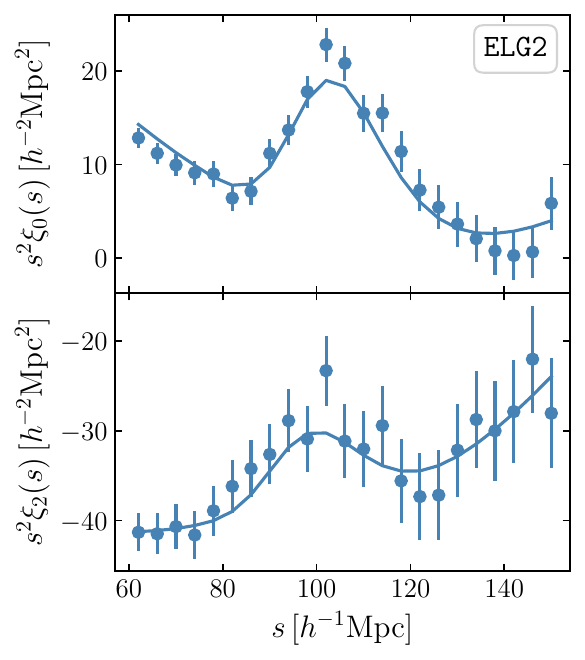} &
        \includegraphics[width=0.32\textwidth]{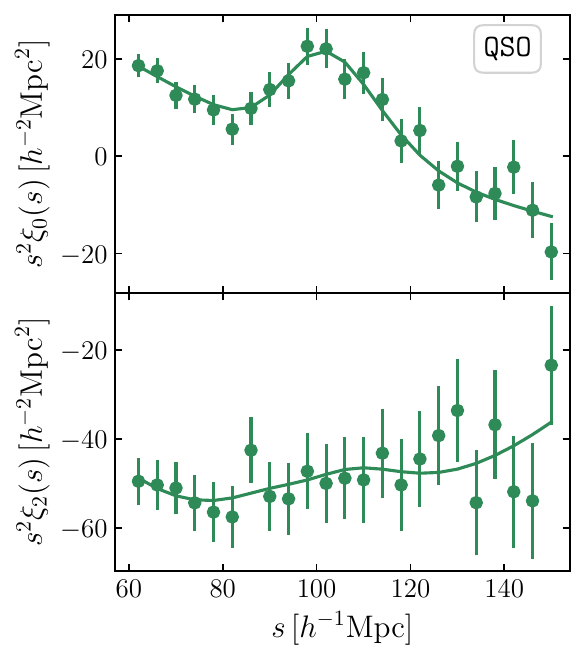} & 
        \includegraphics[width=0.32\textwidth]{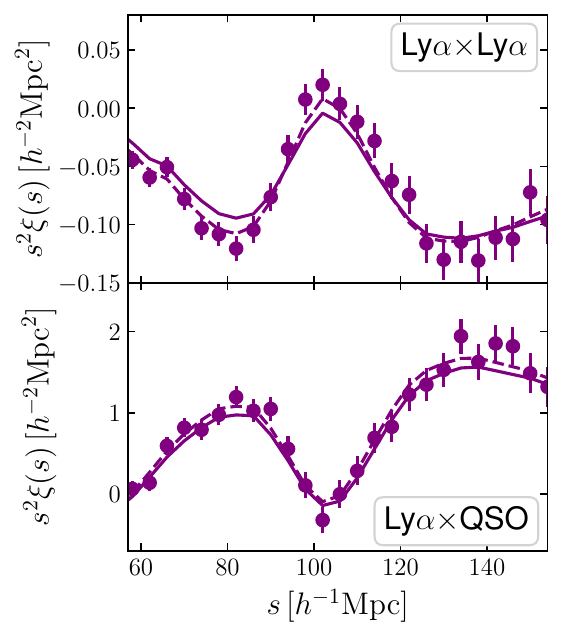}
    \end{tabular}
    \caption{The first eight panels show the multipole moments of the DESI DR2 correlation functions of galaxies and quasars, where the upper and lower subpanels display the monopole and quadrupole moments, respectively. The filled circles correspond to the data measurements and the lines show the best-fit BAO model. We use a solid line for model fits to those samples used in our analysis, and  a dashed line otherwise. Error bars represent 68\% confidence intervals. The last panel (bottom right) shows the autocorrelation of the \lya\ forest (upper sub-panel), and the cross-correlation between the \lya\ forest and the quasars (bottom sub-panel), where the 2D clustering information has been compressed into a single wedge. The solid line in this panel is the baseline model, while the dashed line includes a broad-band polynomial variation that provides a slightly better fit, but does not significantly shift the BAO position (see \cite{DESI.DR2.BAO.lya} for details).}
    \label{fig:clustering_measurements}
\end{figure*}

\cref{fig:clustering_measurements} shows the correlation function multipoles around the BAO scale measured from the reconstructed DESI DR2 catalogs (first eight panels, starting from the upper left). For those data vectors that are used for the cosmological inference presented in the following sections, we also show the best-fit BAO model as a solid line. The acoustic feature is successfully detected as a distinct peak in the correlation around $100 \hinvmpc$ for all tracers. The statistical significance of the detection ranges from 5.6$\sigma$ for the \qso, to 14.7$\sigma$ for \lrgelg, which constitutes the strongest detection of the BAO feature from a galaxy survey to date.  

\begin{table*}
    \centering
    \small
    \resizebox{\textwidth}{!}{
    \begin{tabular}{l|c|c|c|c|c|c|c|c|c}
    \hline
    \hline
     Tracer & $z_{\rm eff}$ & $\alpha_{\rm iso}$   & $\alpha_{\rm AP}$   & $D_{\rm V}/r_{\rm d}$   & $D_{\rm M}/D_{\rm H}$   & $r_{\rm V,M/H}$   & $D_{\rm M}/r_{\rm d}$   & $D_{\rm H}/r_{\rm d}$   & $r_{\rm M,H}$ \\
    \hline
     {\tt BGS}       &          0.295 & $0.9857 \pm 0.0093$  & ---                 & $7.942 \pm 0.075$       & ---                     & ---                & ---                     & ---                     & ---                         \\
     {\tt LRG1}      &          0.510 & $0.9911 \pm 0.0077$  & $0.9555 \pm 0.0261$ & $12.720 \pm 0.099$      & $0.622 \pm 0.017$       & $0.050$            & $13.588 \pm 0.167$      & $21.863 \pm 0.425$      & $-0.459$                    \\
     {\tt LRG2}      &          0.706 & $0.9749 \pm 0.0067$  & $0.9842 \pm 0.0227$ & $16.050 \pm 0.110$      & $0.892 \pm 0.021$       & $-0.018$           & $17.351 \pm 0.177$      & $19.455 \pm 0.330$      & $-0.404$                    \\
     {\tt LRG3+ELG1} &          0.934 & $0.9886 \pm 0.0046$  & $1.0237 \pm 0.0157$ & $19.721 \pm 0.091$      & $1.223 \pm 0.019$       & $0.056$            & $21.576 \pm 0.152$      & $17.641 \pm 0.193$      & $-0.416$                    \\
     {\tt ELG2}      &          1.321 & $0.9911 \pm 0.0071$  & $1.0257 \pm 0.0237$ & $24.252 \pm 0.174$      & $1.948 \pm 0.045$       & $0.202$            & $27.601 \pm 0.318$      & $14.176 \pm 0.221$      & $-0.434$                    \\
     {\tt QSO}       &          1.484 & $1.0032 \pm 0.0153$  & $0.9885 \pm 0.0564$ & $26.055 \pm 0.398$      & $2.386 \pm 0.136$       & $0.044$            & $30.512 \pm 0.760$      & $12.817 \pm 0.516$      & $-0.500$                    \\
     {\tt Lya}       &          2.330 & $0.9971 \pm 0.0082$  & $1.0071 \pm 0.0216$ & $31.267 \pm 0.256$      & $4.518 \pm 0.097$   &   $0.574$                 & $38.988 \pm 0.531$      & $8.632 \pm 0.101$      & $-0.431$                     \\
         \hline
     {\tt LRG3}      &          0.922 & $0.9936 \pm 0.0053$  & $0.9996 \pm 0.0172$ & $19.656 \pm 0.105$      & $1.232 \pm 0.021$       & $0.106$            & $21.648 \pm 0.178$      & $17.577 \pm 0.213$      & $-0.406$                    \\
     {\tt ELG1}      &          0.955 & $0.9888 \pm 0.0091$  & $1.0574 \pm 0.0291$ & $20.008 \pm 0.183$      & $1.220 \pm 0.033$       & $0.420$            & $21.707 \pm 0.335$      & $17.803 \pm 0.297$      & $-0.462$                    \\
     \hline
     \hline
    \end{tabular}
    }
    \caption{Constraints on the BAO scaling parameters and distance ratios at effective redshifts $z_{\rm eff}$ from fits to the clustering measurements of DESI DR2 galaxies and quasars, and the \lya\ forest. The constraints are expressed in terms of the mean and standard deviation of the marginalized posterior of each parameter. We also show the cross-correlation coefficients $r_{\rm V,M/H}$ between $\DVrd$ and $\DM/\DH$ and $r_{\rm M,H}$ between $\DMrd$ and $\DHrd$. For \lya, we quote here the values of $\aiso$ and $\alpha_{\rm AP}$ defined in the same way as for other tracers (\cref{eqn:aisoap_defs}) although this is different to the best-measured combination as discussed in \cref{subsec:lya_clustering}. The results for the \lrgth\ and \elgo\ tracer redshift bins, shown in the bottom two rows, are not used for cosmology inference in this paper since they are correlated with and superseded by the \lrgelg\ results that are used instead. These constraints include the contribution from the systematic error budget described in \cref{sec:bao_measurements}.
    }
    \label{tab:alphas}
\end{table*}

\subsection{Clustering from the Lyman-$\alpha$ Forest}
\label{subsec:lya_clustering}

The DR2 \lya forest BAO measurement \cite{DESI.DR2.BAO.lya} is based on the combination of four correlation functions: the auto-correlation function of the \lya forest in region A ($1040 - 1205$ \AA; Ly$\alpha$(A)$\times$Ly$\alpha$(A)), the auto-correlation of the \lya forest in region A with region B ($920 - 1020$ \AA; Ly$\alpha$(A)$\times$Ly$\alpha$(B)), the cross-correlation of the forest in region A with quasars (Ly$\alpha$(A)$\times$QSO), and of region B with quasars (Ly$\alpha$(B)$\times$QSO). We measure the \lya forest in these two distinct regions because region B has additional astrophysical noise due to higher-order Lyman series absorption, and the spectra of region B are generally lower signal-to-noise ratio because this region is only observed in higher-redshift quasars. We do not measure the auto-correlation of region B nor the auto-correlation of the quasars because these have even lower signal-to-noise ratio. For regions A and B we measure the forest absorption relative to an estimate of the forest continuum with the \texttt{picca} code.\footnote{\url{https://github.com/igmhub/picca}} This code fits a mean continuum model plus two  diversity parameters for each quasar to derive the over or under-density of the forest absorption. 

We measure these four correlation functions with the \texttt{picca} analysis code, along with the covariance and cross-covariance of the four correlation functions. The correlations are calculated from $0 - 200\,h^{-1}$\,Mpc in $4\,h^{-1}$\,Mpc bins in $r_\|$ and $r_\perp$ that correspond to comoving distances along and transverse to the line of sight, respectively. We fit the correlation functions with the \texttt{Vega} package\footnote{\url{https://github.com/andreicuceu/vega}} to measure the two BAO parameters ($\alpha_\|, \alpha_\perp$) and 15 nuisance parameters that include the \lya forest bias, the \lya redshift-space distortion (RSD) parameter, quasar bias, the biases of various metals in the IGM, and several others. All four correlation functions are fit in 2D from $30 < r < 180\,h^{-1}$\,Mpc. The fit has 9306 data points and 17 parameters. Further details are given in \cite{DESI.DR2.BAO.lya} and references therein. Figure~\ref{fig:clustering_measurements} shows the \lya autocorrelation function and the cross-correlation with quasars in the bottom right panel. This panel compresses these 2D measurements into a single correlation, while the analysis in~\cite{DESI.DR2.BAO.lya} uses the whole 2D information without this compression.

We finalized the \lya forest analysis choices with blinded data and mock datasets before we measured the BAO parameters. Unlike for lower-redshift galaxies and quasars, the blinding procedure for the \lya forest analysis shifts the BAO peak location in the correlation function (following the method developed for DR1, see  \cite{DESI2024.IV.KP6}). Briefly, we calculated a correlation function model that sets nuisance parameters to the values in DR1 and then computed a blinding template with \texttt{Vega} for some shift $\Delta \alpha_\|, \Delta \alpha_\perp$ that our analysis pipeline (\texttt{picca}) applied to any calculation of the correlation function. The shift values were not stored anywhere, and instead we stored the random number seed and other information necessary to recover the shift. The virtue of blinding the correlation function, rather than the catalog-level blinding used at lower redshifts, is that it preserves the locations of features in the correlation function due to metal-line contamination, which otherwise would have revealed the magnitude and direction of the blinding.

There are two substantial improvements in the \lya analysis relative to DESI DR1 \cite{DESI2024.IV.KP6} that we describe in the companion paper \cite{DESI.DR2.BAO.lya} and additional, important improvements that we describe in two supporting papers \cite{Y3.lya-s1.Casas.2025,Y3.lya-s2.Brodzeller.2025}. The two substantial improvements are in how we account for the distortion of the correlation function due to the continuum fitting, and the modeling of metal contamination in the forest. Supporting paper \cite{Y3.lya-s1.Casas.2025} describes improvements to the number and quality of mock catalogs used to validate our analysis pipeline. The most significant improvement in the mocks is the inclusion of non-linear broadening of the BAO peak. Another important improvement is the development of a new software package to identify DLAs based on template fitting, as well as a careful characterization of the purity and completeness of our DLA catalog with mock datasets. We now use the DLA template-fitting package, as well as two other codes based on a Convolutional Neural Network \cite{Wang2022} and a Gaussian Process method \cite{Ho2021}, to produce the DLA catalog. The new template-fitting DLA finder, the performance of the three finders, and the construction of the final DLA catalog are described in another supporting paper \cite{Y3.lya-s2.Brodzeller.2025}. As shown in \cite{DESI.DR2.BAO.lya}, the cumulative impact of all these improvements on the BAO results is very small.

The \lya forest measures the BAO signal much better in the line-of-sight direction than in the transverse direction due to the large value of the \lya\ RSD parameter. The optimal combination of $\alpha_\|$ and $\alpha_\perp$ for the \lya dataset is $\alpha_\|^{0.55} \alpha_\perp^{0.45}$, which differs from the best-measured $\aiso$ quantity for the galaxy and quasar BAO in \cref{eqn:aisoap_defs}: we measure this optimal combination with a precision of 0.64\% at an effective redshift of $z_{\rm eff}=2.33$ with the DR2 \lya sample, a significant improvement relative to even the 1.1\% measurement from DESI DR1. Nevertheless, in this paper we still report \lya\ BAO results in the conventional $\aiso$, $\alpha_{\rm AP}$ basis as for the other tracers. 

Several recent theoretical studies have reported $\sim 3\sigma$ measurements of a shift in the BAO position in the \lya forest due to non-linear evolution \cite{Sinigaglia2024,deBelsunce2024,Hadzhiyska2025}. Each of these studies has various strengths and weaknesses, and do not agree on the magnitude and direction of the shift, so we have 
adopted a theoretical systematic uncertainty of $\Delta \alpha_\| = 0.3\%,\,\Delta \alpha_\perp = 0.3\%$, which approximately encompasses the theoretical estimates. We add this systematic uncertainty to the statistical covariance matrix to obtain the total error budget. This systematic error reduces the precision of the isotropic BAO measurement to 0.7\%. We discuss the theoretical studies and our choice of systematic error further in \cite{DESI.DR2.BAO.lya}.

\subsection{DESI Distance Measurements}
\label{subsec:distances}

As implied by \cref{eqn:alpha_defs}, once the BAO scaling parameters are fit through a template approach, we can use the assumed fiducial cosmology to convert our measurements into cosmological distance estimates. \Cref{tab:alphas} presents the distance measurements derived from the DESI DR2 BAO measurements, reported at the effective redshift $z_\mathrm{eff}$ (defined in Eq.~(2.1) of \cite{DESI2024.III.KP4}) for each bin. This definition of $z_\mathrm{eff}$ is chosen to best represent the redshift of the maximum statistical weight of each sample. However, as long as the conversion of the measured BAO $\alpha$ parameters to distances at this redshift is correctly performed, small changes to the definition of $z_\mathrm{eff}$ are inconsequential for cosmology inference.  For \texttt{BGS}, we only report the angle-averaged distance ratio $D_{\rm V}/r_{\rm d}$, since we performed a purely isotropic BAO fit to the monopole.  For the other tracers, we report the marginalized constraints on angle-averaged distance, $D_{\rm V}/r_{\rm d}$, and the anisotropic factor, $D_{\rm M}/D_{\rm H}$, that is inversely proportional to $\alpha_{\rm AP}$. We also convert these into perpendicular and parallel distances. The reported constraints include the contribution from the systematic error budget that is described in \cref{sec:bao_measurements}, and the final distance measurements and uncertainties do not depend on the choice of fiducial cosmology (see Section VII of \cite{Y3.clust-s1.Andrade.2025}).

We note that any calculations of the degree of consistency or otherwise of DESI BAO results with those of any external dataset mathematically cannot depend on whether they are performed in terms of $\left(D_{\rm V}/r_{\rm d}, D_{\rm M}/D_{\rm H}\right)$ or $\left(\DMrd, \DHrd\right)$, or any other rotated basis in this space.

\begin{figure*}
    \centering
    \includegraphics[width=0.8\linewidth]{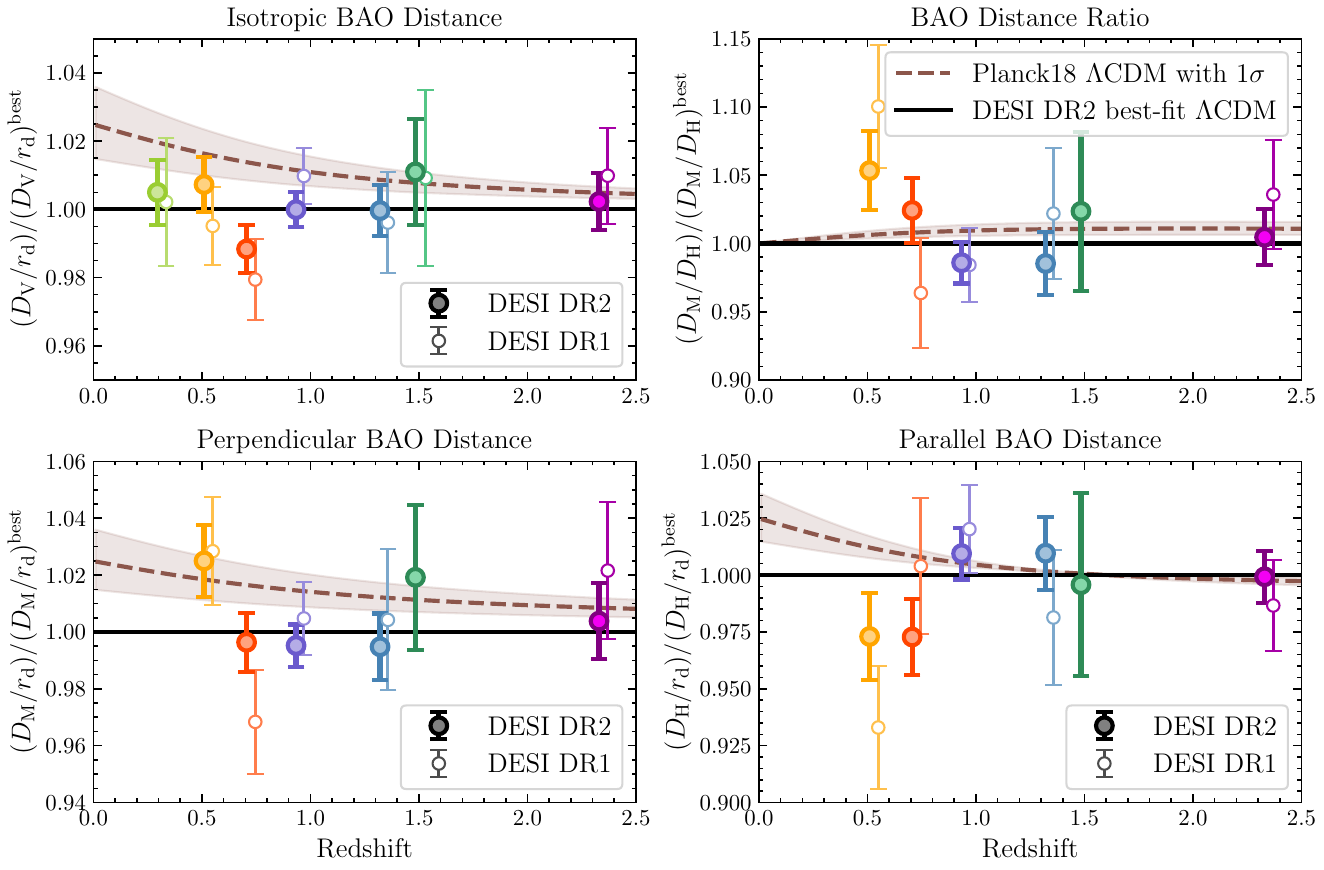}
    \caption{DESI DR2 BAO measurements compared to DR1 measurements. In each panel, DR1 and DR2 results are shown with empty and filled circles, respectively. Results are shown for the following tracers in order of increasing redshift: \bgs\ (yellowgreen points), \lrgo\ (orange), \lrgt\ (orangered), \lrgelg\ (slateblue), \elgt\ (steelblue), \qso\ (seagreen) and \lya\ (purple) \cite[][Appendix D]{DESI2024.II.KP3}. The distances are normalized by the DESI DR2 best-fit $\Lambda$CDM model predictions (black solid lines). All systematic errors are included. The \Planck~$\Lambda$CDM predictions are shown with brown dashed lines with $68\%$ confidence intervals in the brown shaded region. A small artificial offset in redshift has been applied to the DESI DR1 data points for a clearer comparison. }
    \label{fig:HD_BAO}
\end{figure*}

We show a visualization of the cosmological distance estimates in \cref{fig:HD_BAO}, where our measurements have been normalized by the best-fit DESI $\Lambda$CDM predictions. The previous BAO measurements derived from DESI DR1 \cite{DESI2024.III.KP4,DESI2024.IV.KP6} are included for comparison. This plot highlights the improvement in the parameter precision of the new data release, as well as the inclusion of a new measurement of the BAO distance ratio (and thus of the perpendicular and parallel distances individually) at $z \sim 1.5$ from the \qso\ (green points).

\begin{figure}
    \centering
    \includegraphics[width=\columnwidth]{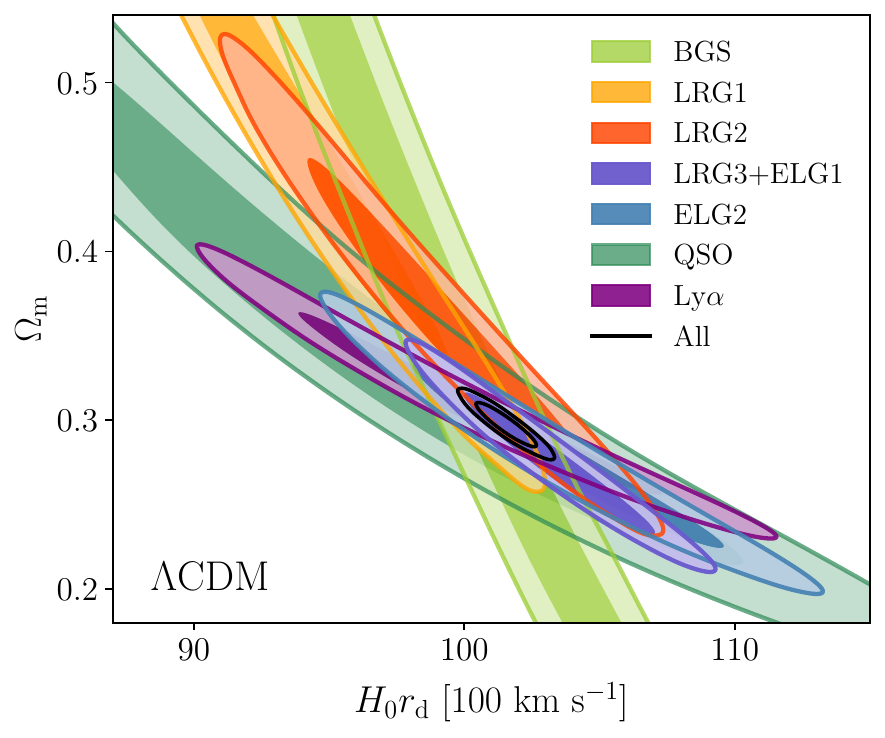}
    \caption{68\% and 95\% confidence contours for $H_0 r_d$ and $\Om$ under the $\Lambda$CDM model. From low redshift to high redshift, the increase on the effective redshift of the sample induces a counter clockwise shift in the degeneracy direction. The results from each individual tracer are mutually consistent and complementary in providing tighter constraints.
    }
    \label{fig:lcdm_bao_tracers}
\end{figure}

\subsubsection{Internal consistency of DESI BAO measurements}
\label{subsec:internal_consistency}

As DR1 is a subset of DR2, and the data reduction and analysis pipelines are very similar, we conservatively assume a perfect correlation between the DR1 and DR2 measurements shown in \cref{fig:HD_BAO} to assess their consistency. We find improvements in the statistical precision between 30\% and 50\% for DR2. We find that the largest discrepancies between results are $1.56\sigma$ in $\aiso$ (for \lrgelg) and $1.93\sigma$ in $\alpha_{\rm AP}$ (for \lrgt), with the joint uncertainty defined as $\sigma\equiv\sqrt{\sigma_{\rm DR2}^2+\sigma_{\rm DR1}^2-2C\sigma_{\rm DR1}}$, where $C$ is the correlation coefficient between the two samples.\footnote{For galaxy and quasar clustering measurements, we use $C=\sqrt{N_{\rm DR1}/N_{\rm DR2}}$ , where $N_{\rm DR1}$ and $N_{\rm DR2}$ are the numbers of tracers in DR1 and DR2, respectively. For Lya, we use $C=0.61$, estimated using the method described in Appendix F of \cite{DESI2024.IV.KP6}.} The Kolmogorov-Smirnov (KS) statistic for the differences is 0.25 for 12 data points, corresponding to a $p$-value of 0.40 and indicating high consistency between DR1 and DR2.

In order to check the internal consistency of the BAO measurements in different redshift bins, we interpret each within the flat \lcdm\ cosmological model, using the methods described below in \cref{sec:inference_method}. In this model, BAO measurements constrain the parameter $\Om$ and the combination $\Hrd$. \cref{fig:lcdm_bao_tracers} shows the results for each individual redshift bin and the combination: the posteriors all have significant overlap, indicating no internal inconsistency within a \lcdm\ model. The largest difference in the recovered parameters is $1.8\sigma$ (between \lrgo\ and \lrgelg). The joint fit to all tracers returns a best-fit $\chi^2/{\rm dof}=10.2/(13-2)$.

\subsubsection{Comparison to SDSS}
\label{subsec:comparison_to_sdss}

In \cite{DESI2024.VI.KP7A} we reported a $\sim3\sigma$ difference between the DESI DR1 value of $\DMrd$ measured in the \lrgt\ redshift bin at $z_{\rm eff}=0.71$ and the corresponding result at $z=0.7$ previously reported by SDSS \cite{2021MNRAS.500..736B, 2020MNRAS.498.2492G, Alam-eBOSS:2021} from the eBOSS LRG sample (although the discrepancy in the two-dimensional $\DMrd$-$\DHrd$ plane was less significant). The new DR2 result in this redshift bin lies between the two in the $\DMrd$-$\DHrd$ plane. 

To investigate the consistency, we compare measurements assuming that the coefficient of correlation between the power spectrum measurements in the two surveys, defined as
\begin{equation}
    \label{eq:corr_coeff}
    C=\frac{\mathrm{Cov}(\hat{P}_1,\hat{P}_2)}{\sqrt{\mathrm{Var}(\hat{P}_1)\mathrm{Var}(\hat{P}_2)}}=\frac{\int dV X_1X_2/X_{12}}{\sqrt{(\int dVX_1)(\int dVX_2)}},
\end{equation}
also describes the correlation between BAO measurements. Here $X_i =(n_iP_i)^2/(1+n_iP_i)^2$ for $i=1,2$, and $X_{12} = \left({n_1n_2P_1P_2}\right)^2\left({n_{12}\sqrt{P_1P_2}+n_1n_2P_1P_2}\right)^{-2}$, 
where $n_1$, $n_2$, and $n_{12}$ respectively are the mean galaxy number densities of DR2 \lrgt, the eBOSS LRGs, and the common sample. $P_1$ and $P_2$ are the corresponding power spectrum amplitudes, for simplicity assumed to be the same here. This calculation gives an estimate $C\sim0.57$ for the \lrgt\ and eBOSS samples---significantly higher than the 0.21 for DR1 due to the larger degree of overlap between the DR2 and SDSS footprints. Assuming this level of correlation between the results, we find the discrepancy between the DR2 and SDSS results has reduced from $3\sigma$ to $\sim2.6\sigma$. Although the assumption of no correlation is less plausible, it sets a lower limit of the discrepancy at the $1.9\sigma$ level (compared to $2.7\sigma$ in DR1). When using the DESI reanalysis of SDSS (as shown in the bottom row of Table. 17 in \cite{DESI2024.III.KP4}), the discrepancy is $2.3\sigma$ when using $C=0.57$ and only $1.5\sigma$ when assuming no correlation (compared to $2.8\sigma$ and $2.5\sigma$ for DR1).

For a systematic comparison between DR2 and SDSS, we compare the results in four redshift bins (\lrgo, \lrgt, \qso\ and \lya) where the effective redshifts of DESI and SDSS tracers are similar ($\Delta z_{\rm eff}<0.03$). We convert the SDSS results into $\alpha_{\rm iso}$ and $\alpha_{\rm AP}$ using the reported correlations between $\aper$ and $\apar$, and then compute the significance of differences between these values and the DR2 results assuming no correlation between the surveys. The biggest difference we find is $1.49\sigma$ for $\alpha_{\rm iso}$ and $1.69\sigma$ for $\alpha_{\rm AP}$. The KS statistic is $0.30$, $p=0.39$. If we conservatively treat SDSS as a pure subsample of DESI DR2 and use the same method to calculate correlation coefficients as in the DR1 vs DR2 comparison above, we find $p=0.15$ from the KS test. We conclude that there is no significant discrepancy between the DR2 measurements and those from SDSS. 

Unlike the case for DR1, in DR2 the effective volume and statistical constraining power of each DESI redshift bin is far larger than that of the corresponding SDSS counterpart. We therefore no longer consider any combination of bins picked from DESI and SDSS data at different redshifts as done in \cite{DESI2024.VI.KP7A}.

\section{External Data}
\label{sec:external_data}

We use results from a number of external experiments together with our BAO measurements in order to obtain the most precise cosmological constraints. These are briefly described below.

\subsection{Big Bang Nucleosynthesis (BBN) prior on $\Ob h^2$}
\label{subsec:bbn}

Absent an external determination of $\rd$, BAO measurements serve as an uncalibrated ruler and therefore measure the degenerate combination $\Hrd$ rather than $H_0$ and $\rd$ individually. Given that BAO also measure $\Om$, from \cref{eqn:rdformula} we can see that, assuming standard neutrino content ($\Neff=3.044$ and $\sumnu=0.06$ eV), knowledge of $\Ob h^2$ is sufficient to break this degeneracy. Careful observational determinations of the primordial deuterium abundance D/H \cite{Cooke:2018} and the helium fraction $Y_{\rm P}$ \cite{Aver:2022} in astrophysical systems can be connected to predictions from BBN for the abundances of D and $^4$He in the early Universe to determine the baryon-to-photon ratio and thus the physical baryon density $\Ob h^2$, when combined with a measurement of the CMB temperature \cite{PDG:2022}.

As in DR1, we use estimates of $\Ob h^2$ determined by \cite{Schoeneberg:2024} based on the \texttt{PRyMordial} code \cite{Burns:2024} including marginalization over uncertainties in nuclear reaction rates. This corresponds to 
\be
\label{eq:BBN_LCDM}
\Ob h^2= 0.02218 \pm 0.00055,
\ee
in the standard \lcdm{} model, and
\be
\label{eq:BBN_Neff}
\Ob h^2=  0.02196 \pm 0.00063
\ee
when allowing $\Neff$ to vary (in which case there is also a covariance between $\Ob h^2$ and $\Neff$). We use this prior when we require calibrated results that are independent of CMB information, and denote it as BBN in plots and tables.

\subsection{Cosmic microwave background (CMB)}
\label{subsec:CMB}

The power spectra of anisotropies in the CMB contain a wealth of information on cosmological parameters that is complementary to that from BAO. The large-scale temperature and polarisation power spectra and CMB lensing power spectrum have been exquisitely measured by \Planck\ \cite{Planck-2018-overview,Planck-2018-cosmology}, with additional information from smaller scales and improved lensing reconstruction provided by the Atacama Cosmology Telescope (ACT; \cite{Madhavacheril:ACT-DR6,Qu:2023,MacCrann:2023}) and the South Pole Telescope (SPT; \cite{SPT}).

When using the full power of the CMB information, our baseline analysis makes use of the temperature ($TT$), polarization ($EE$) and cross ($TE$) power spectra from \Planck, specifically using the \texttt{simall}, \texttt{Commander} (for $\ell<30$) and \texttt{CamSpec} (for $\ell\geq30$) likelihoods, plus the combination of \Planck\ and ACT DR6 CMB lensing from \cite{Madhavacheril:ACT-DR6}.\footnote{We note that the results in \cite{DESI2024.VI.KP7A} used an older version of the \Planck+ACT lensing likelihood code, which has since been updated, leading to a small shift in neutrino mass results. This paper uses the updated version \texttt{v1.2} likelihood.} \texttt{CamSpec} \cite{Efstathiou:2021,Rosenberg:2022} is a new likelihood built on the latest \texttt{NPIPE} PR4 data release from the \Planck\ collaboration, which replaces the original \texttt{Plik} likelihood based on the older PR3 release and includes some important differences in methodology. Another set of likelihoods based on PR4, known as \texttt{LoLLiPoP} (for low $\ell$) and \texttt{HiLLiPoP} (for high $\ell$) have also been independently released \cite{Tristram:2023}, which for brevity we will refer to as \texttt{L-H} in the following. In \cite{DESI2024.VI.KP7A,DESI2024.VII.KP7B} we used \texttt{Plik} as our default high-$\ell$ likelihood but noted that both \texttt{CamSpec} and \texttt{L-H} reduced the so-called $A_L$ anomaly which has a small effect on certain cosmological parameters. They also make use of slightly more data, with larger sky fractions, than \texttt{Plik}. Our choice of \texttt{CamSpec} as the default is based on the speed of evaluation of the likelihood code relative to \texttt{L-H}, but where any results depend on the choice of likelihood (notably for the sum of neutrino masses $\sumnu$) we provide both sets of results.

Fits using the full CMB likelihoods require specification of a full physical model, including computation of perturbations, the late integrated Sachs-Wolfe (ISW) effect and lensing-induced distortions. However, certain aspects of the early Universe can be robustly constrained independently of any modifications of the late-time cosmological model. The most precisely determined quantity from the CMB is the angular scale of the acoustic fluctuations, $\theta_\ast=r_\ast/\DM(z_\ast)$, where $r_\ast$ is the comoving sound horizon at recombination and
$\DM(z_\ast)$ is the transverse comoving distance to that redshift. This quantity is the direct analogue of the measurement of $\DMrd$ that we make using BAO. We can consider `BAO-only' information by combining galaxy BAO with $\theta_\ast$. When doing this, we use a Gaussian prior on $\theta_\ast$,
\begin{equation}
100\theta_\ast = 1.04110 \pm 0.00053.
\label{eq:thetastar_prior}
\end{equation}
The mean value of this prior matches the \Planck\ result for \lcdm, and the width of the prior has been conservatively increased by $\sim75\%$ over the \Planck\ reported uncertainty, to account for the (small) variation seen over different assumed late-time models.

More generally, the CMB determines more information than just $\theta_\ast$ independent of the late-time evolution. Following \cite{EarlyUniverseCompression}, this can be expressed in terms of correlated Gaussian posteriors on the parameters $\left(\theta_\ast,\ob,\obc\right)$ obtained after marginalizing over the ISW and lensing contributions and other possible late-time effects.\footnote{An alternative compression in terms of shift parameters $R$ and $l_A$, and the physical baryon density $\ob$ was introduced by \cite{Shift_parameters_Wang_2007}. We have compared this compression and found that it gives equivalent results to the one described above.} We use this information, in the form of a correlated Gaussian prior, as an alternative to using the full CMB likelihood that is also more model-independent. In practice we determine the correlation between parameters from a set of early-Universe results provided by \cite{EarlyUniverseCompression} based on the \texttt{CamSpec} likelihood\footnote{In the form of MCMC chains that can be downloaded from \url{https://github.com/cmbant/PlanckEarlyLCDM}.} and use this to define the Gaussian prior. Numerical details of the implementation are given in \cref{appendix:cmb_compression}. We refer to this in the following text and plots as $(\theta_\ast,\ob,\obc)_{\rm CMB}$ priors. 

For constraints on dark energy evolution in particular, these CMB priors contain most of the relevant information from the full CMB likelihoods, in the form of a high-redshift calibration of the low-redshift observations like BAO that directly probe the background evolution. In this form, the CMB information is independent of whether or not dark energy evolves. Such evolution would alter the distance to last scattering, which is already allowed for in the definition of $\theta_\ast$ as the angular size of the horizon; the densities $\ob$ and $\ocdm$ are determined via a ratio to the radiation density at $z_\ast$, which is independent of evolution at lower redshifts.

\subsection{Type Ia supernovae (SNe)}
\label{subsec:SN}

Type Ia SNe are luminous standardizable candles that provide another probe of the expansion history of the Universe, especially useful at low redshifts ($0.01<z<0.3$) where BAO measurements are limited by cosmic variance. There are three recent SNe datasets and likelihoods available for cosmology: the Pantheon+ \cite{Scolnic:2021amr,Brout:2022}, Union3 \cite{Rubin:2023} and Dark Energy Survey Year 5 (DESY5; \cite{DES:2024tys}) samples. 

The Pantheon+ and Union3 samples consist of compilations of 1,550 and 2,087 spectroscopically-classified SNe, respectively, drawn from multiple observational programs over the past few decades of observations. While 1,363 of the objects are in common between the two, there are differences in the calibration, modeling and  treatment of systematic uncertainties. The DESY5 sample consists of 1,635 photometrically-classified $z>0.1$ SNe with uniform calibration drawn from a single survey up to $z=1.13$, together with a small set of 194 SNe at $z<0.1$ drawn from historical sources, some of which are in common with the Pantheon+ and Union3 datasets. Despite the larger differences in the underlying data, the Pantheon+ and DESY5 analysis methodologies are quite similar in nature, although DESY5 uses the SALT3 \cite{SALT3} light-curve fitting model as opposed to SALT2 \cite{SALT2}, among other differences. Union3 shares a larger fraction of the data with Pantheon+ but uses a different approach based on the Bayesian hierarchical modeling framework \texttt{Unity1.5} \cite{Unity_Rubin}.

As in our DR1 analysis, we do not choose between these three options but instead present our headline results in combination with each SNe dataset independently, noting the differences in interpretation where appropriate. All three likelihoods are implemented in the \texttt{Cobaya} sampling code \cite{Torrado:2019,Torrado:2021} and the underlying data are publicly available.\footnote{\url{https://github.com/CobayaSampler/sn_data.git}}

In \cref{sec:de_constraints}, we illustrate the distance modulus residuals for the SNe datasets in redshift bins. For visual purposes only, we rebin the SNe in redshift using the same bins for all datasets, and then calculate the weighted average distance moduli and errors using the inverse of the covariance matrix (including both statistical and systematic errors and after removing the weighted mean of the distance moduli). All cosmological fits use the data and covariance as originally provided. It is important to note that in all SN samples the absolute magnitude $M$ of the SNe is completely degenerate with the Hubble parameter $H_0$ and therefore, unless some external calibration of either $M$ or $H_0$ is assumed, the distance moduli can be scaled by any constant offset. For cosmological inference we therefore always marginalize over the value of $M$; for plots we show the residuals of the distance modulus relative to the mean over all SNe. 

\subsection{Dark Energy Survey 3$\times$2pt}
\label{subsec:desy3-3x2pt}

Weak gravitational lensing of galaxies and its combination with galaxy clustering are local probes of structure formation. For models with an evolving dark energy equation of state, we use galaxy-galaxy, galaxy-shear, and shear-shear two-point correlation functions (referred to here as 3$\times$2pt) measurements from the Dark Energy Survey Year-3 (DESY3) analysis~\cite{2022PhRvD.105b3520A}. These data vectors were estimated from observations of over 10 million lens galaxies in the {\tt MagLim} sample covering an area of approximately 4000 square degrees. We adopt the same modeling choices and scale cuts as the corresponding DESY3 analysis~\cite{DES:2022ccp}. We choose the same priors as DESY3 with the exception of the parameters $w_{\rm 0}$, $\wa$, $H_{\rm 0}r_{\rm d}$ and $\Om$, where we impose priors that match those used for the DESI analysis in this paper. In particular, in post-processing we compute $H_{\rm 0}r_{\rm d}$ as a derived parameter and reweight the chains to impose the same prior for free parameters in the DESI chains. We sample the DESY3 likelihood using the publicly available DESY3 {\tt CosmoSIS} pipeline~\cite{Zuntz:2014csq}. To combine these constraints with DESI and SNe data at the posterior level, we use {\tt CombineHarvesterFlow}~\cite{Taylor:2024eqc} to fit normalizing flows to the DESY3 chains and reweight the DESI and DESI+SNe chains.

\section{Cosmological Inference}
\label{sec:inference_method}

\begin{figure*}[t]
    \centering
    \includegraphics[width=\columnwidth]{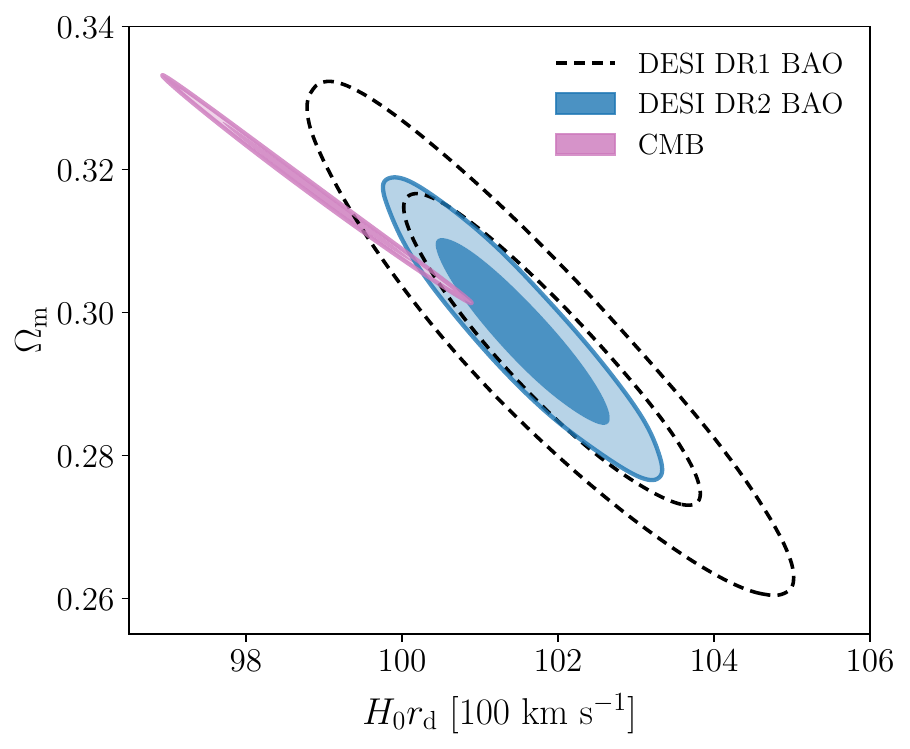}
    \includegraphics[width=\columnwidth]{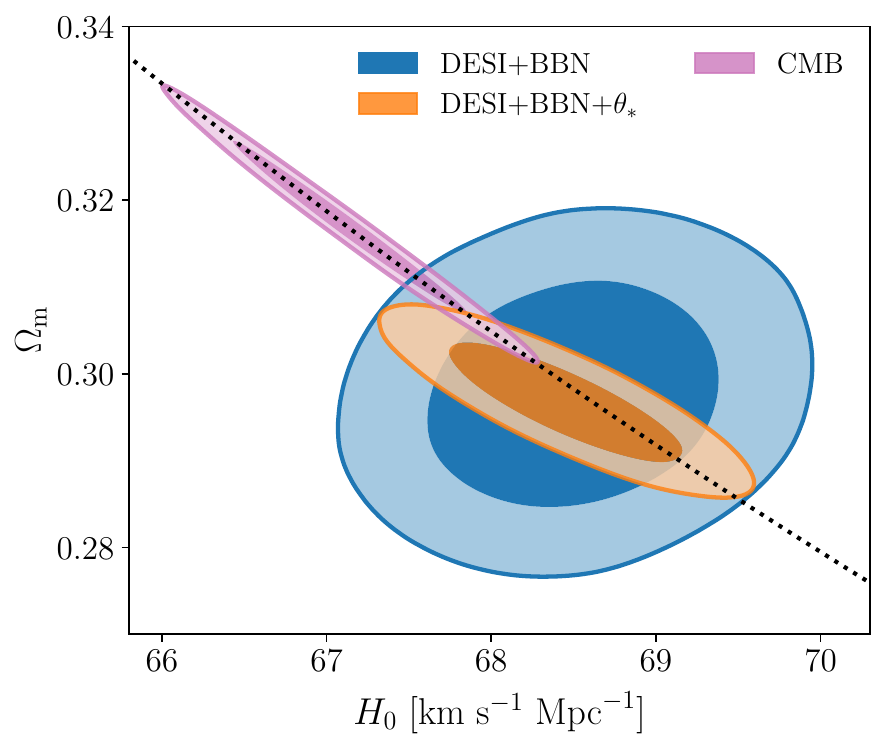}
    \caption{Cosmology results in the \lcdm\ model (\cref{sec:cosmologcal_constraints}). \textit{Left panel}: Marginalized posterior constraints on $H_0r_d$ and $\Om$ from DESI DR1 and DR2 BAO and the CMB (including \Planck\ and \Planck+ACT lensing), computed under the assumption of a $\Lambda$CDM cosmology. Contours show the 68\% and 95\% regions. DESI DR2 is fully consistent with DR1 but as the statistical uncertainty has significantly decreased, the tension with the CMB has increased. \textit{Right panel}: Posterior contours in $\Om$ and $H_0$ for CMB (pink) compared to DESI DR2 BAO calibrated with a BBN prior on $\ob$, with (orange) and without (blue) including the $\theta_\ast$ constraint from the CMB. The dotted black curve shows the direction of constant $\Om h^3$ matching the degeneracy direction of the CMB. 
    }
    \label{fig:lcdm_bao_constraints}
\end{figure*}

We use the cosmological inference code \texttt{Cobaya} \cite{Torrado:2021,Torrado:2019} to sample posteriors in parameter space for cosmological inference, using Metropolis-Hastings Monte Carlo Markov Chain (MCMC) sampling. Theory models are computed using interfaces to \texttt{CAMB} \cite{LewisCAMB:2000} and \texttt{Cobaya} likelihoods. When fitting to BAO data only, we sample over $\Omega_{\rm m}$ and the parameter $h\rd$, where $h = H_0/(100\,\kmsMpc)$. When the BAO data are calibrated through the use of an external prior such as BBN, we sample in $H_0$ and $\ob$ instead of $h\rd$. Runs involving use of the full CMB likelihood sample the base set of six cosmological parameters $\left(\obc,\ob,100\theta_{\rm MC},\ln(10^{10} A_s),n_s,\tau \right)$, where $\theta_\text{MC}$ is an approximation\footnote{$\theta_{\rm MC}$ is used for sampling because it helps approximate the degeneracy direction of the CMB parameter space while being faster to calculate the conversion to $H_0$ and $\Om$. However, any difference between the values of $\theta_{\rm MC}$ and $\theta_\ast$ only slightly affects the efficiency of the sampling while having no effect on the accuracy of the likelihood evaluation.} to the acoustic angular scale $\theta_\ast$ based on fitting formulae given in \cite{Hu:1996}, $\ln(10^{10} A_s)$ and $n_s$ are the amplitude and spectral index of the primordial scalar perturbations, and $\tau$ is the optical depth. Extended cosmological models additionally allow the spatial curvature $\Ok$, the dark energy equation of state parameters $w_0$ (or $w$) and $\wa$, or the sum of neutrino masses $\sumnu$ to vary in addition to the base parameters mentioned above. When $\sumnu$ is varied, we assume three degenerate mass eigenstates unless testing specific mass ordering scenarios; when it is not varied, it is fixed to a default value of 0.06 eV assuming a single non-zero mass eigenstate. Prior ranges on all sampled parameters match those given in Table 2 of \cite{DESI2024.VI.KP7A}.

The convergence criterion for MCMC sampling is that the Gelman-Rubin statistic \cite{GelmanRubin} satisfies $R-1<0.01$, or that the chains have an effective sample size of $\gtrsim10^3$, whichever is longer. Summary statistics for our chains as well as plots are obtained with \texttt{getdist}\footnote{\url{https://github.com/cmbant/getdist}} software package \cite{Lewis:2019xzd}. For 1D marginalized posterior results we quote the mean and standard deviation when the distributions are symmetric and the 68\% minimal credible interval when they are not. Where only limits on parameter values can be determined, upper or lower bounds are quoted at the 68\% level, \emph{except} for $\sumnu$ where we quote the 95\% upper bound to ensure comparability with previous work. 

In order to determine the best fit points and the corresponding $\chi^2$ we use the \texttt{iminuit} \cite{iminuit} algorithm starting from the maximum a posteriori (MAP) points of each of the chains in the MCMC sampling. When comparing the fits of two different models we compare the quantity $\dchisq\equiv-2\Delta\ln \mathcal{L}$ representing twice the difference in the negative log posteriors at the maximum posterior points for each model. This measure accounts for the contribution of any non-uniform priors when evaluating the differences between two MAP points. For likelihood combinations including DESY3 ($3\times2$pt), when computing the best-fit points and $\chi^2$, we directly use the version of the DESY3 likelihood \cite{2022PhRvD.105b3520A} implemented in the {\tt CosmoSIS} \cite{Zuntz:2014csq} pipeline.\footnote{When using the {\tt CosmoSIS} implementation of this likelihood, we sample using the same priors as in \cite{2022PhRvD.105b3520A}, except on the parameters $w_0$ and $w_a$, where the priors are chosen to match those in the rest of this paper, and the neutrino mass sum $\sumnu$, which is fixed to 0.06 eV as in the default case here.} For these combinations with DESY3, when calculating the deviance information criterion values, we separate the calculation of the mean $\langle \chi^2 \rangle$ via the sum $\langle \chi^2_{\text{DESI(+SNe)}}\rangle +\langle \chi^2_{\text{DESY3}} \rangle$ and compute each term with the corresponding weighted {\tt CombineHarvesterFlow} chains. This enables the use of the values of $\chi^2$ from each individual chain, while the distributions being averaged over are equivalent.

\section{Cosmological constraints in the $\Lambda$CDM model } \label{sec:cosmologcal_constraints}

\begin{table*}
    \centering
    \resizebox{\linewidth}{!}{
    \begin{tabular}{lccccc}
    \toprule
    Model/Dataset & $\Om$ & $H_0$ [km s$^{-1}$ Mpc$^{-1}$] & $10^3\Omega_\mathrm{K}$ & $w$ or $w_0$ & $w_a$ \\
    \midrule
    $\bm{\Lambda}$\textbf{CDM} &  &  &  &  &  \\
    CMB & $0.3169\pm 0.0065$ & $67.14\pm 0.47$ & --- & --- & --- \\
    DESI & $0.2975\pm 0.0086$ & --- & --- & --- & --- \\
    DESI+BBN & $0.2977\pm 0.0086$ & $68.51\pm 0.58$ & --- & --- & --- \\
    DESI+BBN+$\theta_*$ & $0.2967\pm 0.0045$ & $68.45\pm 0.47$ & --- & --- & --- \\
    DESI+CMB & $0.3027\pm 0.0036$ & $68.17\pm 0.28$ & --- & --- & --- \\
    \hline
    $\bm{\Lambda}$\textbf{CDM+}$\bm{\Omega_\mathrm{K}}$ &  &  &  &  &  \\
    CMB & $0.354^{+0.020}_{-0.023}$ & $63.3\pm 2.1$ & $-10.7^{+6.4}_{-5.3}$ & --- & --- \\
    DESI & $0.293\pm 0.012$ & --- & $25\pm41$ & --- & --- \\
    DESI+CMB & $0.3034\pm 0.0037$ & $68.50\pm 0.33$ & $2.3\pm1.1$ & --- & --- \\
    \hline
    $\bm{w}$\textbf{CDM} &  &  &  &  &  \\
    CMB & $0.203^{+0.017}_{-0.060}$ & $85^{+10}_{-6}$ & --- & $-1.55^{+0.17}_{-0.37}$ & --- \\
    DESI & $0.2969\pm 0.0089$ & --- & --- & $-0.916\pm 0.078$ & --- \\
    DESI+Pantheon+ & $0.2976\pm 0.0087$ & --- & --- & $-0.914\pm 0.040$ & --- \\
    DESI+Union3 & $0.2973\pm 0.0091$ & --- & --- & $-0.866\pm 0.052$ & --- \\
    DESI+DESY5 & $0.2977\pm 0.0091$ & --- & --- & $-0.872\pm 0.039$ & --- \\
    DESI+CMB & $0.2927\pm 0.0073$ & $69.51\pm 0.92$ & --- & $-1.055\pm 0.036$ & --- \\
    DESI+CMB+Pantheon+ & $0.3047\pm 0.0051$ & $67.97\pm 0.57$ & --- & $-0.995\pm 0.023$ & --- \\
    DESI+CMB+Union3 & $0.3044\pm 0.0059$ & $68.01\pm 0.68$ & --- & $-0.997\pm 0.027$ & --- \\
    DESI+CMB+DESY5 & $0.3098\pm 0.0050$ & $67.34\pm 0.54$ & --- & $-0.971\pm 0.021$ & --- \\
    \hline
    $\bm{w_0w_a}$\textbf{CDM} &  &  &  &  &  \\
    CMB & $0.220^{+0.019}_{-0.078}$ & $83^{+20}_{-6}$ & --- & $-1.23^{+0.44}_{-0.61}$ & $<-0.504$ \\
    DESI & $0.352^{+0.041}_{-0.018}$ & --- & --- & $-0.48^{+0.35}_{-0.17}$ & $<-1.34$ \\
    DESI+Pantheon+ & $0.298^{+0.025}_{-0.011}$ & --- & --- & $-0.888^{+0.055}_{-0.064}$ & $-0.17\pm 0.46$ \\
    DESI+Union3 & $0.328^{+0.019}_{-0.014}$ & --- & --- & $-0.70\pm 0.11$ & $-0.99\pm 0.57$ \\
    DESI+DESY5 & $0.319^{+0.017}_{-0.011}$ & --- & --- & $-0.781^{+0.067}_{-0.076}$ & $-0.72\pm 0.47$ \\
    DESI+$(\theta_*,\omega_\mathrm{b},\omega_\mathrm{bc})_\mathrm{CMB}$ & $0.353\pm 0.022$ & $63.7^{+1.7}_{-2.2}$ & --- & $-0.43\pm 0.22$ & $-1.72\pm 0.64$ \\
    DESI+CMB (no lensing) & $0.352\pm 0.021$ & $63.7^{+1.7}_{-2.1}$ & --- & $-0.43\pm 0.21$ & $-1.70\pm 0.60$ \\
    DESI+CMB & $0.353\pm 0.021$ & $63.6^{+1.6}_{-2.1}$ & --- & $-0.42\pm 0.21$ & $-1.75\pm 0.58$ \\
    DESI+CMB+Pantheon+ & $0.3114\pm 0.0057$ & $67.51\pm 0.59$ & --- & $-0.838\pm 0.055$ & $-0.62^{+0.22}_{-0.19}$ \\
    DESI+CMB+Union3 & $0.3275\pm 0.0086$ & $65.91\pm 0.84$ & --- & $-0.667\pm 0.088$ & $-1.09^{+0.31}_{-0.27}$ \\
    DESI+CMB+DESY5 & $0.3191\pm 0.0056$ & $66.74\pm 0.56$ & --- & $-0.752\pm 0.057$ & $-0.86^{+0.23}_{-0.20}$ \\
    DESI+DESY3 (3$\times$2pt)+Pantheon+ & $0.3140\pm 0.0091$ &---&---& $-0.870\pm 0.061$ & $-0.46^{+0.33}_{-0.29}$ \\
    DESI+DESY3 (3$\times$2pt)+Union3 & $0.333\pm 0.012$ &---&---& $-0.68\pm 0.11$ & $-1.09^{+0.48}_{-0.39}$ \\
    DESI+DESY3 (3$\times$2pt)+DESY5 & $0.3239\pm 0.0092$ &---&---& $-0.771\pm 0.068$ & $-0.82^{+0.38}_{-0.32}$ \\
    \hline
    $\bm{w_0w_a}$\textbf{CDM+}$\bm{\Omega_\mathrm{K}}$ &  &  &  &  &  \\
    DESI & $0.357^{+0.041}_{-0.030}$ & --- & $-2\pm56$ & $-0.45^{+0.33}_{-0.17}$ & $<-1.43$ \\
    DESI+CMB+Pantheon+ & $0.3117\pm 0.0056$ & $67.62\pm 0.60$ & $1.1\pm1.3$ & $-0.853\pm 0.057$ & $-0.54\pm 0.22$ \\
    DESI+CMB+Union3 & $0.3273\pm 0.0086$ & $65.98\pm 0.86$ & $0.6\pm1.3$ & $-0.678\pm 0.092$ & $-1.03^{+0.33}_{-0.29}$ \\
    DESI+CMB+DESY5 & $0.3193\pm 0.0056$ & $66.82\pm 0.58$ & $0.8\pm1.3$ & $-0.762\pm 0.060$ & $-0.81\pm 0.24$ \\
    \bottomrule
    \end{tabular}
    }
    \caption{Summary table of cosmological parameter constraints from DESI DR2 BAO (labelled in the table as `DESI') in combination with external datasets and priors, in \lcdm\ and various extended models. Results quoted for all parameters are the marginalized posterior means and 68\% credible intervals in each case where two-sided constraints are possible, or the 68\% upper limits when only one-sided constraints are possible.
    \label{tab:cosmo_constraints}
    }
\end{table*}

We start by presenting cosmological constraints in the base \lcdm\ model, and examining tensions between different datasets in this scenario, before introducing the freedom of extended models in \cref{sec:de_constraints,sec:constraints_neutrinos}.

As discussed in \cref{subsec:distances} and shown in \cref{fig:lcdm_bao_tracers}, the values of $\Om$ and $\Hrd$ inferred from each DESI tracer individually within the \lcdm\ framework are consistent with each other. The result of a combined fit to the BAO results from all redshift bins together is shown in the left panel of \cref{fig:lcdm_bao_constraints} compared to those obtained from DR1 and from the CMB. We find
\twoonesig[2cm]
{\Omega_{\mathrm{m}} &= 0.2975\pm 0.0086,}
{hr_\mathrm{d} &= (101.54\pm 0.73) \text{ Mpc},}{DESI DR2, \label{eq:constraints_lcdm_desi}}
with a correlation coefficient of $r=-0.92$. This represents a $\sim$40\% improvement in the precision on $\Om$ and $h r_\mathrm{d}$ compared to the DR1 results, while being perfectly consistent with them as well as with the SDSS and DESI+SDSS results reported in \cite{DESI2024.VI.KP7A}. In the following text and figures, we refer to DESI DR2 simply as DESI, unless explicitly comparing to DESI DR1.

\Cref{fig:lcdm_bao_constraints} shows that the overlap with the CMB posterior has decreased, a sign of the increased discrepancy between the results from DESI BAO and CMB probes when interpreted in base \lcdm. We calculate the relative $\chi^2$ between the two datasets as
\begin{equation}
    \chi^2 = (\mathbf{p}_A - \mathbf{p}_B)^T(\mathrm{Cov}_A+\mathrm{Cov}_B)^{-1}(\mathbf{p}_A - \mathbf{p}_B),
\end{equation}
where $\mathbf{p}_A$ and $\mathbf{p}_B$ are the ($\Om, \rd h$) values obtained from DESI and the CMB respectively, and $\mathrm{Cov}_A$ and $\mathrm{Cov}_B$ are the corresponding $2\times 2$ posterior parameter covariances. Converting this $\chi^2$ into a probability-to-exceed (PTE) value, we find it is equivalent to a $2.3\sigma$ discrepancy between BAO and CMB in \lcdm, increased from $1.9\sigma$ in DR1.  However, we note that this reduces to 2.0$\sigma$ if CMB lensing is excluded. This discrepancy is part of the reason why more models with a more flexible background expansion history than \lcdm, such as the evolving dark energy considered in \cref{sec:de_constraints} below, may provide a statistically preferable fit. 

By calibrating the BAO relative distance measurements using external information we are able to determine the Hubble constant $H_0$. The right panel of \cref{fig:lcdm_bao_constraints} compares the constraints obtained in the $\Om$-$H_0$ plane from calibrated BAO measurements to those from the CMB. We show DESI BAO results calibrated using the BBN $\ob$ prior in the blue contours, while those in orange illustrate the additional combination with the $\theta_\ast$ BAO-like measurement from the CMB. The overlap with the full CMB posterior, shown in pink, is small and has decreased since DR1, showing a 2.3$\sigma$ discrepancy with DESI after BBN calibration. The centers of both DESI+BBN and DESI+BBN+$\theta_\ast$ posteriors remain quite well aligned with the degeneracy direction of the CMB, which is $\propto\Om h^3$ \cite{Percival2002:astro-ph/0206256} and shown by the black dotted line in the figure, while the degeneracy direction of the DESI+BBN+$\theta_\ast$ contour instead follows the line of constant $\Om h^2$.

\begin{figure}
    \centering
    \includegraphics[width=\columnwidth]{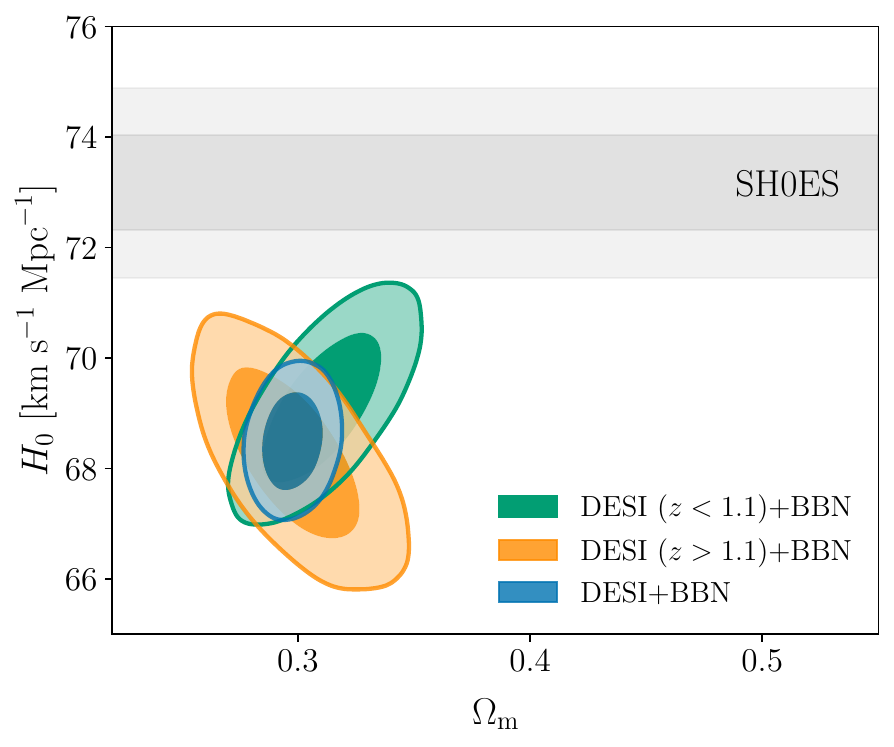}
    \caption{Comparison of our $H_0$ constraints with respect to SH0ES, assuming a $\Lambda$CDM model (\cref{sec:cosmologcal_constraints}). We show the combination DESI+BBN for our low redshift and high redshift samples. For $z>1.1$, only the \elgt, \qso, and \lya\ tracers are included, while DESI($z<1.1$) includes \bgs, \lrgs\ and the \lrgelg\ tracer combinations. Both subsets are individually in $>3\sigma$ tension with SH0ES measurements. 
    }
    \label{fig:LCDM_DESI_tension_SHOES}
\end{figure}

Using the conservative BBN prior on $\ob$, we obtain
\oneonesig[6cm]
{H_0 = (68.51\pm 0.58)\;\kmsMpc
}
{DESI+BBN}{,\label{eq:H0_BAO+BBN}}
a 0.8\% precision measurement that is independent of any CMB anisotropy information and comparable to the result from the CMB \cite{Planck-2018-cosmology}. This value is 28\% more precise than the corresponding result from DR1, $H_0=(68.53\pm 0.80)\;\kmsMpc$, but very consistent with it. While the result is clearly more in agreement with the low values obtained from early-Universe measurements than with those from the Cepheid-calibrated local distance ladder \cite{Riess:2021jrx}, it is also noticeably higher than the \Planck\ value, a reflection of the growing tension between DESI and \Planck\ when interpreted in flat \lcdm. Adding information on the very well-measured acoustic angular scale further improves this result to 
\oneonesig[6cm]
{H_0 = (68.45\pm 0.47)\;\kmsMpc}{DESI+BBN+$\theta_\ast$}{,\label{eq:H0_BAO+BBN+thetastar}}
now slightly more precise than from the CMB alone \cite{Tristram:2023}.

\Cref{fig:LCDM_DESI_tension_SHOES} shows the DESI+BBN result for $\Om$ and $H_0$ relative to the SH0ES result \cite{Breuval:2024lsv}. The contours also show how the constituent tracers of the DR2 sample at different redshifts contribute to the final constraint, with the degeneracy directions of the contours changing as the best measured combination of transverse and line-of-sight BAO changes with redshift. In \lcdm\ the tension between the DESI+BBN and SH0ES $H_0$ results now stands at $4.5\sigma$ independent of the CMB. Note that the DESI+BBN result does assume standard pre-recombination physics to determine $\rd$ through \cref{eqn:rdformula}.

We have highlighted the tension between DESI and CMB in \lcdm\ in order to provide context to the results for extended models in the following sections. However, given that this tension is not close to $3\sigma$, it is still valid to combine the two datasets within the \lcdm\ model and obtain joint constraints. In this case we find
\twoonesig[2cm]
{\Omega_{\mathrm{m}} &= 0.3027\pm 0.0036,}
{H_0 &= (68.17\pm 0.28)\;\kmsMpc
,}
{DESI+CMB, \label{eq:constraints_lcdm_desi_cmb}}
with a correlation coefficient of $r=-0.975$.

We also allow for spatial curvature to vary in our cosmological fits and we do not find a significance preference for a non-flat $\Lambda$CDM model. \Cref{tab:cosmo_constraints} summarizes the cosmological parameter results from DESI alone as well as in combination with  external datasets, in both \lcdm\ and extended models.

\begin{figure}
    \centering
    \includegraphics[width=\columnwidth]{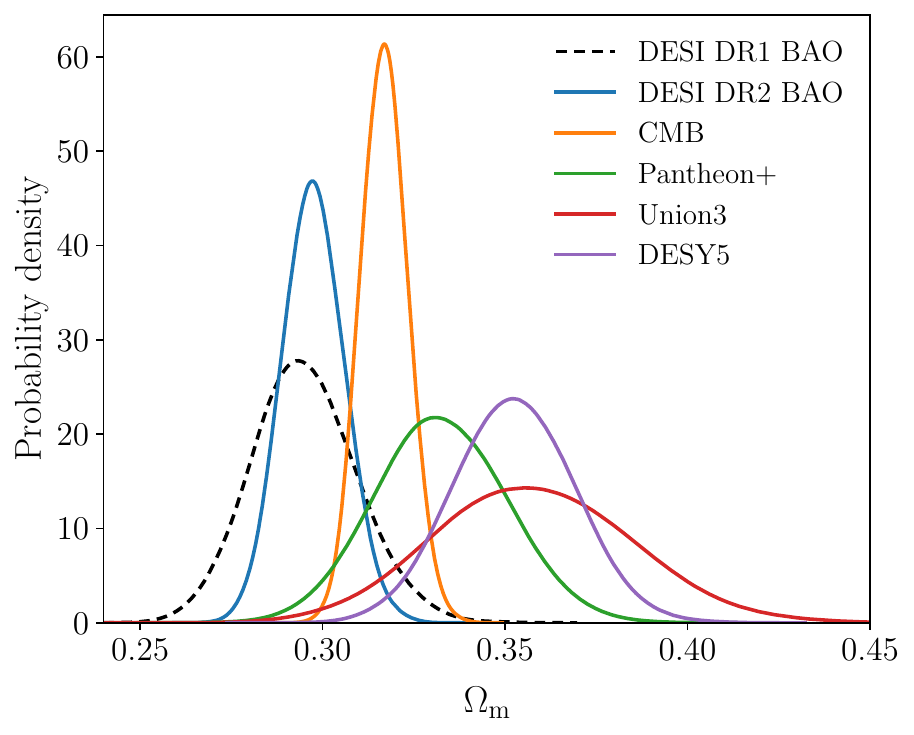}
    \caption{Marginalized 1D posteriors for $\Om$, when fixing the background model to \lcdm\ (\cref{sec:cosmologcal_constraints}). We show the probability distributions for DESI DR1 and DR2, as well as the measurements from CMB, and the three SNe datasets used throughout this paper.}
    \label{fig:lcdm_omegam_discrepancy}
\end{figure}

Finally, as in \cite{DESI2024.VI.KP7A}, we note a mild to moderate discrepancy between the recovered values of $\Om$ from DESI and SNe in the context of the \lcdm\ model. This is shown in the marginalized posteriors in \cref{fig:lcdm_omegam_discrepancy}: the discrepancy is $1.7\sigma$ for Pantheon+, $2.1\sigma$ for Union3, and $2.9\sigma$ for DESY5, with all SNe samples preferring higher values of $\Om$ though with larger uncertainties. For \lcdm\ we do not report joint constraints on parameters from any combination of DESI and SNe data. However, as with the CMB,  these apparent parameter differences potentially indicate that DESI and at least some of the SNe datasets cannot be consistently fit except with models that have greater freedom in the background evolution, as described in the next section (see also \cite{Tang:2024}). 

\section{Dark Energy} 
\label{sec:de_constraints}

Probing the behavior and nature of dark energy is the primary goal of DESI. The question of perhaps greatest interest, and the one that BAO measurements can best illuminate, is the value of the equation-of-state parameter $w=P/(\rho c^2)$, and its possible evolution with time. To examine this we will primarily use the so-called Chevallier-Polarski-Linder (CPL) parametrization \cite{Chevallier:2001,Linder2003} of \cref{eqn:w0wa}. While this form of $w(a)$ does not arise directly from an underlying physical model, it is a flexible parametrization that is capable of matching the predictions for observable quantities obtained in a wide range of models that are physically motivated \cite{dePutter2008}. The accompanying paper \cite{Y3.cpe-s1.Lodha.2025} explores various other parametrizations of $w(z)$, as well as non-parametric reconstruction methods.

For certain ranges of parameters $w_0$ and $\wa$, the parametrization of \cref{eqn:w0wa} allows so-called `phantom' behavior of dark energy, in which the equation of state crosses to the regime $w(z)<-1$ \cite{Caldwell:2002} where the null energy condition (NEC)---which requires that the energy density of dark energy not increase with the expansion of the Universe---is violated. For single scalar-field models of dark energy, this phantom crossing presents severe theoretical difficulties \cite[e.g.,][]{Carroll:2003,Vikman:2005}. However, more complex models of dark energy, with multiple fields, other dark energy internal degrees of freedom, or non-minimal coupling, can evade these difficulties, as can some modified gravity models, see, e.g., \cite{Hu:2005,2023PTEP.2023k3E01K,Wolf:2024,Clifton:2011jh,Ishak:2018his}. 
We therefore adopt wide uniform priors on the parameters, $w_0\in\mathcal{U}[-3, 1]$ and $\wa\in\mathcal{U}[-3, 2]$, together with imposing the condition $w_0+\wa<0$ to enforce early matter domination. While other justifiable choices are possible, and the values of Bayesian quantities such as the model evidence will always depend on the particular choice used, we consider this the minimal empirical approach. Whenever the equation of state crosses the $w=-1$ boundary we use the parametrized post-Friedmann (PPF) approach of \cite{Hu:2007,2008PhRvD..78h7303F} to include dark energy perturbations when calculating CMB power spectra---however, as shown below, the method of accounting for dark energy perturbations does not play a major role, since simply applying an early-Universe CMB prior on $(\theta_\ast,\ob,\ocdm)$ largely reproduces the same results on $w_0$ and $\wa$.

Our primary measure of the statistical significance of preference for evolving dark energy from a given data combination is based on $\Delta\chi^2_\mathrm{MAP}$ between the best-fit \lcdm\ and \wowacdm\ models for that combination.  Because \lcdm\ is nested within \wowacdm, corresponding to $w_0=-1$, $\wa=0$, Wilks' theorem \cite{Wilks:1938dza} implies that $\Delta\chi^2_\mathrm{MAP}$ should follow a $\chi^2$ distribution with two degrees of freedom under the assumption the null hypothesis (\lcdm\ model) holds, and assuming that errors are Gaussian and correctly estimated.  To translate $\Delta\chi^2_\mathrm{MAP}$ into familiar terms, we quote the corresponding frequentist significance $N\sigma$ for a 1D Gaussian distribution,
\begin{equation}
    \mathrm{CDF}_{\chi^2}\left(\Delta\chi^2_\mathrm{MAP} |\, 2\,\mathrm{dof}\right) = \frac{1}{\sqrt{2\pi}}\int_{-N}^{N} e^{-t^2/2} dt~, 
\end{equation}
where the left hand side denotes the cumulative distribution of $\chi^2$.  We also compute the Deviance Information Criterion (DIC) \cite{DIC-Spiegelhalter2,Trotta:2008qt,Grandis:2016fwl,kass1995bayes}, which takes into account the Bayesian complexity of the model and penalizes including extra parameters.

\begin{figure}
    \centering
    \includegraphics[width=\columnwidth]{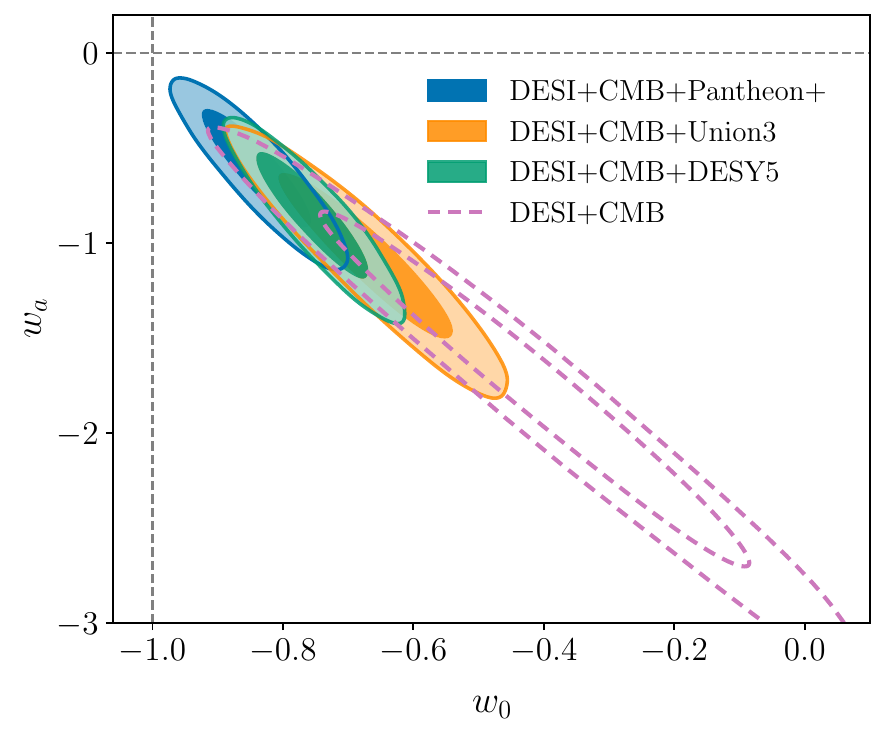}
    \caption{Results for the posterior distributions of $w_0$ and $w_a$, from fits of the \wowacdm\ model to DESI in combination with CMB and three SNe datasets as labelled. We also show the contour for DESI combined with CMB alone. The contours enclose 68\% and 95\% of the posterior probability. The gray dashed lines indicate $w_0=-1$ and $\wa=0$; the \lcdm\ limit ($w_0=-1$, $w_a=0$) lies at their intersection. The significance of rejection of \lcdm\ is 2.8$\sigma$, 3.8$\sigma$ and 4.2$\sigma$ for combinations with the Pantheon+, Union3 and DESY5 SNe samples, respectively, and $3.1\sigma$ for DESI+CMB without any SNe.}
    \label{fig:w0wa_SN}
\end{figure}

\subsection{Results}
\label{sec:DE_results}

From DESI DR2 BAO alone, we obtain rather weak constraints on the parameters
\twoonesig[2cm]{w_0 &=-0.48^{+0.35}_{-0.17}}{\wa &<-1.34}{DESI BAO, \label{eq:w0wa_DESI}}
which mildly favor the $w_0>-1$, $\wa<0$ quadrant. The upper bound on $\wa$ here is the 68\% limit, and $\wa=0$ is not excluded at 95\%. As was the case in DR1, BAO data alone define a degeneracy direction in the $w_0$-$\wa$ plane, but they do not show a strong preference for dark energy evolution: the improvement in $\chi^2_\mathrm{MAP}$ relative to the \lcdm\ case of $w_0=-1$, $\wa=0$ is equivalent to a preference of just 1.7$\sigma$. Note that the posteriors in this poorly constrained case are cut off by the priors, so the marginalized means and limits quoted above are prior-dependent.

The minimal extension we consider, beyond BAO data alone,  is to add a high-redshift constraint from the early universe. This can be achieved by imposing CMB-derived priors on $\theta_\ast$, $\ob$ and $\obc$, as described in \cref{sec:external_data}. These priors are independent of the late-time dark energy, and also marginalize over contributions such as the late ISW effect and CMB lensing. Therefore, they provide us with an early time physics prior that can help us set the sound horizon and is based solely on early-Universe information. The result from this data combination is
\twoonesig[3cm]{w_0 &=-0.43\pm 0.22}{\wa &=-1.72\pm 0.64}{DESI+$(\theta_\ast,\ob,\obc)_{\rm CMB}$. \label{eq:w0wa_DESI_compCMB}}
While this is still bounded by the $\wa>-3$ prior at the lower end, the posterior already clearly disfavors \lcdm. The $\dchisq$ value decreases to $-8.0$, indicating a preference for an evolving dark energy equation of state at the 2.4$\sigma$ level. 

Replacing these minimal early-Universe priors with the full CMB information leads to only a small shift in the maginalized posteriors \twoonesig[2cm]{w_0 &=-0.42\pm0.21}{\wa &=-1.75\pm0.58}{DESI+CMB, \label{eq:w0wa_DESI_CMB}}
showing that most of the information that the CMB provides on $w(z)$ comes from its role in anchoring early-Universe values of $(\theta_\ast,\ob,\obc)$ and thus limiting the freedom for models to fit the low-redshift data without an evolving dark energy component. Nevertheless, when including the full CMB information the $\dchisq$ decreases to $-12.5$, corresponding to a 3.1$\sigma$ preference for evolving dark energy. This change in the $\dchisq$ is driven primarily by the inclusion of CMB lensing, the effect of which is (by construction) not captured in the minimal early-Universe priors (see  \cref{appendix:cmb_compression} for further discussion and a comparison of posteriors with different choices of CMB likelihoods).

SNe data alone provide a complementary degeneracy direction in the $w_0$-$\wa$ plane, as they measure $w_0$ well independently of $\wa$, which is only weakly constrained. The combination of SNe data with DESI BAO can therefore measure $w_0$ and $\wa$ without having the posteriors cut off by the prior ranges we assumed. The marginalized posterior results are listed in \cref{tab:cosmo_constraints} and depend on the choice of SNe dataset, with the significances of the preference for the model over \lcdm\ ranging from $1.7\sigma$ to $3.3\sigma$ as summarized in \cref{tab:dchisq}. 

However, as discussed in \cite{DESI2024.VI.KP7A}, the posterior for the combination of DESI BAO and SNe alone allows quite a wide range of posterior values of $\omega_\mathrm{m}$ (or, equivalently, $\omega_\mathrm{c}$). CMB information places extremely tight constraints on $\omega_\mathrm{m}$ that are largely independent of the late-time background model. Therefore, the full statistical power of the data is achieved through the combination of the BAO, CMB and SNe datasets, giving the maginalized posterior results
\twoonesig[2cm]{w_0 &=-0.838\pm 0.055}{\wa &=-0.62^{+0.22}_{-0.19}}{DESI+CMB+ Pantheon+, \label{eq:w0wa_DESI_CMB_P+}}
for the combination with the Pantheon+,
\twoonesig[2cm]{w_0 &=-0.667\pm 0.088}{\wa &=-1.09^{+0.31}_{-0.27}}{DESI+CMB +Union3, \label{eq:w0wa_DESI_CMB_U3}}
with Union3, and
\twoonesig[2cm]{w_0 &=-0.752\pm 0.057}{\wa &=-0.86^{+0.23}_{-0.20}}{DESI+CMB +DESY5, \label{eq:w0wa_DESI_CMB_DY5}}
with DESY5. The posteriors in these three cases, along with the DESI+CMB posterior, are shown in \cref{fig:w0wa_SN}. The $\dchisq$ values are $-10.7$, $-17.4$, and $-21.0$, corresponding to preferences for the \wowacdm\ model over \lcdm\ at the $2.8\sigma$, $3.8\sigma$, and $4.2\sigma$ levels, for combination with Pantheon+, Union3 and DESY5 respectively. These significances have all increased compared to the values reported in \cite{DESI2024.VI.KP7A} based on the DESI DR1 BAO results.

\begin{table}

\centering
\resizebox{\columnwidth}{!}{
    \begin{tabular}{lccc}
    \toprule
    Datasets & $\dchisq$ & Significance & $\Delta$(DIC) \\
    \midrule
    DESI & $-4.7 $& $1.7\sigma$  & $-0.8$ \\
    DESI+$(\theta_\ast,\ob,\obc)_{\rm CMB}$ & $-8.0$ & $2.4\sigma$ & $-4.4$ \\
    DESI+CMB (no lensing) & $-9.7$ & $2.7\sigma$ & $-5.9$ \\
    DESI+CMB & $-12.5$ & $3.1\sigma$  & $-8.7$ \\
    DESI+Pantheon+ & $-4.9 $& $1.7\sigma$  & $-0.7$ \\
    DESI+Union3 & $-10.1$ & $2.7\sigma$  & $-6.0$ \\
    DESI+DESY5 & $-13.6$ & $3.3\sigma$  & $-9.3$ \\
    DESI+DESY3 (3$\times$2pt) & $-7.3 $& $2.2\sigma$ & $-2.8$ \\
    DESI+DESY3 (3$\times$2pt)+DESY5 & $-13.8$ & $3.3\sigma$ & $-9.1$ \\
    DESI+CMB+Pantheon+ & $-10.7$ & $2.8\sigma$  & $-6.8$ \\
    DESI+CMB+Union3 & $-17.4$ & $3.8\sigma$ & $-13.5$ \\
    DESI+CMB+DESY5 & $-21.0$ & $4.2\sigma$ & $-17.2$ \\
    \bottomrule
    \end{tabular}
}
\caption{Summary of the difference in the effective $\chi^2_{\rm MAP}$ value (defined as twice the negative log posterior at the maximum posterior point) for the best-fit \wowacdm\ model relative to the best \lcdm\ model with  $w_0=-1$, $\wa=0$, for fits to different combinations of datasets as indicated. The third column lists the corresponding (frequentist) significance levels given 2 extra free parameters, and the final column shows the results for $\Delta(\mathrm{DIC})=\mathrm{DIC}_{w_0\wa\mathrm{CDM}}-\mathrm{DIC}_{\Lambda\mathrm{CDM}}$. As a rule of thumb, $\Delta(\mathrm{DIC})$ values $<-5$ indicate a `strong' preference for \wowacdm\ and values $<-10$ a `decisive' preference \cite{Grandis:2016fwl}. 
\label{tab:dchisq}
}
\end{table}

The deviance information criterion (DIC) values for the combination of DESI+CMB with Pantheon+, Union3 and DESY5 SNe are $\Delta(\mathrm{DIC})\equiv\mathrm{DIC}_{w_0\wa}-\mathrm{DIC}_{\Lambda\mathrm{CDM}}=-6.8$, $-13.5$, $-17.2$, respectively. These indicate preferences for the \wowacdm\ model consistent with those obtained from the $\dchisq$ values above. Again, the changes in the $\Delta(\mathrm{DIC})$ values obtained here with DESI DR2 BAO data compared to the DR1 values reported in \cite{DESI2024.VI.KP7A} show that the preference for \wowacdm\ has increased with the additional data. Further details on the calculation of DIC values are given in \cite{Y3.cpe-s1.Lodha.2025}.

The pivot redshift $\zpiv$ at which $w(z)$ in this parametrization is best constrained by the data depends on the particular combination of datasets used. For DESI+CMB, $\zpiv=0.53$ and $\wpiv=w(\zpiv)=-1.024\pm0.043$: this is a lower pivot redshift and a tighter constraint on $\wpiv$ than that found for the same combination with DR1 BAO in \cite{DESI2024.VI.KP7A}, reflecting the additional constraining power of the DR2 BAO results. For the DESI+CMB+DESY5 combination, we find $\zpiv=0.31$ and $\wpiv=-0.954\pm0.024$, indicating a mild preference for a deviation from $w=-1$ at the best-measured redshift. As the other two SNe datasets are slightly less constraining than DESY5, the pivot redshifts for combinations with them are slightly larger, as are the uncertainties on $\wpiv$, but the results for all choices of DESI+CMB+SNe are mutually consistent.

\begin{figure}
    \centering
    \includegraphics[width=\columnwidth]{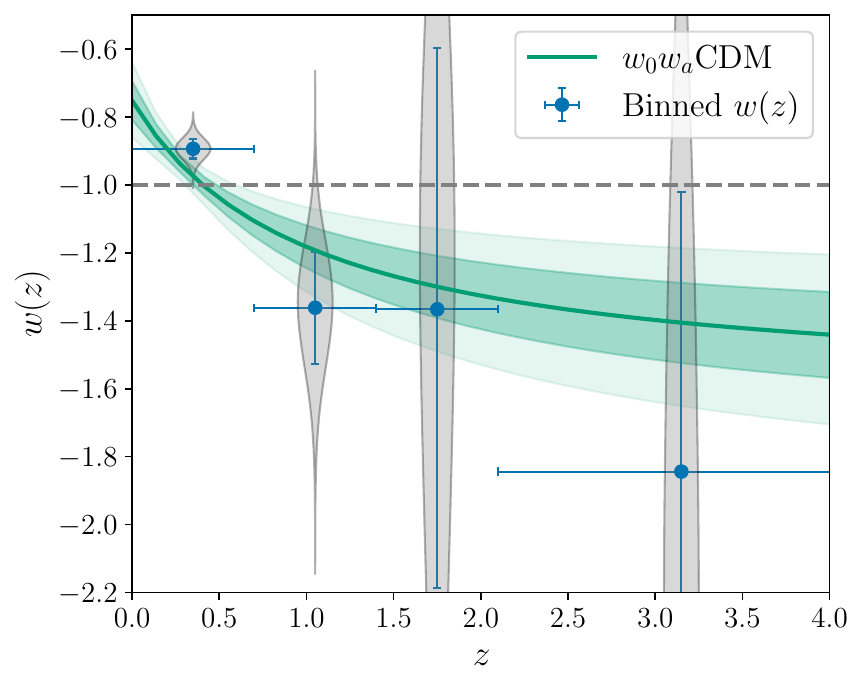}
    \caption{Comparison of the constraints on the equation of state of dark energy using the CPL parametrization and a binning reconstruction approach, using DESI+CMB+DESY5. The solid green line shows the best-fit $w(z)$ based on $w_0$ and $\wa$ inference and the green contours around it represent the 68\% and 95\% confidence intervals. We also show in blue the constraints from the binning approach \cite{Y3.cpe-s1.Lodha.2025}, with the horizontal bars indicating the bin width (which is fixed) and the vertical bars representing the 1$\sigma$ error. Additionally, we include in gray the 1D posterior for each binning parameter. The $\Lambda$CDM limit corresponds to the horizontal gray dashed line. 
    }
    \label{fig:binning_w_of_z}
\end{figure}

These results sharpen the preference already seen in \cite{DESI2024.VI.KP7A} for an evolving equation of state for dark energy: although the statistical significances from all data depend somewhat on the choice of SNe dataset included, even the weakest of them (the DESI+CMB+Pantheon+ combination) is still nearly $3\sigma$, and the significance is $3.1\sigma$ even when excluding all SNe data altogether (DESI+CMB). In all cases, the favored $w(z)$ shows a phase of $w>-1$ at low redshifts and a phantom crossing to $w<-1$ above redshifts $z\simeq0.4$. Within the \wowacdm\ model, the necessity of such a crossing and the redshift at which it occurs is determined by the requirement to match the precise CMB measurement of $\theta_\ast$ together with $\Omega_m$ \cite{Linder2007}. 
The details of the recovered form of $w(z)$ and the $w_0$, $\wa$ parameter values naturally depend on the choice of parametrization. The accompanying paper \cite{Y3.cpe-s1.Lodha.2025} explores various other parametrizations of $w(z)$ beyond \cref{eqn:w0wa}, and non-parametric reconstruction methods, that exhibit a similar behavior. Reference \cite{Y3.cpe-s1.Lodha.2025} also performs binned reconstruction of $w(z)$ without assuming a functional form for the equation of state and finds a consistent picture, as shown in \cref{fig:binning_w_of_z}. The lowest redshift bin shown in the figure favors a value of $w>-1$ at high significance \cite{Y3.cpe-s1.Lodha.2025}, inconsistent with the \lcdm\ expectation $w=-1$.

The apparent preference for phantom crossing to $w(z)<-1$ at intermediate redshifts, and the consequent violation of the NEC, is thus rather independent of parametrization choices made in the analysis (see, e.g., \cite{DESI:2024aqx} using DR1 data for non-parametric results). However, the equation of state is not directly observable, and we only observe quantities, such as distances, that depend indirectly on $w(z)$. In some circumstances it may therefore be possible to construct particular models that provide reasonable fits to the low-redshift data while still respecting $w(z)\geq-1$ at all $z$ (\cite{Shlivko:2024,DESI:2024kob} provide explicit examples, but \cite{Lewis:2024} discusses the general limitations of such models in fitting DESI+CMB data). 
The supporting paper \cite{Y3.cpe-s1.Lodha.2025} examines several NEC-respecting models of dark energy evolution and finds that while they can somewhat outperform \lcdm, they have low $\Delta(\text{DIC})$ values relative to $w_0\wa$CDM for the combination of DESI, CMB and SNe data, indicating a preference for phantom crossing. 

\Cref{tab:cosmo_constraints} gives a more complete list of the results for other parameters when fitting this model to different combinations of data. It is worth noting that allowing $w(z)$ to vary does not help resolve the so-called Hubble tension \cite{DivalentinoHtension}, since in \wowacdm\ the recovered $H_0$ is lower than the \Planck\ \lcdm\ value. Although not discussed here, \cite{DESI2024.VII.KP7B} showed that allowing evolving dark energy also does not affect the value of the $S_8$ parameter determined from DESI data, which remains consistent with values from the CMB.

\subsection{The nature of the evidence for evolving dark energy}
\label{sec:understanding_DE}

\begin{figure*}
    \centering
    \includegraphics[width=1\linewidth]{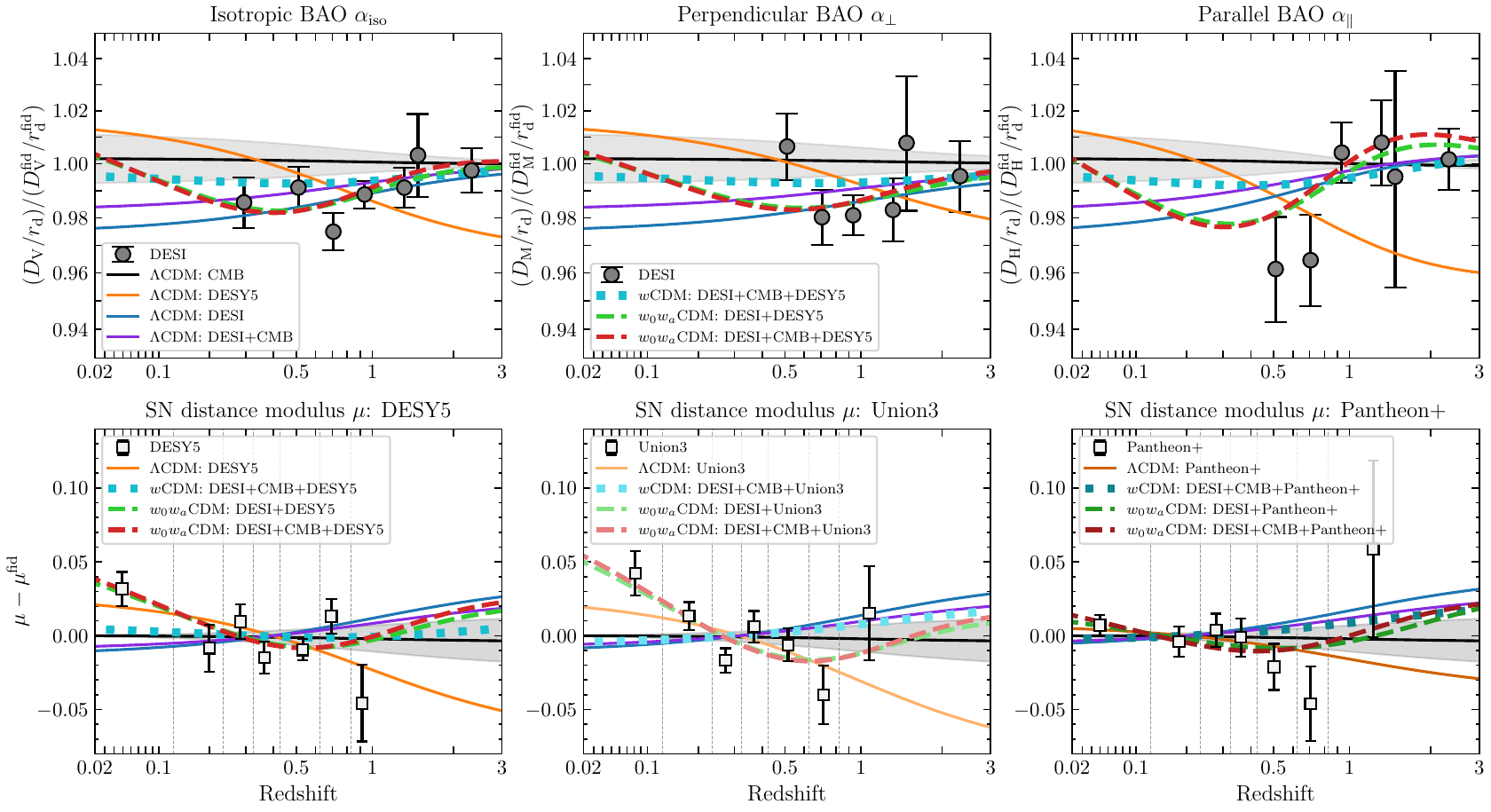}
    \caption{Hubble diagrams showing comparisons of DESI BAO and SNe data to models. In the top panels, DESI BAO measurements relative to distances in the fiducial cosmology based on \Planck\ 2018 results are shown with black circles, and DESY5 is used as the SNe dataset for determining model fits. In the bottom row, binned DESY5, Union3 and Pantheon+ SNe distance modulus residuals are shown with black squares in the three panels. The SNe binning method is described in \cref{subsec:SN}. The bin edges for the SNe bins are indicated by vertical gray dashed lines. \lcdm\ predictions from \Planck\ CMB (including the updated \Planck+ACT CMB lensing), DESI, SNe and DESI+CMB are shown in black, blue, orange and purple solid lines, respectively, and the gray shaded region indicates the extrapolated uncertainty in the CMB results. The $w$CDM predictions from best-fit DESI+CMB+SNe are shown in cyan dotted lines. $w_0\wa$CDM predictions from fits to DESI+SNe and DESI+CMB+SNe are shown in green and red dashed lines; in the top panels DESY5 is used as the SNe sample, while in the bottom panels, the model curves are adjusted to use the corresponding SNe sample in the title of each panel. In the SNe panels, the distance moduli can be arbitrarily shifted in a vertical direction depending on the chosen calibration value for $H_0$, so all models are pinned at the average of the SNe data. }
    \label{fig:HD_BAO_SN}
\end{figure*}

\begin{figure*}
    \centering
    \includegraphics[width=\textwidth]{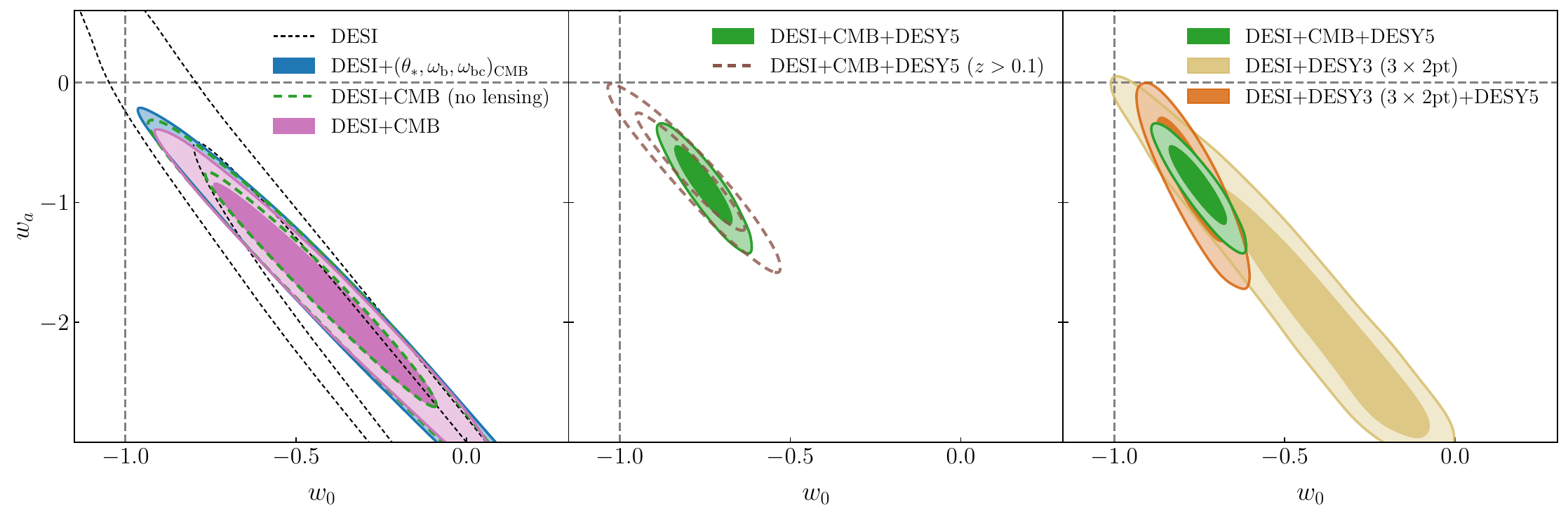}
    \caption{Tests of the robustness of dark energy results to different data selections. All contours shown contain 68\% and 95\% of the posterior probability. \emph{Left panel}: Constraints from DESI alone (black dashed thin contours), and combinations of DESI with $(\theta_\ast,\ob,\obc)_\mathrm{CMB}$ early-Universe priors (blue), CMB without CMB lensing (green dashed) and the full CMB (pink). The CMB tightens constraints in the $w_0$-$\wa$ plane primarily by helping fix $\Om$ to break degeneracies (cf. \cref{sec:understanding_DE}). Most of the CMB information is therefore already contained in the simple priors, although the significance of preference over \lcdm\ is increased by including CMB lensing (\cref{tab:dchisq}). \emph{Center}: The change to the DESI+CMB+DESY5 posterior (green) when excluding low-redshift SNe at $z<0.1$ from the DESY5 sample (brown dashed). The uncertainties are much larger when excluding these SNe but the shift in the best-fit values of $w_0$ and $\wa$ is small. \emph{Right}: Constraints obtained when \emph{replacing} the CMB with DESY3 (3$\times$2pt) information, which is also able to break the degeneracy with $\Om$. We show results for DESI+DESY3 (3$\times$2pt) (mustard) and DESI+DESY3 (3$\times$2pt)+DESY5 (brick red), the latter of which also excludes \lcdm\ at high significance. These constraints use only low-redshift data and include no early-Universe information.}
    \label{fig:w0wa_desi_with_variations}
\end{figure*}

The Hubble diagrams in \cref{fig:HD_BAO_SN} illustrate the nature of the evidence for evolving dark energy and its dependence on the adopted datasets.  The upper panels show the isotropic, perpendicular, and parallel BAO measurements ($\aiso$, $\aper$, and $\apar$), which are normalized to the predictions of the \Planck\ \lcdm\ cosmology. The lower panels plot $\mu-\mu_\mathrm{fid}$, the distance modulus relative to the fiducial \Planck\ \lcdm\ prediction, for the three SNe datasets (noting again that for uncalibrated SNe, the distance modulus is only known to an arbitrary constant offset). Our procedure for creating binned data points from the SNe data is described in \cref{subsec:SN}. Because the fiducial SNe absolute magnitude is unknown, all data points are free to move up or down together by the same amount in $\mu$, and we have chosen the normalization such that error-weighted mean of $\mu-\mu_\mathrm{fid}$ is equal to zero.  Equivalently, any model curve in these panels can be shifted up or down by a constant $\Delta\mu$, and we have normalized them to match the weighted mean of the data.

In all panels, the horizontal black line represents the prediction of the \lcdm\ model best fit by the CMB, with the $1\sigma$ range of these predictions shown by the shaded gray region. This differs very slightly from the fiducial cosmology based on the Planck 2018 results, due to the switch from \texttt{Plik} to \texttt{CamSpec} and the addition of ACT lensing data, explaining the slight offset from unity. The blue solid curve shows predictions for the \lcdm\ model that best fits the DESI BAO data, while purple represents the model fit to DESI+CMB, which closely matches the best fit to DESI with just the  $(\theta_\ast,\ob,\obc)$ early-Universe priors from the CMB (see \cref{tab:cosmo_constraints}). The orange curves in each of the lower panels represent the \lcdm\ models that, from left to right, best fit the DESY5, Union3 and Pantheon+ SNe data respectively. The orange curves in the top row are all those for the \lcdm\ DESY5 best fit.

While the statistical significance of disagreement cannot be judged accurately from this plot alone, the DESI measurements clearly prefer lower distances (by 1-2\%) than the \Planck\ \lcdm\ prediction at redshifts $z \leq 1$.  There \emph{is} a \lcdm\ model that fits the DESI data well (blue curve), but it has a lower $\Om$ than the \Planck\ model (0.297 vs. 0.317), as shown previously in \cref{fig:lcdm_bao_constraints}. The joint-fit model (purple curve) has an intermediate $\Om = 0.303$, and consequently has a worse fit to both DESI BAO and the CMB (top panels) and also fails to describe the SNe data (lower panels). Similarly, the DESY5 data in the lower left panel exhibit a tension with \Planck\ \lcdm, primarily because of the contrast between the low redshift ($z<0.1$) data at $\mu-\mu_\mathrm{fid} \approx 0.03$ and the points at higher redshift. The story is similar for Union3, but for Pantheon+ the value of $\mu-\mu_\mathrm{fid}$ at low redshift is smaller, only 0.01. 

As is the case for BAO, there exist \lcdm\ models that can reasonably fit the SNe data (orange curves), but these have large $\Om$ that do not well match CMB constraints and are also strongly inconsistent with DESI data, which prefer an $\Om$ value that is lower than \Planck, not higher. Conversely, the conflict with the DESI-constrained \lcdm\ models (purple and blue curves) and DESY5 SNe is 0.04 in $\mu$ at low redshifts, $z<0.1$, which is worse than the 0.03 offset from the \Planck\ fiducial model. All these observations point to the fact that the \lcdm\ model struggles to consistently fit all three datasets: BAO, CMB, and SNe.

One way to address this tension is to adopt a model with more flexibility in the background expansion. However, the \wcdm\ model with a constant equation of state lacks sufficient flexibility. The cyan dotted curves in the top row and bottom left show a model with constant $w=-0.971$ and $\Om=0.310$, which is the \wcdm\ model that best fits the DESI+CMB+DESY5 data combination; in the remaining two panels the cyan curves show the predictions of very similar models obtained by substituting Union3 or Pantheon+ for DESY5. The high-$z$ anchor does not allow enough redshift evolution for this model to provide a good fit to the BAO and SNe data at low redshifts, performing only slightly better than \Planck\ \lcdm. If instead the \wcdm\ model were chosen to fit the DESI+DESY5 data, it would necessarily have an $\Om$ value that would fail to match the CMB constraints on $(\theta_\ast,\ob,\obc)$.
 
On the other hand, the \wowacdm\ model does have sufficient flexibility to simultaneously achieve good fits to all three datasets. The green dashed and red dashed curves in the top panels show predictions for \wowacdm\ models with parameters matching the best fits to DESI+DESY5 and DESI+CMB+DESY5 respectively. These are barely distinguishable in the plots, showing that the \wowacdm\ model that best fits the BAO and SNe automatically also provides a good fit to the CMB. Over the range of redshifts covered by DESI BAO, these model predictions provide a better fit to BAO data than the best such \lcdm\ model---in particular, in the fit to the parallel BAO distances $\apar$---while also simultaneously resolving the mismatch in $\Om$ between DESI and CMB in the \lcdm\ framework. The bottom left panel also shows that they are also equally good at fitting the $\mu$ offsets between low-$z$ and high-$z$ DESY5 data. These observations qualitatively help to explain why these models are strongly preferred over \lcdm. 

In the equivalent curves in the center and right panels on the bottom row, DESY5 is replaced in all fits by Union3 and Pantheon+, respectively. While the qualitative picture is the same for Union3 as DESY5, the value of $\mu-\mu_\mathrm{fid}$ at low redshift is smaller for Pantheon+ and so using this dataset does not strengthen the preference for evolving dark energy relative to that already provided by the joint fit to DESI and the CMB.

\subsection{Are alternative explanations possible?}
\label{sec:robustness_DE}

Given the surprising results from our analysis---our best-fit evolving dark energy models apparently imply phantom dark energy at some redshifts, and order unity changes in $w$ over the redshift range spanned by the observations considered---it is worth considering alternative explanations for the data.

The simplest of these alternatives is simply a statistical fluctuation. Based on the $\dchisq$ values presented in \cref{sec:DE_results} above, the statistical preference for the best-fit \wowacdm\ model over \lcdm\ ranges from around $3\sigma$ to over $4\sigma$, depending on the combination of datasets considered.
This level of significance is not trivially dismissed, even if it does not yet rise to the $5\sigma$ threshold commonly accepted for establishing new physics. The variations in the statistical significance within this range with the choice of SNe sample also highlight the importance of calibrating the low-to-intermediate redshift SNe distance scale.

The constraining power of SNe in measuring the equation of state comes primarily from the comparison of low-redshift ($z<0.1$) and high redshift ($z>0.1$) supernovae. For supernovae at $z>0.1$, which partially overlap the redshift range of DESI, the \lcdm\ model that best fits the DESI data is also a good fit to the SNe data. Relative to models that best fit each of the DESY5, Union3 and Pantheon+ SNe samples alone, over the full redshift range, the DESI best-fit model gives only small shifts in the quality of the fit to the SNe data, with $\Delta\chi^2=-1.2$, 1.5 and 2.3 respectively. Unfortunately, no SNe compilation yet exists in which objects from both redshift regimes were collected as part of a uniform observational program. The Pantheon+ and Union3 datasets are compilations of SNe observed in many individual datasets, each with different calibrations and selection functions. Even for the DESY5 dataset, while 1635 SNe with $z>0.1$ come from the DES survey program with homogeneous calibration, 194 SNe at $z<0.1$ are drawn from a mix of historical observational programs with the best-controlled calibrations. The consistent calibration and processing of these data is therefore a delicate operation. Recently, \cite{Efstathiou:2024xcq} noted an offset in the differences between the standardized brightnesses of low-$z$ and high-$z$ SNe for objects in common between the DESY5 and Pantheon+ samples; however, \cite{DES:2025_Efstathiou_response} explain the causes for this and argue that the differences are well justified. 

If we exclude the low-z sample entirely and take only the DESY5 SNe, which is the most uniformly calibrated sample, naturally the constraining power and thus the statistical significance of the preference for evolution is reduced, but the best-fit values of $w_0$ and $\wa$ remain far from \lcdm\ (central panel of \cref{fig:w0wa_desi_with_variations}). Even if SNe at all redshifts are excluded altogether, the statistical significance of the preference from DESI+CMB alone still exceeds $3\sigma$. Future cosmology analyses of homogeneous SNe samples from ZTF \cite{ZTFSNDR2,ZTFoverview} and LSST \cite{LSSTWFD, LSSTDD} may shed further light on the relative calibration across redshifts.

Using the simple early-Universe prior on $(\theta_\ast,\ob,\obc)$ that marginalizes over information dependent on late-time models in addition to DESI gives very similar posterior constraints on $w_0$ and $\wa$ to the full DESI+CMB combination (left panel of \cref{fig:w0wa_desi_with_variations}), and the central $w_0$, $\wa$ values are very stable. However, the tension with \lcdm\ drops somewhat (from 3.1$\sigma$ to $2.4\sigma$) when using these simple priors instead of the full CMB: this is because the (\Planck+ACT) CMB lensing likelihood makes a sizable contribution to the $\dchisq$ calculation.

Although recent independent results from SPT \cite{Ge:2024SPT} and ACT \cite{Madhavacheril:ACT-DR6}  find cosmological parameter values that are consistent with those from \Planck, another possibility is that the CMB constraints used here suffer from some unknown systematic error. It is therefore interesting to consider the constraints that can be obtained entirely independent of the high-redshift CMB anchor. As argued in \cref{sec:understanding_DE}, the primary role of this CMB anchor is that it limits the freedom to vary $\Om$ when fitting to the low-redshift BAO and SNe distance-redshift measurements. This can be achieved instead by replacing the CMB with the DESY3 (3$\times$2pt) information to obtain a constraint coming entirely from low-redshift cosmological probes. As shown in the right panel of \cref{fig:w0wa_desi_with_variations}, this combination favors the same region of parameter space, although with somewhat larger uncertainties. 

We also consider the possibility of an undetected systematic error in our BAO measurements that would have evaded the many internal checks we have performed (cf. details in \cite{DESI2024.III.KP4,Y3.clust-s1.Andrade.2025,DESI.DR2.BAO.lya}). For instance, a coherent systematic shift of 1.5\% to all the measured $\alpha_{\rm iso}$ values (\cref{eqn:alpha_defs}), applied in a redshift-independent manner to all tracers, would decrease the tension between DESI and \Planck\ in the \lcdm\ model, and thus decrease the evidence for evolving dark energy.\footnote{Such a shift would result in the DESI contours in \cref{fig:lcdm_bao_constraints} shifting $\sim1.5\%$ to the left and thus overlapping the \Planck\ CMB results in \lcdm. However, any such hypothetical shift could not be explained by a simple change in the physical value of $\rd$, as that would also shift the CMB constraints.}~However, the required magnitude of such a hypothetical shift exceeds the total systematic error budget indicated by our tests by a factor of 6 (for QSO) to 10 (for BGS and LRGs), and we consider this highly unlikely.

Physical models that can satisfactorily explain the data without requiring dark energy evolution are more challenging. \cref{tab:cosmo_constraints} confirms the finding, previously reported in \cite{DESI2024.VI.KP7A}, that allowing the curvature $\Ok$ to vary freely, while increasing uncertainties, does not shift the posterior in $w_0$ and $\wa$ towards \lcdm. While we have fixed $\sumnu=0.06$ eV for our fiducial analysis, it is allowed to vary in \cref{sec:constraints_neutrinos} below, and we find that this has almost no effect on the dark energy constraints. This is because fitting the CMB power spectrum imposes a positive correlation between $\sumnu$ and $\Om$, thus given a physical prior that $\sumnu>0$ eV, in most of the available parameter space neutrino masses can only increase the CMB value of $\Om$, which exacerbates rather than relieves the tension with DESI BAO, and in the region where $0\leq\sumnu\leq0.06$ eV they have very limited effect. However, the more exotic scenario where $\sumnu$ is treated as an `effective' neutrino mass such that values $\sum m_{\nu,{\rm eff}}<0$ eV are allowed \cite{Green24,Craig24,Elbers24} can allow CMB fits to be consistent with lower $\Om$. This has the effect of improving the \lcdm\ fit and thus reducing the preference for evolving dark energy from DESI+CMB+SNe, but not below the level that is found from the relevant DESI+SNe combination alone without any CMB information (which can be up to 3.3$\sigma$, \cref{tab:dchisq}), as shown in \cref{sec:constraints_neutrinos}. That is, allowing negative effective neutrino masses still does not fully remove the preference for non-cosmological constant dark energy. At the same time, such dark energy models do permit  $\sum m_{\nu,{\rm eff}}>0$ values that are consistent with terrestrial experiments. More standard extensions, such as varying the number of effective neutrino species, $\Neff$, have no effect on the dark energy results.

Arguably, the least contorted way to reconcile our findings with \lcdm\ is to exclude all constraints from SNe data (on the grounds that the significance depends on the choice of SNe dataset and we have no \emph{a priori} reason to choose one over the others), and then dismiss the remaining $3.1\sigma$ discrepancy with DESI+CMB (or $2.7\sigma$ when also not including CMB lensing) as a statistical fluke.

\section{Neutrinos} 
\label{sec:constraints_neutrinos}

Cosmological observations are sensitive to the sum of the neutrino masses. However, this measurement depends on the choice of a cosmological model. So far, our baseline analysis has assumed the sum of neutrino masses to be fixed to the minimum value allowed by terrestrial experiments under a normal mass ordering, $\sumnu=0.06$ eV, modeled as a single massive and two massless states. In this section we allow $\sumnu$ to vary. When doing so,  we assume by default three degenerate mass eigenstates except when testing specific physical mass ordering scenarios. This model produces a good approximation of the observable effects of both normal and inverted physical mass ordering scenarios and, in the case of a positive detection of neutrino masses, would recover the correct $\sumnu$ value without bias \cite{Lesgourgues:2006,CORE:2018}. We also impose a minimal physical (uniform) prior $\sumnu>0$ eV, while noting that scenarios with negative `effective' neutrino mass have also recently been discussed (e.g., \cite{Green24,Craig24,Elbers24}). Supporting paper \cite{Y3.cpe-s2.Elbers.2025} explores the implications of different modelling assumptions in greater detail.

\begin{figure}
    \centering
    \includegraphics[width=\columnwidth]{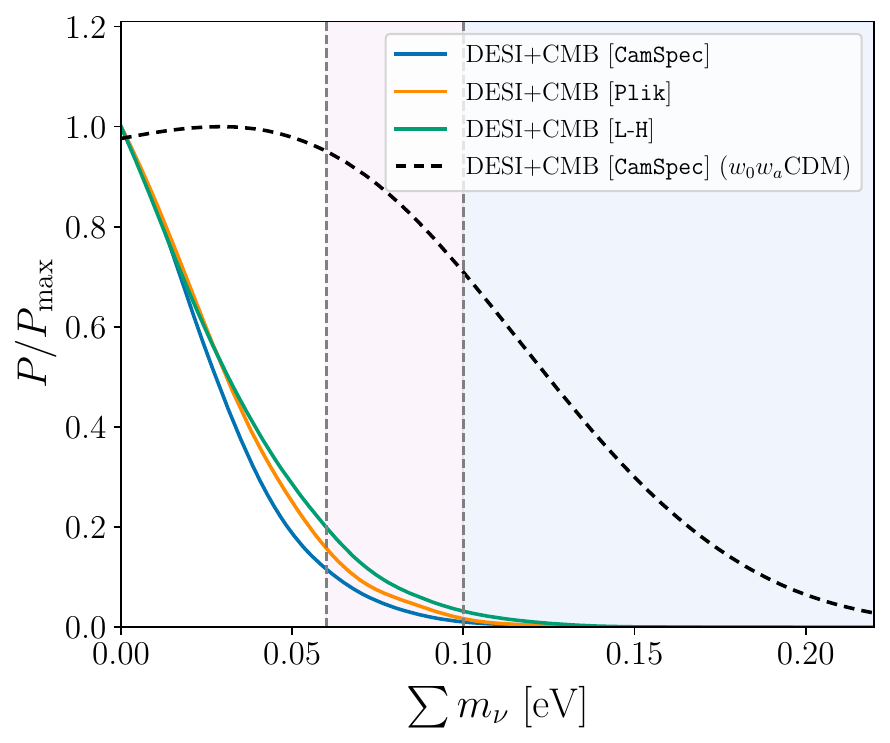}
    \caption{1D marginalized posterior constraints on $\sumnu$ from DESI DR2  BAO measurements combined with different CMB likelihoods, assuming the \lcdm +$\sumnu$ model. We show the 1D posteriors for the \texttt{CamSpec} CMB likelihood (leading to the tightest constraint) as well as the \texttt{Plik} and \texttt{L-H} CMB likelihoods. We also show the posterior for the \wowacdm +$\sumnu$ model, using DESI and the \texttt{CamSpec} CMB. Other models and datasets are presented in Table~\ref{tab:neutrino_constraints}. The vertical dashed lines and shaded regions indicate the minimum allowed $\sumnu$ values for (from left to right) the normal and inverted mass ordering scenarios, respectively.}
    \label{fig:neutrino_mass_constraints}
\end{figure}

\begin{table*}
    \centering
    \resizebox{\linewidth}{!}{
\begin{tabular}{lcccccc}
\toprule
Model/Dataset & $\Om$ & $H_0$ [km s$^{-1}$ Mpc$^{-1}$] & $H_0 r_\mathrm{d}$ [100 km s$^{-1}$] & $\sum m_\nu$ [eV] & $w$ or $w_0$ & $w_a$ \\
\midrule
$\bm{\Lambda}$\textbf{CDM+}$\bm{\sum m_\nu}$ &  &  &  &  &  &  \\
DESI BAO+CMB [\texttt{Camspec}] & $0.3009\pm 0.0037$ & $68.36\pm 0.29$ & $100.96\pm 0.48$ & $< 0.0642$ & --- & --- \\
DESI BAO+CMB [\texttt{L}-\texttt{H}] & $0.2995\pm 0.0037$ & $68.48\pm 0.30$ & $101.16\pm 0.49$ & $< 0.0774$ & --- & --- \\
DESI BAO+CMB [\texttt{Plik}] & $0.2998\pm 0.0038$ & $68.56\pm 0.31$ & $101.09\pm 0.50$ & $< 0.0691$ & --- & --- \\
\hline
$\bm{w}$\textbf{CDM+}$\bm{\sum m_\nu}$ &  &  &  &  &  &  \\
DESI BAO+CMB & $0.2943\pm 0.0073$ & $69.28\pm 0.92$ & $102.3\pm 1.3$ & $< 0.0851$ & $-1.039\pm 0.037$ & --- \\
DESI BAO+CMB+Pantheon+ & $0.3045\pm 0.0051$ & $67.94\pm 0.58$ & $100.35\pm 0.84$ & $< 0.0653$ & $-0.985\pm 0.023$ & --- \\
DESI BAO+CMB+Union3 & $0.3047\pm 0.0059$ & $67.93\pm 0.69$ & $100.33\pm 0.99$ & $< 0.0649$ & $-0.985\pm 0.028$ & --- \\
DESI BAO+CMB+DESY5 & $0.3094\pm 0.0049$ & $67.34\pm 0.53$ & $99.49\pm 0.78$ & $< 0.0586$ & $-0.961\pm 0.021$ & --- \\
\hline
$\bm{w_0w_a}$\textbf{CDM+}$\bm{\sum m_\nu}$ &  &  &  &  &  &  \\
DESI BAO+CMB & $0.353\pm 0.022$ & $63.7^{+1.7}_{-2.2}$ & $93.8^{+2.5}_{-3.2}$ & $< 0.163$ & $-0.42^{+0.24}_{-0.21}$ & $-1.75\pm 0.63$ \\
DESI BAO+CMB+Pantheon+ & $0.3109\pm 0.0057$ & $67.54\pm 0.59$ & $99.62\pm 0.86$ & $< 0.117$ & $-0.845\pm 0.055$ & $-0.57^{+0.23}_{-0.19}$ \\
DESI BAO+CMB+Union3 & $0.3269\pm 0.0088$ & $65.96\pm 0.84$ & $97.3\pm 1.2$ & $< 0.139$ & $-0.674\pm 0.090$ & $-1.06^{+0.34}_{-0.28}$ \\
DESI BAO+CMB+DESY5 & $0.3188\pm 0.0058$ & $66.75\pm 0.56$ & $98.43\pm 0.83$ & $< 0.129$ & $-0.758\pm 0.058$ & $-0.82^{+0.26}_{-0.21}$ \\
\bottomrule
\end{tabular}  }
    \caption{Cosmological parameter constraints where the neutrino mass parameter is allowed to vary assuming a $\sumnu>0$ prior. Additionally, we include models with more general dark energy backgrounds beyond $\Lambda$CDM. While we quote the 95\% upper limit for the neutrino mass parameter in eV units, we refer to the 68\% credible interval for the rest of the parameters. We quote the constraints for DESI and three different CMB likelihoods for $\Lambda$CDM+$\sumnu$; in all other rows the label `CMB' refers to use of the baseline \texttt{CamSpec} likelihood.}
    \label{tab:neutrino_constraints}
\end{table*}

Primordial neutrinos were copiously produced before the nucleosynthesis era and decoupled at a temperature of about $1\,\text{MeV}$; see, e.g., \cite{Lesgourgues:2006} for a review of neutrino cosmology. As the Universe expanded, neutrinos gradually lost kinetic energy, behaving as radiation in the early Universe and transitioning to non-relativistic matter around redshifts of $z \sim 100$ for realistic neutrino masses, thereafter influencing the late-time expansion history by contributing to the matter component. The main effect of massive neutrinos on the CMB is to impact the angular diameter distance to last scattering, which is degenerate with the effects of other cosmological parameters such as $\Om$ and $H_0$ (see, e.g., \cite{Loverde:2024} for a recent discussion). Neutrinos also affect the lensing of CMB anisotropies by suppressing the growth of structure below the free-streaming scale. BAO are not sensitive to the latter effect at all, and only probe the background geometry by constraining the total matter density $\Om$ and the parameter combination $H_0r_\mathrm{d}$, so DESI BAO alone cannot constrain the neutrino masses. Nevertheless by breaking geometrical degeneracies, BAO significantly enhance the ability of the CMB to constrain this parameter.

The upper limits on $\sumnu$ that we obtain from the combination of DESI and CMB depend on the particular choice of the CMB likelihood used, since the various likelihoods differ slightly in the amount of lensing power they infer from the lensed $TT$, $TE$ and $EE$ power spectra. This can be incorporated into a phenomenological parameter $A_{\rm L}$ that scales the model lensing power used to compute the lensed power spectra (but not the power reconstructed from the 4-point function), such that values $A_{\rm L}>1$ indicate an excess of lensing power, often referred to as the `$A_{\rm L}$ anomaly'. Increasing $\sumnu$ above 0.06 eV decreases the expected lensing power, while smaller values would boost it.  

The baseline CMB likelihood we have adopted through this paper combines the \texttt{CamSpec} likelihood \cite{Efstathiou:2021,Rosenberg:2022} together with the combined CMB likelihood from \Planck\ \cite{Carron:2022} and ACT \cite{Madhavacheril:ACT-DR6}, both of which are based on the PR4 data release. Combining these with the DESI DR2 BAO, and assuming a \lcdm\ background, we find the marginalized posterior limit
\onetwosig[5.5cm]{\sumnu < 0.064 \, {\rm eV}}{DESI+CMB}{. \label{eq:mnu_base_lcdm}}
Although the \texttt{CamSpec} likelihood appears to prefer a lower lensing power and reduces the $A_{\rm L}$ anomaly \cite{Rosenberg:2022}, the tighter constraints on other cosmological parameters degenerate with $\sumnu$ mean that this upper bound is tighter than the one obtained with the original \texttt{Plik} likelihood:
\onetwosig[5.5cm]{\sumnu < 0.069 \, {\rm eV}}{DESI+CMB[\texttt{Plik}]}{. \label{eq:mnu_plik_lcdm}}
The result of \cref{eq:mnu_plik_lcdm} is directly comparable to the result obtained from BAO and CMB in the DESI DR1 analysis \cite{DESI2024.VI.KP7A,DESI2024.VII.KP7B}, which used the \texttt{Plik} likelihood as the default and obtained the corresponding constraint $\sumnu<0.082$ eV (95\%).\footnote{Note that \cite{DESI2024.VI.KP7A} originally reported a slightly tighter constraint due to a bug in the version of the ACT lensing likelihood used in that analysis. This was corrected in \cite{DESI2024.VII.KP7B}, where the quoted result was obtained using \texttt{v1.2} of the ACT likelihood code that is also used here.} This close to $20\%$ reduction in the upper bound reflects the higher precision of the DESI DR2 BAO data compared to DR1.

\begin{figure}
    \centering
    \includegraphics[width=\columnwidth]{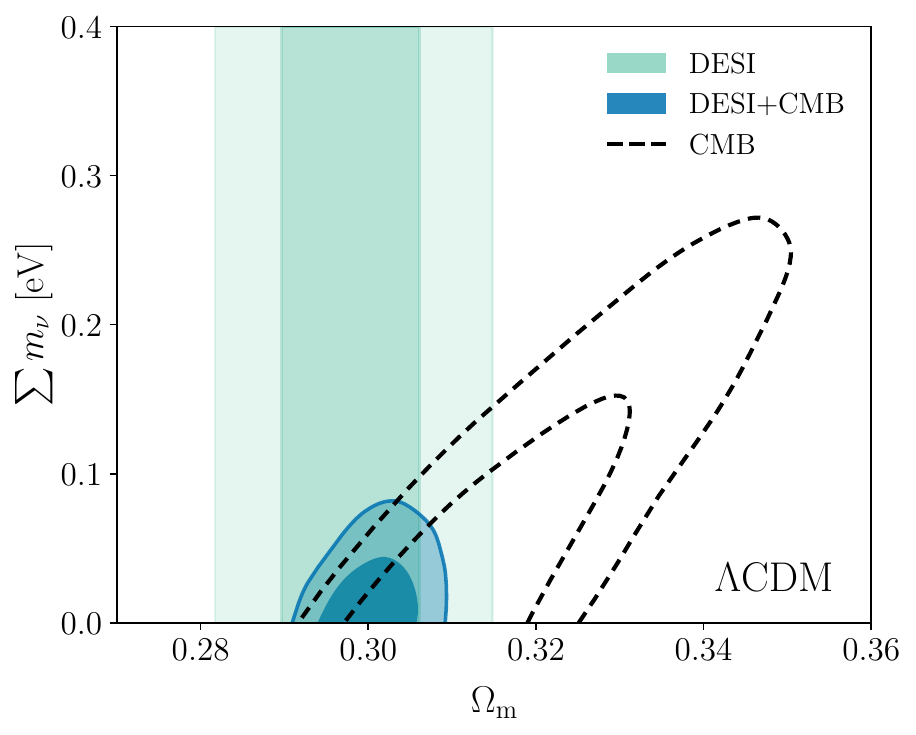}
    \caption{68\% and 95\% confidence contours for the sum of the neutrino masses and $\Om$ in \lcdm. The constraints assume a prior $\sumnu>0$ eV. The CMB constraints show a high degree of correlation between $\sumnu$ and $\Om$. DESI BAO constraints are insensitive to the neutrino mass parameter but measure $\Om$, thus helping to break the geometric degeneracy and tightening the upper bound on $\sumnu$. The constraint on $\sumnu$ from DESI+CMB is particularly tight because of the DESI preference for lower $\Om$ values.
    }
    \label{fig:neutrinos_lcdm}
\end{figure}

 \begin{figure*}
    \centering
    \includegraphics[width=\textwidth]{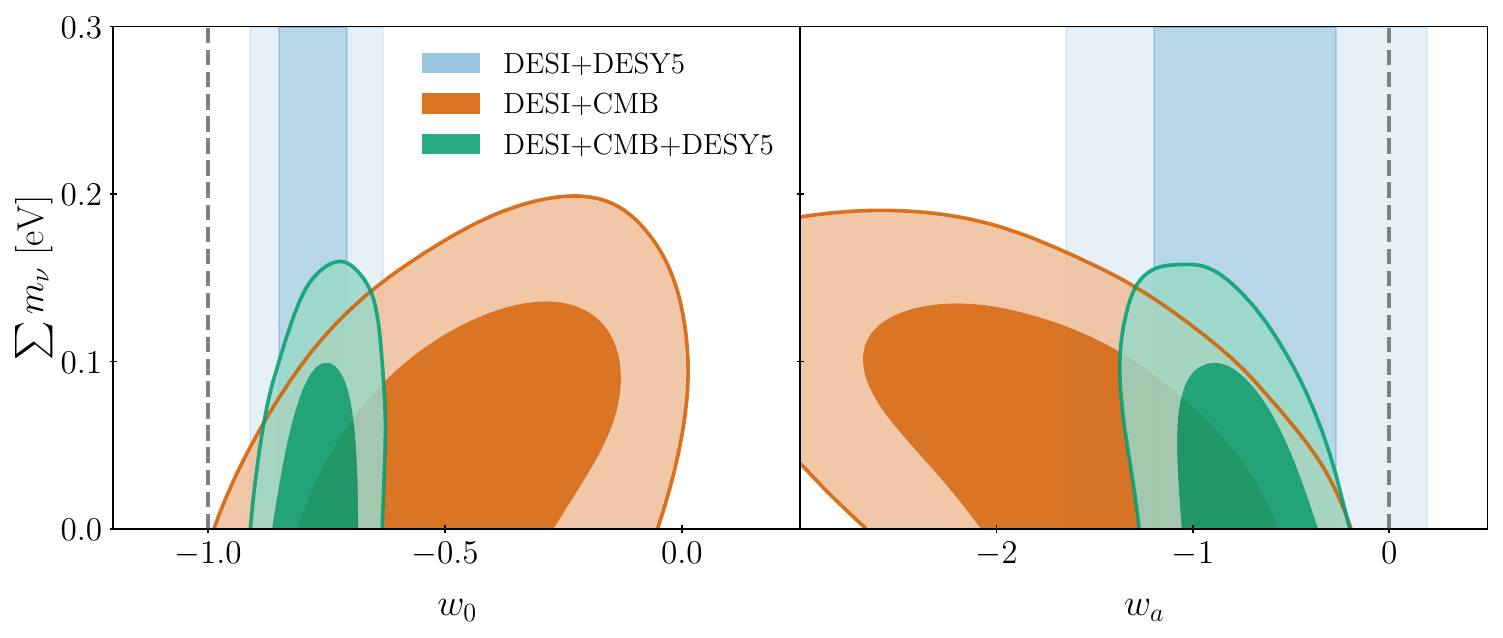}
    \caption{Constraints on the sum of the neutrino masses assuming a \wowacdm\ model. All contours enclose 68\% and 95\% of the posterior probability, and all cases use the prior $\sumnu>0$ eV. The gray dashed lines represent the \lcdm\ limit. We show the degeneracy between $\sumnu$ and the dark energy parameters in the DESI+CMB contours (dark orange contours) and the constraints from the combination DESI+CMB+DESY5. The vertical blue bands show the geometrical constraints from DESI+DESY5, that do not measure $\sumnu$ but break degeneracies helping to further limit the upper bound on $\sumnu$. An extended plot allowing an effective neutrino mass parameter that can take negative values is shown in the companion paper \cite{Y3.cpe-s2.Elbers.2025}.}
    \label{fig:neutrinos_w0wa}
\end{figure*}

The use of the alternative \texttt{LoLLiPoP} and \texttt{HiLLiPoP} (\texttt{L-H}) likelihoods of \cite{Tristram:2023} also show a lower excess lensing power than \texttt{Plik}. In this case, this effect is also reflected in significantly looser upper bounds on the neutrino mass,
\onetwosig[5.5cm]{\sumnu < 0.077 \, {\rm eV}}{DESI+CMB[\texttt{L-H}]}{. \label{eq:mnu_LH_lcdm}}
\cref{fig:neutrino_mass_constraints} shows the marginalized 1D posterior constraints on $\sumnu$ for the combination of DESI with each of the CMB likelihoods.

Irrespective of which CMB likelihood is used, all three upper bounds are very close to the lower bound $\sumnu>0.059$ eV set by neutrino oscillation experiments \cite{NuFitV6}. The posteriors shown in \cref{fig:neutrino_mass_constraints} all resemble the tails of distributions that would peak at negative neutrino masses if this were allowed by the prior. This behavior has been explored in a few recent studies \cite{Green24,Noriega:2024lzo,Elbers24,Naredo-Tuero:2024sgf}, as well as in our companion work \cite{Y3.cpe-s2.Elbers.2025}, which also shows that it holds for profile likelihoods based on extrapolating the $\chi^2$ likelihood surface to $\sumnu<0$ eV. In the context of a model with effective neutrino masses that are allowed to assume negative values, the DESI+CMB likelihood peaks at $\sum m_{\nu,{\rm eff}}<0$, and the tension with the minimal mass bound from terrestrial experiments increases to $3.0\sigma$ \cite{Y3.cpe-s2.Elbers.2025}.

This discrepancy can be related to the tension between DESI and CMB results for the parameters $\Om$ and $\Hrd$ in a \lcdm\ background discussed in \cref{sec:cosmologcal_constraints} above, and particularly in \cref{fig:lcdm_bao_constraints}. For fixed angular acoustic scale $\theta_\ast$, the effects on the CMB of changing $\sumnu$ and changing $\Om$ are positively correlated, as can be seen from the dashed black contour in \cref{fig:neutrinos_lcdm} (similar degeneracies can also be observed between $\sumnu$ and $\Hrd$, but with a negative correlation). In \lcdm, DESI tightly constrains $\Om$, as shown by the vertical shaded contour, to a value close to the low-end tail of the marginalized posterior distribution for CMB measurements. Consequently, the combination of DESI and CMB rules out all but the smallest neutrino masses. This result is thus related to the preference of DESI for lower values of $\Om$ compared to the CMB discussed in the previous Sections.

As the upper limits on $\sum m_\nu$ from cosmology approach the lower limits from neutrino oscillations, the distinction between the different mass orderings becomes increasingly relevant. The baseline constraint \cref{eq:mnu_base_lcdm}, obtained assuming a $\sum m_\nu>0$ prior, already appears to rule out the inverted ordering. However, to determine the evidence in support of the normal ordering, one should also account for the fact that much of the posterior volume violates the constraints for both mass orderings. To do this we use a physical model for the total mass $\sumnu$ in terms of lightest neutrino mass $m_l\geq0$, which includes information on the mass squared splittings $\Delta m_{21}^2$ and $|\Delta m_{31}^2|$ and allows for either mass ordering, as described in \cite{Loureiro:2019}. We then determine the upper bounds on this model from a combined analysis of DESI, CMB (with the baseline \texttt{CamSpec} likelihood) and 3-flavor oscillation constraints from NuFit-6.0 \cite{NuFitV6}, assigning equal prior probabilities to either mass ordering. The result obtained is
\begin{equation}
    \begin{split}
    \sumnu < 0.112 \, {\rm eV}\quad \text{(95\%, DESI+CMB\dataplus} & \\ \text{NuFit-6.0)}.
    \end{split}
    \label{eq:NuFit result}
\end{equation}
The difference between the limits in \cref{eq:mnu_base_lcdm} and \cref{eq:NuFit result} arises because the former only requires $\sumnu > 0$ while the latter incorporates oscillation data that effectively require $\sumnu > 0.06\eV$, therefore pushing the upper limit into territory disfavored by the DESI+CMB data alone.
In the supporting paper \cite{Y3.cpe-s2.Elbers.2025}, we also compare the two mass orderings and find a Bayes factor of $K=10$ providing strong evidence in support of the normal mass ordering under the assumption of the \lcdm +$\sumnu$ model.

Given the discrepancy between DESI and CMB under the \lcdm\ model, and the strong preference for a \wowacdm\ model discussed in \cref{sec:de_constraints}, we consider the effect of allowing the equation-of-state parameters $w_0$ and $\wa$ to vary. In this scenario and assuming a $\sumnu>0$ prior, we find
\begin{equation}
    \begin{split}
    \sumnu < 0.163 \, {\rm eV}\quad \text{(95\%, \wowacdm: DESI+}  & \\ \text{CMB),} 
    \end{split}
\end{equation}
from combining DESI with CMB, and 
\begin{equation}
    \begin{split}
    \sumnu < 0.129 \, {\rm eV}\quad \text{(95\%, \wowacdm: DESI+} & \\ \text{CMB+DESY5)}
    \end{split}
\end{equation}
when also using DESY5 SNe to better constrain the equation of state of dark energy. More details of parameter posteriors in this scenario, including those obtained with other data combinations and the \wcdm +$\sumnu$ model, are presented in \cref{tab:neutrino_constraints}. Through comparison to \cref{tab:cosmo_constraints}, we find that allowing $\sumnu$ to vary with a conventional physical prior $\sumnu>0$ eV has little effect on the $w(z)$ constraints, but that allowing $w_0$ and $\wa$ to vary significantly relaxes the neutrino mass bound in all cases. Moreover, in the case of DESI and CMB, the peak of the 1D marginalized posterior is recovered in the positive mass range, as shown in \cref{fig:neutrino_mass_constraints}.

In \lcdm, BAO measurements from DESI set a geometric constraint on $\Om$ (or, equivalently, on $\Hrd$) that helps break the degeneracy in the CMB results for $\sumnu$. In the \wowacdm\ model, these geometric measurements primarily help by constraining the background $w_0$ and $\wa$ values, thus limiting the available $\sumnu$ range from the CMB alone, which is itself weaker than in \lcdm\ due to the additional freedom in the expansion history. This effect is even clearer when also adding SNe to DESI and CMB, as the constraints on the background become significantly tighter. This is illustrated in \cref{fig:neutrinos_w0wa}, where the vertical blue bands show the purely geometrical constraints from DESI+DESY5. An extension of this plot where these degeneracies extend further in parameter space by allowing for negative neutrino masses through and effective parameter $\sum m_{\nu,{\rm eff}}>0$ is shown in \cite{Y3.cpe-s2.Elbers.2025}.

Finally, constraints on the number of effective relativistic degrees of freedom, $\Neff$, are not discussed in detail here, but are consistent with the particle physics value $N_{\rm eff}=3.044$ as described in our supporting paper \cite{Y3.cpe-s2.Elbers.2025}.

\section{Conclusions}
\label{sec:conclusions}

We have presented BAO measurements from over 14 million discrete galaxy and quasar tracers drawn from the first three years of operation of DESI and which will be included in the second data release (DR2). These results use samples of nearby bright galaxies, LRGs, ELGs and quasars over the redshift range $0.1<z<2.1$, and cover a cumulative effective volume of over $42\;{\rm Gpc}^3$. Complementary BAO measurements from correlations of the \lya\ forest and high-redshift quasars at effective redshift $z=2.33$ are presented in the companion paper \cite{DESI.DR2.BAO.lya}.

Our BAO analysis largely follows the methods used for the previous DESI DR1 analysis and presented in \cite{DESI2024.III.KP4}, but with improved statistical precision as the effective volume of the data has increased by more than a factor of two. Some particular differences include the use of a fainter limiting magnitude cut to define the bright galaxy sample, resulting in a higher number density, and the inclusion of quadrupole information in BAO fits to the quasar sample. As for the DR1 analysis, we applied a strict catalog-level blinding to our data while initial data checks were carried out and the analysis pipeline was being finalized. The validation tests and the criteria that were required to be met before the data were unblinded are described in detail in the supporting publication \cite{Y3.clust-s1.Andrade.2025}.

The final BAO results provide a precision on the isotropic distance scale measurement of $\DVrd$ that ranges from 1.54\% (for the QSO) down to just 0.45\% for our most constraining composite \lrgelg\ sample; other than for the QSO, the precision is sub-percent for every tracer and redshift range. We also obtained a precision of a few percent on the measurement of the Alcock-Paczy\'nski distance ratio for every tracer except the lowest redshift BGS. Together these results are the most precise BAO measurements ever made at all redshifts covered, including at $z<0.8$ where DESI DR2 now greatly exceeds the precision of the Sloan Digital Sky Survey (SDSS). 

The DR2 results are very consistent with those previously reported from the DR1 data that form a subset of DR2. The $p$-value determined from the Kolmogorov-Smirnov test for the distribution of the differences is 0.40. The DR2 results are also consistent with SDSS. \cite{DESI2024.III.KP4} previously reported a $\sim3\sigma$ discrepancy in the measurement of the transverse BAO scale $\DMrd$ in the \lrgt\ sample at effective redshift $z_{\rm eff}=0.706$ compared to the equivalent result from SDSS. For DR2, the difference has decreased in significance, lying within the range $1.5\sigma$ to $2.6\sigma$, depending on assumptions about the degree of correlation between the two samples.

The combination of all BAO measurements from DR2 is well fit by a flat \lcdm\ cosmological model with matter density parameter $\Om=0.2975\pm0.0086$ and product of the scaled Hubble constant and sound horizon at the drag epoch $h\rd=(101.54\pm0.73)$ Mpc. This represents a 40\% improvement in precision compared to the equivalent results from DR1, but with excellent consistency between the two. However, while a \lcdm\ model provides a good fit to DESI data, the model parameters obtained from this fit are now in 2.3$\sigma$ tension with those derived from the CMB, increased from 1.9$\sigma$ in DR1. This tension is present despite DESI being consistent with the acoustic angular scale $\theta_*$ measured by CMB. 

When calibrated with an external prior on $\Ob h^2$ from BBN, our BAO measurements correspond to a value $H_0=(68.50\pm0.58)\;\kmsMpc$ in \lcdm, a value that is independent of any information on CMB anisotropies. In the $\Om$-$H_0$ plane, the results from BBN-calibrated BAO are now more discrepant with those from CMB; the offset of the results is again along the $\Om h^3={\rm constant}$ degeneracy direction of the CMB that is determined by the very precisely measured acoustic angular scale $\theta_\ast$. Combining DESI BAO with BBN and the \Planck\ $\theta_\ast$ result gives $H_0=(68.45\pm 0.47)\;\kmsMpc$, a 0.7\% precision measurement competitive with that from the CMB itself, and in strong disagreement with SH0ES \cite{Riess:2021jrx}.

Within the \lcdm\ framework, DESI results are also somewhat in tension with the high $\Om$ values preferred by SNe datasets, which---contrary to DESI---prefer larger $\Om$ than \Planck. While not individually rising to the $3\sigma$ significance threshold, these results point to an incompatibility between different cosmological datasets when interpreted in the \lcdm\ model. Interestingly, the relative levels of $\Om$ and $H_0$ values currently measured by SN, DESI and CMB datasets match what would be expected if data from a true evolving dark energy model were analyzed in the restrictive \lcdm\ model, as pointed out recently by \cite{Tang:2024}.

Assuming a \lcdm\ background, the combination of DESI and CMB data give the tightest upper bound on the neutrino mass sum to date, $\sumnu<0.064$ eV (95\% limit) in our baseline analysis. Although this relaxes to $\sumnu<0.078$ eV when using an alternative \Planck\ likelihood \cite{Tristram:2023}, both results are approaching the lower bound set by terrestrial neutrino oscillation experiments, $\sumnu\geq0.059$ eV. Indeed, if the model is extended to allow for negative `effective' neutrino masses, most of the posterior mass and the peak of the likelihood lie at $\sum m_{\nu,{\rm eff}}<0$ eV, another possible sign of growing tensions within the \lcdm\ model with DESI DR2 \cite{Y3.cpe-s2.Elbers.2025}.

The evidence for a departure from \lcdm\ in the form of evolving dark energy has increased with the DR2 BAO data. Comparing the evolving dark energy model parametrized by $w(a)=w_0+\wa(1-a)$ to \lcdm, we find a $3.1\sigma$ evidence in favor of dynamical dark energy from DESI+CMB alone. When we add the recent Pantheon+, Union3 or DESY5 SNe datasets to this combination, the preference for \wowacdm\ over \lcdm\ is $2.8\sigma$, $3.8\sigma$ or $4.2\sigma$ respectively, with all three giving results for $w_0$ and $\wa$ consistent with each other within their 68\% credible regions. The preferences for \wowacdm\ under these dataset combinations have increased compared to our DR1 results in \cite{DESI2024.VI.KP7A}. All combinations of available datasets we study favor mutually consistent results with $w_0>-1$ and $\wa<0$, apparently indicating a weakening dark energy today and a phantom crossing at some point in the past (in this parametrization). The supporting paper \cite{Y3.cpe-s1.Lodha.2025} examines this behavior using a wider range of models and non-parametric methods. We note that the degeneracy direction for the constraints in the $w_0$-$\wa$ plane approximately points towards the \lcdm\ solution, although this is neither exact nor consistent between fits to different combinations of datasets. This constitutes a weak coincidence as there is no \emph{a priori} reason for the best-fit values of $w_0$ and $\wa$ to have the values that lead to this observation.

In \cref{sec:understanding_DE,sec:robustness_DE}, we examined the contributions to the preference for evolving dark energy in detail. For DESI+CMB, the discrepancy in preferred $\Om$ values in \lcdm\ is resolved in \wowacdm, leading to the $3.1\sigma$ ($\dchisq=-12.5$) preference for the latter. If we omit CMB lensing the best-fit \wowacdm\ parameters are unchanged, though the significance drops to $2.7\sigma$ ($\dchisq=-9.7$). If we also marginalize over any late-time dependence in the CMB and only impose a prior on $(\theta_\ast,\ob,\obc)$, there is a further slight reduction to $2.4\sigma$ ($\Delta\chi^2_\mathrm{MAP} = -8.0$). The \wowacdm\ and `compromise' \lcdm\ models predict different evolution of the distance-redshift relation at $z<0.4$, with a $\sim 2\%$ difference in distances at $z<0.1$. BAO do not provide high precision in this regime, but supernovae do, and the DESY5 and Union3 samples clearly favor the \wowacdm\ prediction. However, the distance-redshift evolution in the Pantheon+ analysis is closer to the \lcdm\ prediction, so adding Pantheon+ to DESI+CMB does not strengthen the evidence for \wowacdm. \cref{appendix:cmb_compression} shows that our results are robust to changes in the choice of CMB likelihood. We also show that {\it replacing} the CMB likelihood with the weak gravitational lensing constraints from the DESY3 (3$\times$2pt) analysis retains a clear preference for evolving dark energy, with, e.g., a $\gtrsim3\sigma$ preference for \wowacdm\ from DESI+DESY3 (3$\times$2pt)+DESY5 SNe. Removing the preference for \wowacdm\ through changes to the DESI measurements themselves would require systematic errors to be far larger than any found in our tests.

In the \wowacdm\ background model, when the neutrino mass scale is allowed to vary with physical non-negative prior bounds, the upper limit on $\sumnu$ is significantly relaxed to $\sumnu<0.16$ eV (95\%, from DESI+CMB) or $\sumnu<0.13$ eV (95\%, from DESI+CMB+DESY5, the middle result from the three possible SNe samples), while the constraints on $w_0$ and $\wa$ do not materially change from the case where $\sumnu$ is fixed. These limits are entirely consistent with neutrino oscillation experiment results. The supporting paper \cite{Y3.cpe-s2.Elbers.2025} shows that even when allowing for effective masses $\sum m_{\nu,{\rm eff}}$ with a wide prior that allows negative values, the marginalized posterior in \wowacdm\ from DESI+CMB is consistent with the oscillation lower limit and its peak lies in the positive region. 

In \cite{DESI2024.VI.KP7A} we characterized the results from DESI DR1 and external datasets as providing ``a tantalizing suggestion of deviations from the standard cosmological model". The DR2 data presented here have sharpened this evidence, although significance levels still vary depending on the external data used---particularly the choice of SNe sample---and no combination exceeds $4.2\sigma$. Nevertheless, it is becoming clear that unless some unidentified systematic error affects one or several of the different cosmological datasets used, the challenge to the \lcdm\ model has increased. Sharper measurements from future DESI analyses and from other experiments will show whether these challenges to the standard cosmological model herald yet another radical transformation in our understanding of the evolution and energy content of the Universe.

\section{Data Availability}
The data used in this analysis will be made public along with Data Release 2 (details in \url{https://data.desi.lbl.gov/doc/releases/}). BAO likelihoods for DESI DR2, integrated in the \texttt{cobaya} code, are available at \url{https://github.com/CobayaSampler/bao_data}. The data points corresponding to the figures from this paper are available in Ref. \cite{DR2KPZenodo}.

\acknowledgments

This material is based upon work supported by the U.S.\ Department of Energy (DOE), Office of Science, Office of High-Energy Physics, under Contract No.\ DE–AC02–05CH11231, and by the National Energy Research Scientific Computing Center, a DOE Office of Science User Facility under the same contract. Additional support for DESI was provided by the U.S. National Science Foundation (NSF), Division of Astronomical Sciences under Contract No.\ AST-0950945 to the NSF National Optical-Infrared Astronomy Research Laboratory; the Science and Technology Facilities Council of the United Kingdom; the Gordon and Betty Moore Foundation; the Heising-Simons Foundation; the French Alternative Energies and Atomic Energy Commission (CEA); the National Council of Humanities, Science and Technology of Mexico (CONAHCYT); the Ministry of Science and Innovation of Spain (MICINN), and by the DESI Member Institutions: \url{https://www.desi. lbl.gov/collaborating-institutions}. 

The DESI Legacy Imaging Surveys consist of three individual and complementary projects: the Dark Energy Camera Legacy Survey (DECaLS), the Beijing-Arizona Sky Survey (BASS), and the Mayall z-band Legacy Survey (MzLS). DECaLS, BASS and MzLS together include data obtained, respectively, at the Blanco telescope, Cerro Tololo Inter-American Observatory, NSF NOIRLab; the Bok telescope, Steward Observatory, University of Arizona; and the Mayall telescope, Kitt Peak National Observatory, NOIRLab. NOIRLab is operated by the Association of Universities for Research in Astronomy (AURA) under a cooperative agreement with the National Science Foundation. Pipeline processing and analyses of the data were supported by NOIRLab and the Lawrence Berkeley National Laboratory. Legacy Surveys also uses data products from the Near-Earth Object Wide-field Infrared Survey Explorer (NEOWISE), a project of the Jet Propulsion Laboratory/California Institute of Technology, funded by the National Aeronautics and Space Administration. Legacy Surveys was supported by: the Director, Office of Science, Office of High Energy Physics of the U.S. Department of Energy; the National Energy Research Scientific Computing Center, a DOE Office of Science User Facility; the U.S. National Science Foundation, Division of Astronomical Sciences; the National Astronomical Observatories of China, the Chinese Academy of Sciences and the Chinese National Natural Science Foundation. LBNL is managed by the Regents of the University of California under contract to the U.S. Department of Energy. The complete acknowledgments can be found at \url{https://www.legacysurvey.org/}.

Any opinions, findings, and conclusions or recommendations expressed in this material are those of the author(s) and do not necessarily reflect the views of the U.S.\ National Science Foundation, the U.S.\ Department of Energy, or any of the listed funding agencies.

The authors are honored to be permitted to conduct scientific research on I’oligam Du’ag (Kitt Peak), a mountain with particular significance to the Tohono O’odham Nation.



\bibliographystyle{mod-apsrev4-2} 
\bibliography{Y1KP7a_references,references, DESI_supporting_papers}


\appendix

\section{Early-Universe priors and robustness to choice of CMB likelihoods}
\label{appendix:cmb_compression}

\begin{figure*}
    \centering
    \includegraphics[width=\columnwidth]{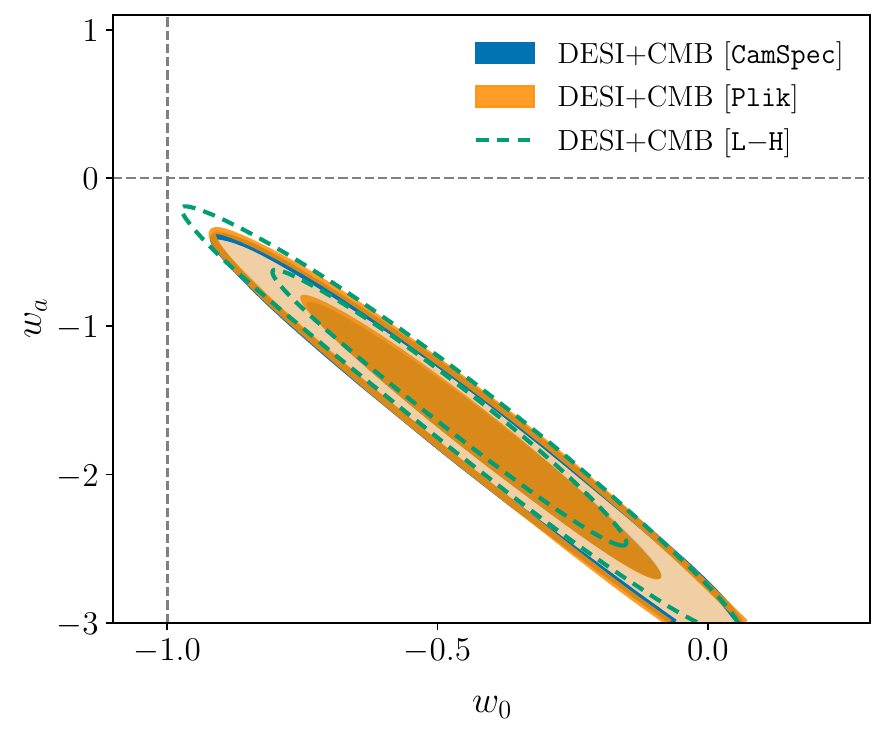}
    \includegraphics[width=\columnwidth]{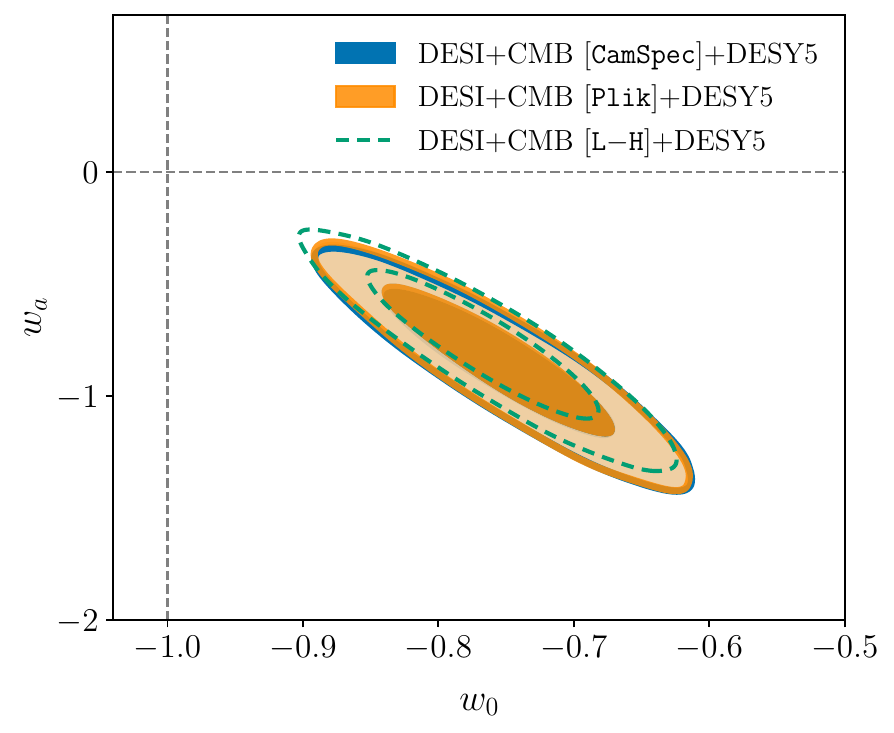}
    \caption{The effect of changing the CMB likelihood between our default \texttt{CamSpec}, \texttt{Plik}, and \texttt{L-H} (see \cref{subsec:CMB} for details), on results in the $w_0$-$\wa$ plane. \emph{Left}: Posteriors in the DESI+CMB case. \emph{Right}: The case when also combining with DESY5 SNe information. The \texttt{L-H} likelihood shows a small shift compared to the other two likelihoods, but this difference is suppressed when also including SNe data.
    }
    \label{fig:CMB_robustness}    
\end{figure*}

As discussed in \cref{sec:external_data}, for some purposes it is convenient to make use of a compression of the full CMB information into a multivariate correlated Gaussian prior on the quantities $\theta_\ast$, $\obc$ and $\ob$, which are early-Universe quantities that can be determined independently of assumptions about the late-time evolution by maginalizing over late-time effects such as the ISW effect and CMB lensing. This process was performed by \cite{EarlyUniverseCompression} based on the \texttt{CamSpec} CMB likelihood. We use their results to define a Gaussian prior with mean
\begin{equation}
\boldsymbol{\mu} (\theta_\ast, \ob, \obc) = 
\begin{pmatrix}
0.01041 \\
0.02223 \\
0.14208
\end{pmatrix}
\label{eq:CMB_compression_mean}
\end{equation}
and covariance
\begin{equation}
\mathbf{C} = 10^{-9} \times 
\begin{pmatrix}
0.006621 &  0.12444 & -1.1929 \\
0.12444 &  21.344 & -94.001 \\
-1.1929 & -94.001 & 1488.4
\end{pmatrix}\,.
\label{eq:CMB_compression_cov}
\end{equation}
The use of this prior has been indicated throughout the paper by the use of the $(\theta_\ast, \ob, \obc)_{\rm CMB}$ notation. As shown in the left panel of \cref{fig:w0wa_desi_with_variations} the posterior results for the \wowacdm\ model are remarkably similar in all cases when comparing fits to DESI+$(\theta_\ast, \ob, \obc)_{\rm CMB}$, DESI+CMB (with full CMB information) and DESI+CMB (no lensing), indicating that most of the information added by the CMB in this scenario comes from the anchoring constraint it provides on $\Om$, limiting the freedom of the model to absorb background expansion effects of $w_0$ and $\wa$ into other parameters. We also tested a separate compression of the CMB information based on the shift parameters $R$ and $l_A$ together with $\ob$ \cite{Shift_parameters_Wang_2007} and found that it gave very consistent results as well (see also \cite{Bansal:2025ipo}). Nevertheless, while the addition of CMB lensing information does not noticeably shift the $w_0$-$\wa$ posteriors, it does appreciably change the $\dchisq$ values and thus the significance of rejection of \lcdm.

\cref{fig:CMB_robustness} shows the differences in the constraints obtained from the DESI+CMB+DESY5 data combination in the \wowacdm\ case, arising from the use of the three different CMB likelihoods, namely \texttt{CamSpec} (our baseline in this paper), \texttt{Plik} and \texttt{L-H} (cf. \cref{sec:external_data} for descriptions). The results using \texttt{CamSpec} and \texttt{Plik} are remarkably consistent, though we see a small shift at the $0.1\sigma$ level when using \texttt{L-H} PR4.

\section{Robustness to assumption about correlation in BAO systematic errors}\label{sec:corr-sys}

Our baseline approach, described briefly in \cref{sec:bao_measurements} and in more detail in \cite{Y3.clust-s1.Andrade.2025}, assumes that the systematic error contributions to BAO measurements are not correlated with each other across redshift bins. As a robustness test, we evaluated the impact of introducing off-diagonal terms in the systematics covariance matrix of the BAO measurements. To do this, we assumed that the theoretical modelling systematic error contributions, which are estimated to be 0.1\% in $\aiso$ and 0.2\% in $\alpha_{\rm AP}$, are correlated across all redshift bins, with a correlation coefficient of 0.5. This leads to redshift-dependent correlations in the final covariance matrix. We found that including these correlations had no detectable effect on the cosmological constraints obtained in any model or combination of DESI with any external dataset. This result is unsurprising, since the systematic errors are subdominant to statistical errors in all redshift bins, so correlations in the systematic errors cannot play a large role.

\end{document}